\documentclass{PBCreport}
\usepackage{wrapfig}
\usepackage{booktabs}
\usepackage{afterpage}
\usepackage[T1]{fontenc}
\usepackage{textcomp}
\usepackage{color, colortbl}
\usepackage{booktabs}
\usepackage{afterpage}
\usepackage{tabularx}
\usepackage{siunitx}
\usepackage{hyperref}
\usepackage{graphicx,mathptmx,amsmath,amsfonts,amscd,amsthm,amssymb,eucal,psfrag,color,subfig,url,multirow,gensymb}
\usepackage{rotating}
\usepackage{soul}
\usepackage{appendix}
\usepackage{placeins}

\hypersetup{
     colorlinks = true,
     linkcolor = blue,
     anchorcolor = blue,
     citecolor = blue,
     filecolor = blue,
     urlcolor = blue
     }

\definecolor{LightCyan}{rgb}{0.88,1,1}
\definecolor{LightYellow}{rgb}{1,0.97,0.9}

\documentlabel{CERN-PBC-Notes-2026-001}
\begin{document}

\title{%A Long-Baseline Atom Interferometer in the Multi-Function Site of the Gotthard Base Tunnel at Sedrun - Environmental Measurement Results\\
Environmental Measurements in the Sedrun Access Shaft to the Gotthard Base Tunnel - a Promising Site for a Long-Baseline Atom Interferometer
}

\author{\parbox{\textwidth}{\it 
M. Guinchard$^{1,*}$, O.~Buchmueller$^{2,3}$,
S.~Calatroni$^{1}$, J.~Ellis$^{1,4}$, S.~Hoell$^{1}$,
M.~Jaussi$^{1}$, L.~Lombriser$^{5,6}$, M.~Pentella$^{1}$, D.~Thuliez$^{7}$, D.~Valuch$^{1}$} \
~~\\
~~\\
\small{$^1$ CERN, Geneva, Switzerland}\\
\small{$^2$ Imperial College London, United Kingdom}\\
\small{$^3$ University of Oxford, United Kingdom}\\
\small{$^4$ King’s College London, United Kingdom}\\
\small{$^5$ University of Applied Sciences of the Grisons, Switzerland}\\
\small{$^6$ University of Geneva, Switzerland}\\
\small{$^7$ FOSELEV Suisse, Switzerland}\\
\small{$^*$ Editor}}

\abstract{
%Atom interferometer (AI) experiments offer interesting prospects for searches for the interactions of ultralight bosonic dark matter with Standard Model particles as well as detection of gravitational waves in a frequency band inaccessible to experiments that are operating or under construction. Ideal locations for the next generation of such experiments are provided by vertical shafts, such as that providing access to the Gotthard base railway tunnel from the Sedrun locality in canton Graub\"unden, Switzerland.
%called MFS Sedrun for Multi Function Station. 
%This document presents the results of an exploratory environmental measurement campaign at this location to evaluate the ground motion activity and the background electromagnetic field quality in view of a potential installation of an Atom Interferometer.
Atom interferometer (AI) experiments offer interesting prospects for searches for the interactions of ultralight bosonic dark matter with Standard Model particles as well as detection of gravitational waves in a frequency band inaccessible to experiments that are operating or under construction. Ideal locations for the next generation of such experiments are provided by long vertical shafts, such as that providing access to the Gotthard base railway tunnel from the Sedrun locality in the Canton Grisons of Switzerland. We present the results of an exploratory environmental measurement campaign at this location to evaluate the ground motion activity and the background electromagnetic field quality. We find that the backgrounds due to both ground motion and electromagnetic fields, including those due to passing trains, are low enough for successful operation of a 800-m AI experiment.

%\textit{Experimente mit Atominterferometern (AI) bieten interessante Perspektiven für die Suche nach den Wechselwirkungen ultraleichter bosonischer Dunkler Materie mit Teilchen des Standardmodells sowie für den bahnbrechenden Nachweis von Gravitationswellen in einem Frequenzband, das für laufende und im Bau befindliche Experimente unzugänglich ist. Ideale Standorte für die nächste Generation solcher Experimente bieten vertikale Schächte, wie zum Beispiel der Zugang zum Gotthard-Basistunnel vom Ort Sedrun aus. Das vorliegende Dokument enthält einen Vorschlag zur Durchführung einer Umweltmesskampagne an diesem Ort, um die Bodenbewegungsaktivität und die Qualität des elektromagnetischen Feldes im Hinblick auf eine mögliche Installation eines Atominterferometers zu bewerten.\\}

}

\maketitle

\tableofcontents

\newpage

%\Section{Introduction (S. Calatroni, J. Ellis)}
\section{Introduction} 
\label{sec:Intro}

Atom interferometer (AI) experiments offer interesting prospects for
searches for the interactions of ultralight bosonic dark matter with
Standard Model particles as well as detection of gravitational waves in a frequency band inaccessible to experiments that are operating and under construction, via high-precision quantum interference measurements.
Ideal locations for the next generation of such experiments are provided by vertical shafts of height 100 m or more~\cite{Badurina:2019hst,Proceedings:2024foy}, such as those providing access to the Large Hadron Collider (LHC) at CERN or other civil engineering infrastructures such as the Sedrun Multi-Function Site (MFS) of the Gotthard base railway tunnel (also known for the "Porta Alpina" underground train station project) in the municipality of Sedrun.
The Physics Beyond Collider program at CERN~\cite{bib:PBC_public} supports exploratory studies in this field and, following an initial visit to the Sedrun MFS, it was concluded that the site seems suitable to host a long-baseline AI experiment in the future.
It was also concluded that the site's suitability should be confirmed by conducting an exploratory environmental measurement campaign in a vertical shaft of the Sedrun MFS to evaluate the ground motion activity and the background electromagnetic field quality prior to discussions with the Swiss Federal Railways (SBB) about the potential installation of a long-baseline AI.
In this document we summarize the results of the environmental measurement campaign performed in 2025 to explore the site's suitability.

% Explanations of the acronyms used in this document are listed in the Appendix~\ref{sec:Acronyms}. 

%\Section{Introduction (S. Calatroni, J. Ellis)}
% Lucas Lombriser

\section{The Sedrun Multi-Function Site of the Gotthard Base Tunnel} 
\label{sec:Intro}

%The Gotthard Base Tunnel is a railway tunnel that lies about 1 km below the Swiss Alps. Opened in 2016, it enables passenger and freight trains to travel rapidly between central and southern Switzerland, offering fast travel into Italy. In addition to the low-gradient, twin-bore tunnel itself, there are several access shafts, as illustrated in Fig.XX(a). In particular, there are a pair of vertical 800 m shafts that provide direct access to the railway tracks from the community of Sedrun in the Graub{\" u}nden canton in western Switzerland. One of these shafts has a concrete lining and contains an elevator that is used to transport personnel and material between the surface and the Base Tunnel, and the other has a shotcrete lining and is used is for ventilation. Illustrated in Fig.XX(b), the lift shaft is of potential interest for an 800 m atom interferometer experiment. The lift shaft was originally designed as part of a Multi-Function Site (MFS) that could be used to transport railway passengers between the Base Tunnel and Sedrun - the ``Porta Alpina” project - but this project is currently shelved indefinitely. 

\begin{figure}[!h]
\centering % \begin{center}/\end{center} takes some additional vertical space
\includegraphics[width=1.0\textwidth]{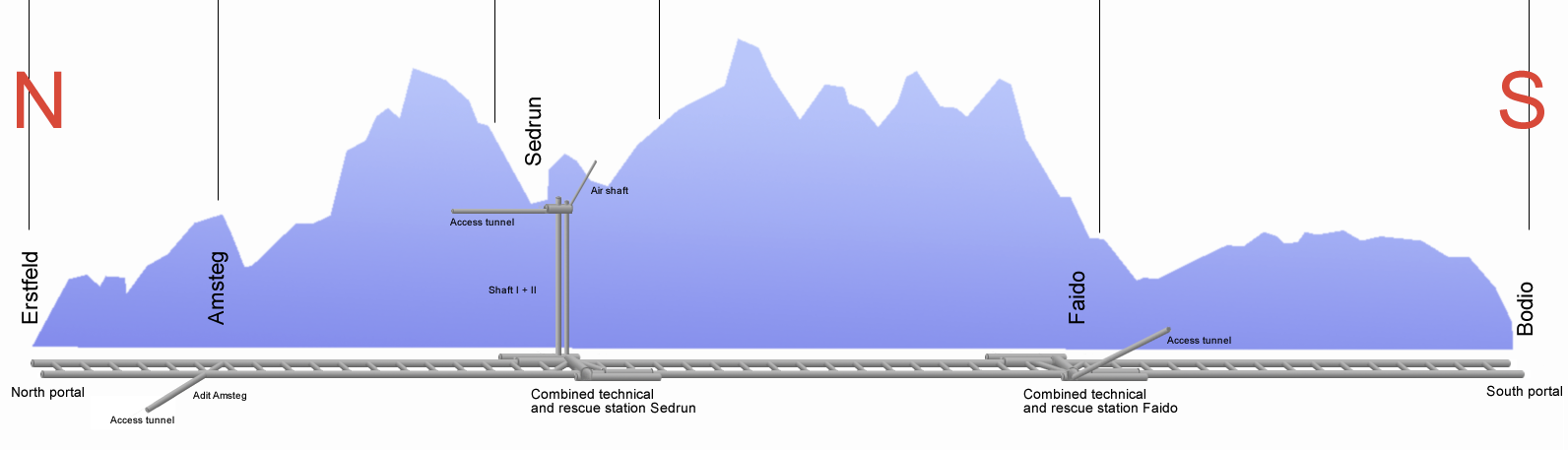}
\qquad
\caption{\label{fig:gotthard} Sketch of a North-South section through the Swiss Alps, illustrating the twin-bore Gotthard Base Tunnel together with its access tunnels and shafts. Adapted from~\cite{Cooper:2005}.}
\end{figure}

The Gotthard Base Tunnel is the longest railway and deepest traffic tunnel in the world, extending 57~km at a maximum depth of 2.5~km under the Swiss Alps. It is also the first low-level connection through a major mountain range at a maximum elevation of 549~m above sea level with a northern and southern entrance at 460~m and 312~m, respectively. Opened in 2016, it enables passenger and freight trains to travel rapidly between northern and southern Switzerland, offering fast travel into Italy. In addition to the low-gradient, twin-bore tunnel itself, there are several access tunnels and shafts, as illustrated in Fig.~\ref{fig:gotthard}. These were built to enable construction from four different sites to expedite construction time and to facilitate maintenance and repairs.

\begin{figure}%[!h]
\centering % \begin{center}/\end{center} takes some additional vertical space
\includegraphics[width=0.9\textwidth]{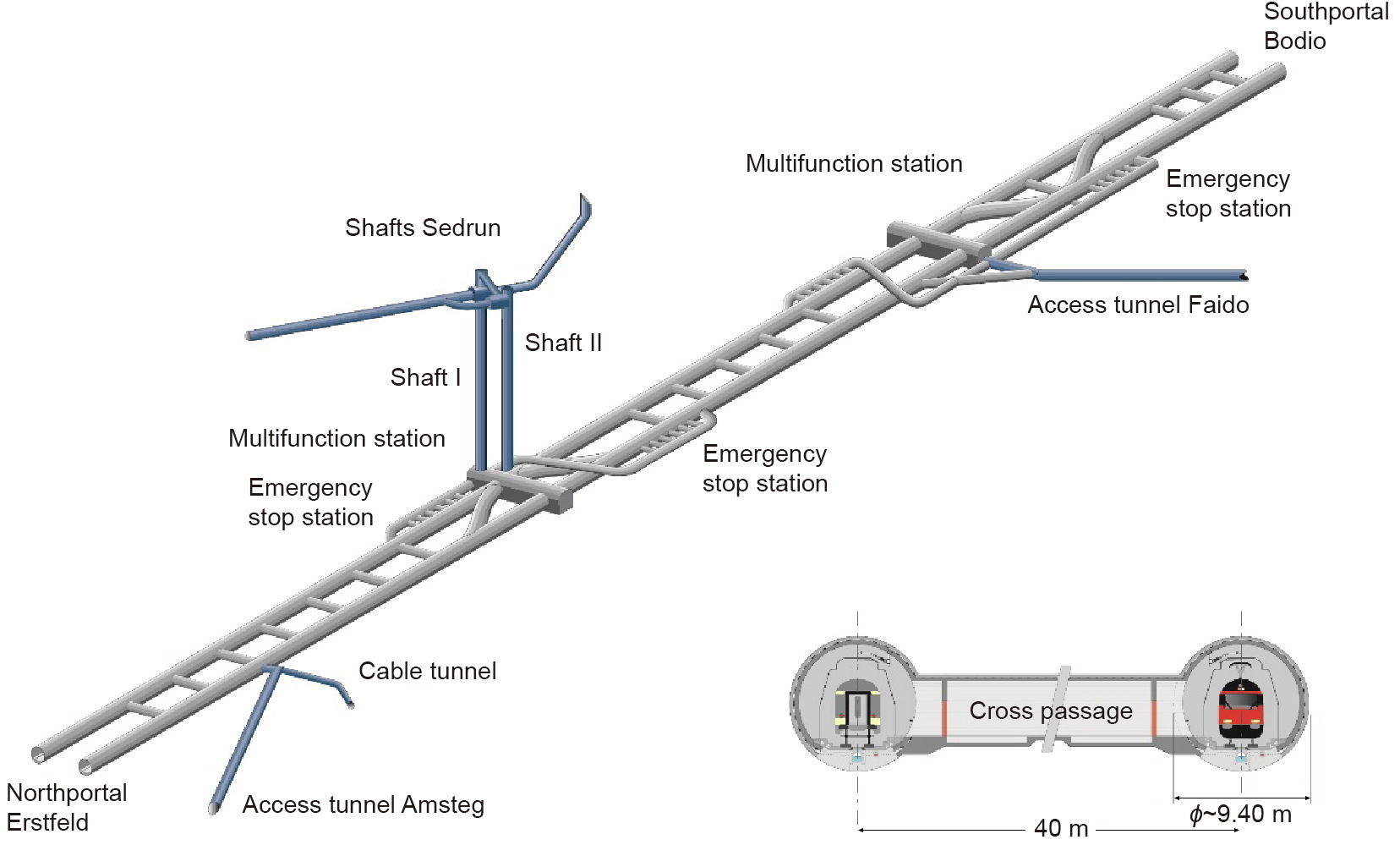}
\qquad
\caption{\label{fig:sedrun} Diagram of the Sedrun Multi-Function Site (MFS) of the Gotthard Base Tunnel, illustrating the horizontal access gallery and the two vertical access shafts~\cite{FABBRI2019379}.}
\end{figure}

In particular, the Multi-Function Site (MFS) near the village of Sedrun in the
%Graub{\"u}nden canton of
Canton Grisons in
Southeastern Switzerland features a pair of vertical 800~m shafts that provide direct access from the Anterior Rhine Valley to the base tunnel railway tracks crossing beneath, as illustrated in Fig.~\ref{fig:sedrun}. The heads of the shafts are located inside the mountain at the end of a 1~km horizontal gallery with road access from Sedrun, at an underground depth of 550~m. The bottom of the shafts, at 1.4~km depth, lies at the base tunnel floor, which at 547~m above sea level is close to its maximal elevation. The larger Shaft~I has an inner diameter of 8.6~m and a concrete lining, and contains an elevator that is occasionally used to transport personnel and material between the surface and the base tunnel for maintenance, whereas Shaft~II has an inner diameter of 7~m and a shotcrete lining, and is used for ventilation.
The head of Shaft~I is accessed through a pressure lock. A pressure equalization system compensates for atmospheric pressure differences between the north and south ends of the base tunnel. The foot of the shaft is connected to the railway tracks through a system of three large, massive steel doors, shielding it from the train traffic.
Shaft~I is of potential interest for an 800~m atom interferometer experiment, and was the focus of our environmental studies.

The large complex of tunnels and caverns at "MFS Sedrun" also features crossover tunnels for changing tracks between the two tubes. Furthermore, a side tunnel of length 1.8~km connects the bottoms of the shafts to four caverns of dimension $38\times10\times5.5~{\rm m}^3$ each. These were built for the envisioned ``Porta Alpina'' underground passenger railway station, from which passengers would be transported via the elevator in Shaft~I to Sedrun. The project was placed on hold in 2007 with new negotiations initiated in 2020 but shelved indefinitely in 2023. Since it offers an attractive alternative use of the site, influential national, cantonal, and municipal politicians in Switzerland have expressed interest and support for the vision of an atom interferometer experiment in the "MFS Sedrun" site making use of the "Porta Alpina" infrastructure.

% * a sketch of the layout

%\Section{Introduction (S. Calatroni, J. Ellis)}
\section{Physics motivations} 
\label{sec:Phys}

Most of the matter in the Universe is invisible Dark Matter (DM) that is not described by the Standard Model of particle physics. DM may be provided by weakly-interacting massive particles (WIMPs) that are being searched for by experiments at the LHC and elsewhere, or by coherent waves of ultralight bosons. The LHC experiments have not yet found any evidence for WIMPs, though searches will continue at the High-Luminosity LHC (HL-LHC). In parallel, searches for Ultra-Light Dark Matter (ULDM) are attracting growing interest, and one of the principal scientific objectives of AI experiments such as AION~\cite{Badurina:2019hst} is to look for the effects of ULDM fields coupled to the constituents of atoms.

%The nature of Dark Matter (DM) is one of the greatest puzzles in fundamental physics, and lies beyond the scope of the Standard Model. The favoured hypothesis is that it is composed of non-relativistic particles such as weakly-interacting massive particles (WIMPs) or coherent waves of ultralight bosons. Experiments at the LHC and elsewhere have not yet found any evidence for WIMPs, though searches will continue during Run 3 of the LHC and at the High-Luminosity LHC (HL-LHC). However, in the meantime the search for Ultra-Light Dark Matter (ULDM) is attracting growing interest, and this is one of the principal scientific objectives of AI experiments such as AION~\cite{Badurina:2019hst}.

The other principal objective of AI experiments is to search for Gravitational Waves (GWs) in a range of frequencies around 1~Hz. Present terrestrial experiments such as LIGO~\cite{LIGOScientific:2014pky}, Virgo~\cite{VIRGO:2014yos} and KAGRA~\cite{Aso:2013eba} (LVK) employ laser interferometry, as will space-borne experiments such as LISA~\cite{LISA:2017pwj}. However, these laser interferometers lose sensitivity at frequencies $\sim1$~Hz, where there may be signals of mergers of black holes with masses intermediate between those whose mergers have been detected by LVK and the supermassive black holes detected in the centres of galaxies~\cite{EventHorizonTelescope:2019dse,EventHorizonTelescope:2022wkp}.  Observations of such mergers of intermediate-mass black holes may help understand how these supermassive black holes have been formed. AI experiments in the intermediate frequency range around 1~Hz may also be sensitive to GWs produced by fundamental physics mechanisms such as the evolution of a network of cosmic strings or a first-order phase transition in the early Universe~\cite{Badurina:2019hst}. 

%The other principal objective of such experiments is the search for Gravitational Waves (GWs) in the range of frequencies around 1~Hz that is intermediate between the peak sensitivities of present terrestrial experiments such as LIGO~\cite{LIGOScientific:2014pky}, Virgo~\cite{VIRGO:2014yos} and KAGRA~\cite{Aso:2013eba}, and the approved space-borne experiment LISA~\cite{LISA:2017pwj}. Among the targets of experiments in this frequency range are mergers of black holes with masses intermediate between those whose mergers have been detected by LIGO and Virgo~\cite{LIGOScientific:2016aoc} and the supermassive black holes detected in the centres of galaxies~\cite{EventHorizonTelescope:2019dse,EventHorizonTelescope:2022wkp}. Detectors in the intermediate frequency range may also be sensitive to a stochastic background of GWs produced by fundamental physics processes such as first-order phase transitions in the early Universe or the evolution of a network of cosmic strings~\cite{Badurina:2019hst}. 

\begin{figure}[t!]
\centering
%~~\\
\vspace{-0.7cm}
\includegraphics[width=0.43\textwidth]{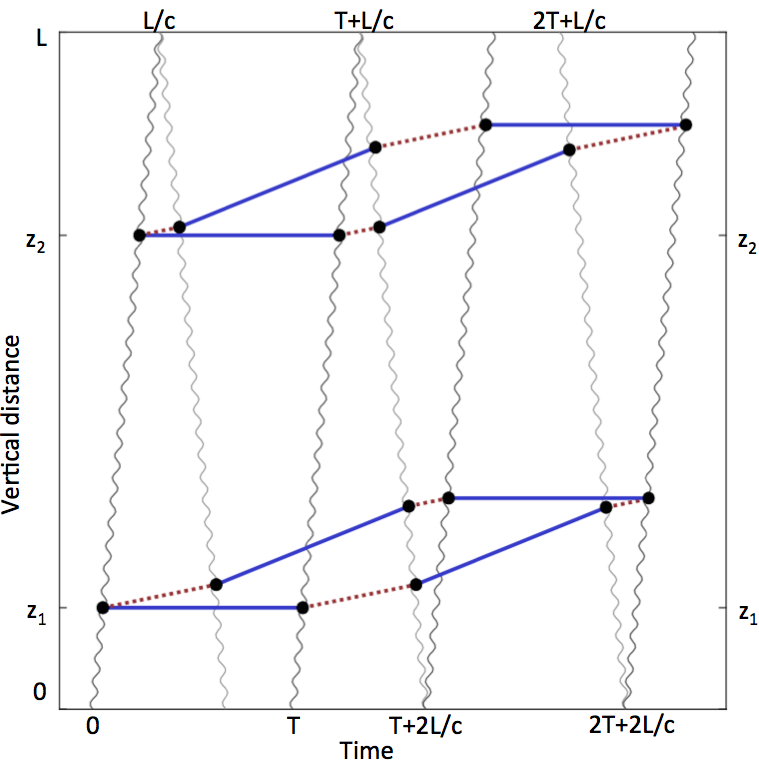}
\caption{%The left plot shows a s
Space-time diagram of a pair of cold-atom interferometers based on single-photon transitions between the ground state (blue) and the excited state (red dashed). Height is shown on the vertical axis and the time axis is horizontal. The laser pulses (wavy lines) travelling across the baseline from opposite ends divide, redirect, and recombine the atomic de Broglie waves, yielding interference patterns that are sensitive to the modulation of the atomic transition frequency caused by coupling to ULDM, or the space-time distortions caused by GWs.}
%Space-time diagram of the operation of a pair of cold-atom interferometers based on single-photon transitions between the ground state (blue) and the excited state (red dashed). Height is shown on the vertical axis and the time axis is horizontal.
%The laser pulses (wavy lines) travelling across the baseline from opposite sides are used to divide, redirect, and recombine the atomic de Broglie waves, yielding interference patterns that are sensitive to the modulation of the atomic transition frequency caused by coupling to ULDM, or the space-time distortions caused by GWs.}
\label{fig:space-time}
%\end{wrapfigure}
\end{figure}

The basic principle of an AI is analogous to that of laser interferometry, as illustrated in Figure~\ref{fig:space-time}. A laser pulse is used to split a cloud of cold atoms into spatially-separated populations of ground-state and excited atoms, and a subsequent ‘mirror’ pulse then interchanges the populations of ground-state and excited atoms. These are then recombined using another laser pulse, allowing the wave functions of the atomic populations to interfere. Interactions of a coherent wave of ULDM with the atoms would alter the atomic excitation energy level, modifying the quantum-mechanical phase and hence the interference pattern. In practice, AI experiments expose two or more sources of atoms to the same laser beam, as also shown in Figure~\ref{fig:space-time}, so as to minimise the effects of laser noise as demonstrated in~\cite{AION:2025igp} and mitigate the effects of ground vibrations. The differences between the interference patterns exhibited by the different sources are sensitive to the space-time dependence of the ULDM field density and the distortions of space-time caused by the passage of a GW. 

%The basic principle of an AI is illustrated in Figure~\ref{fig:space-time}, and is analogous to that of laser interferometry as employed by current GW detectors. A cloud of cold atoms is split by a laser pulse into populations of ground-state and excited atoms, a second `mirror' pulse interchanges the populations of excited and ground-state atoms, which are finally recombined using another laser pulse, and the wave functions of the atomic populations interfere. Each laser interaction with an atom imparts momentum, so the two populations follow different space-time trajectories, as seen in Figure~\ref{fig:space-time}. Interactions of the atoms with a coherent wave of ULDM may alter the excited atomic energy level, modifying the atomic phase and hence the interference pattern. In practice, AI experiments use two or more sources of atoms that are exposed to the same laser beam, as also shown in Figure~\ref{fig:space-time}, thereby minimising the effects of laser noise, and measure the differences between the interference patterns they exhibit, which are sensitive to the space-time dependence of the ULDM field density. Such differential measurements are also sensitive to the distortions of space-time caused by the passage of a GW. 

\begin{figure}[h!]
\centering
%~~\\
%\vspace{-0.6cm}
\includegraphics[width=0.43\textwidth]{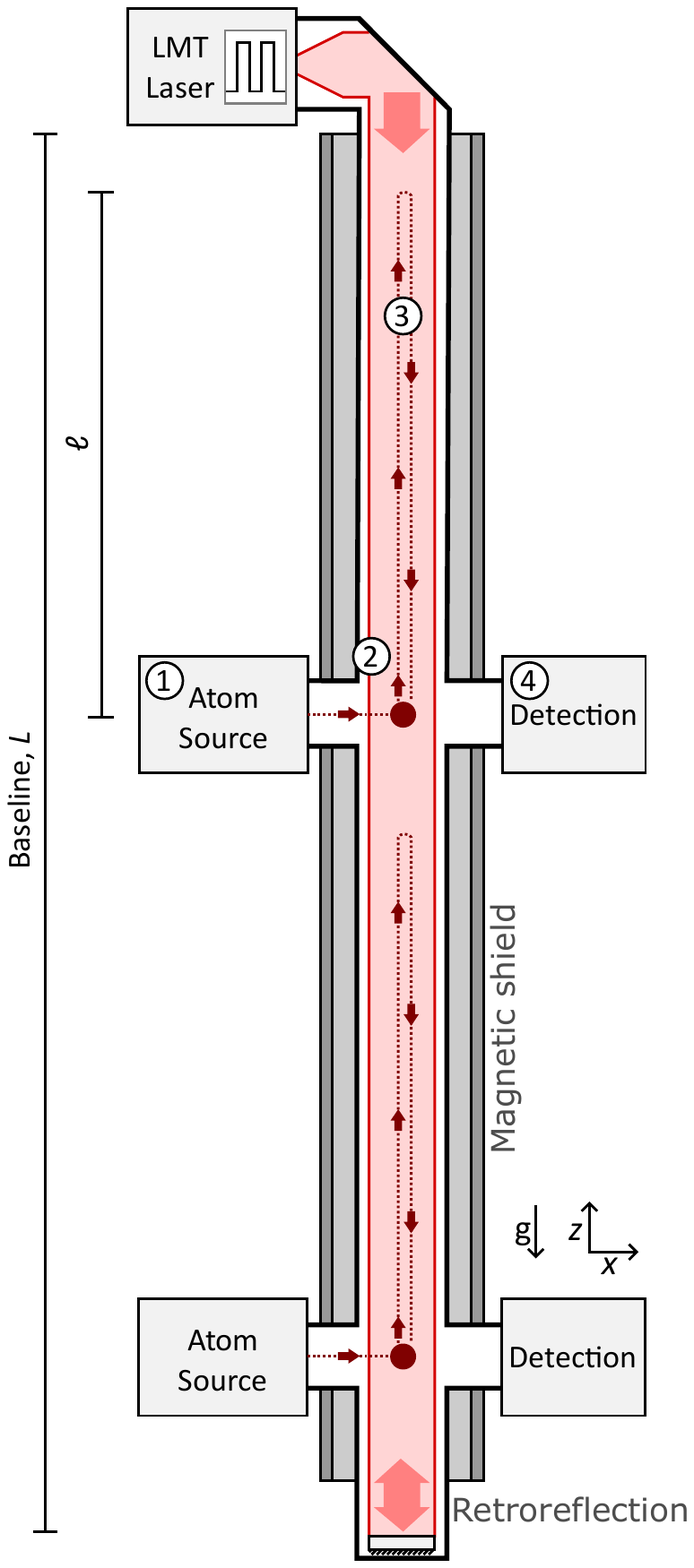}
%\vspace{-0.4cm}
\caption{Conceptual scheme of an Atom Interferometer (AI) experiment with two atom sources that project clouds vertically and are manipulated by a single laser beam (this diagram is not to scale)~\cite{Buchmueller:2023nll}.}
\label{fig:Schematic}
\end{figure}

For the purposes of this study we consider an AI with two or more atom sources in a vertical vacuum tube as illustrated in Figure~\ref{fig:Schematic}~\cite{Buchmueller:2023nll}. The atoms are manipulated by a laser beam generated in a laboratory located at either the top or the bottom of the vacuum tube, though a location at the bottom could also be considered. One of the atom sources is located at the bottom of the vacuum tube, whereas the location of the upper source is only illustrative.  This upper atom source is labelled by (1), the trajectories of the atoms it launches are labelled by (2,3), and the detector to measure interference patterns is labelled by (4). As also indicated, the vacuum pipe would be surrounded by a magnetic shield to suppress electromagnetic interference.

%The conceptual scheme we consider for a vertical AI with two atom sources is shown in Figure~\ref{fig:Schematic}~\cite{Buchmueller:2023nll}. 
%\begin{wrapfigure}[24]{t}{0.45\textwidth}
%\end{wrapfigure}
%There is a single laser source, shown here as located at the top of the vertical vacuum tube, though a location at the bottom could also be considered. One of the atom sources is located at the bottom of the vacuum tube, whereas the location of the upper source is only indicative. This upper atom source is labelled by (1), the trajectories of the atoms it launches are labelled by (2,3), and the detector to measure interference patterns is labelled by (4). Additional sources and detectors may be added as desired. We note that the vacuum pipe is surrounded by a magnetic shield.

%~~\\
%~~\\
%~~\\

%\Section{Introduction (S. Calatroni, J. Ellis)}
\section{Environmental Requirements for an Atomic Interferometer} 
\label{sec:Overview}

The design of an AI is guided by balancing engineering requirements, technical limitations, and ambient sources of systematic noise. Many of the systematics affecting atom interferometers can be found in reference~\cite{MAGIS-100:2021etm} and references therein.

The ambient sources of systematic noise include ambient seismic activity, infrasound, electromagnetic noise and atmospheric temperature fluctuations. As a preliminary site investigation, the local ground motion and the ambient magnetic fields/Radio-Frequency (RF) fields must be investigated as a first step prior to further investigation and assessment of the suitability of a site for the installation of an AI.

%\subsubsection{Vibration and seismic noise (M. Guinchard)}
\subsection{Vibration and seismic noise requirements}
\label{subsubsec:Vibrations}

\begin{figure}[!h]
\centering % \begin{center}/\end{center} takes some additional vertical space
\includegraphics[width=0.7\textwidth]{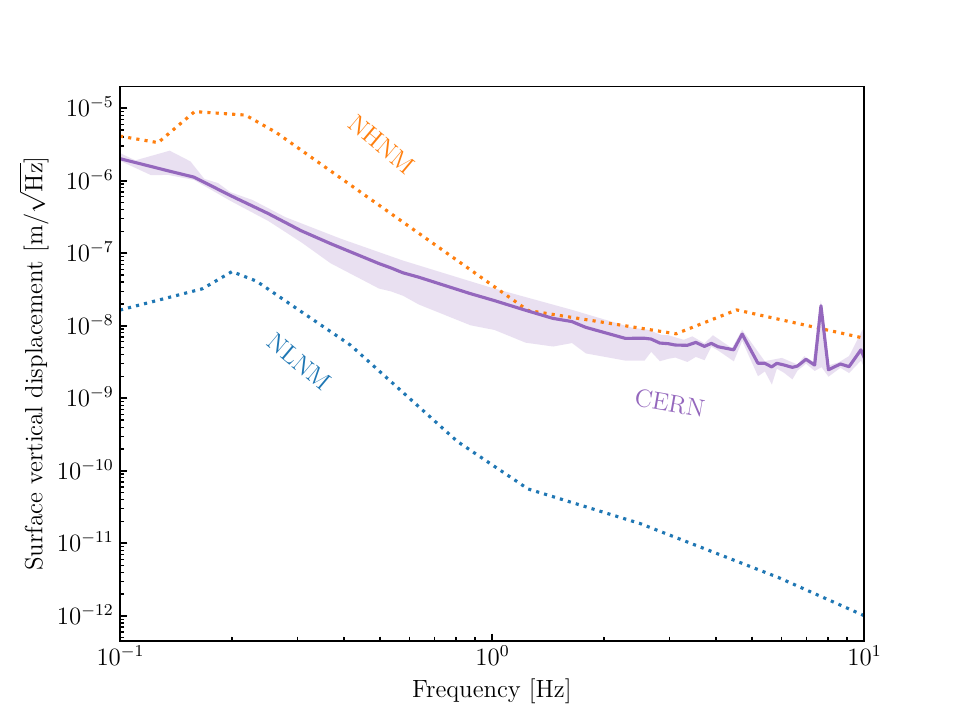}
\qquad
\caption{\label{fig:noisemodels} The RMS spectral density of surface  vertical displacement measurements at CERN~\cite{Arduini:2023wce}, compared with the New High and Low Noise Models (NHNM and NLNM)~\cite{peterson1993observations}. The shaded band corresponds to the difference between the minimum and maximum daily measurements.}
\end{figure}

Seismic activity from natural sources such as earthquakes, or cultural (human-made) sources including civil engineering
works, excite ground vibrations that can be transmitted to the atom sources or the detector components in an AI experiment. Estimates of the effect of ground vibrations on the main interferometry laser beam indicate that local vibrations of the main optics should be kept to acceleration fluctuations below $\delta a \leq \SI{e-4}{(m/s^2)/\sqrt{Hz}}$ for both surface and underground infrastructures\cite{TVLBAI2023WorkshopSummary}\cite{AION100feasibilitystudy}. An additional effect is the Gravity Gradient Noise (GGN) imparted directly to the atoms by fluctuations in the local gravitational field, which should be below the New High-Noise Model (NHNM)~\cite{peterson1993observations} if the experimental measurements are to be competitive. 
Figure~\ref{fig:noisemodels} shows as an illustrative example measurements at CERN~\cite{Arduini:2023wce}, where the surface  vertical displacement measurements at frequencies $< 1$~Hz generally lie below the NHNM. 

%\subsubsection{EM noise (D. Valuch, M. Pentella)}
\subsection{Electromagnetic noise requirements}
\label{subsubsec:EMnoise}

% to be prepared by Mariano Pentella and Daniel Valuch 

AIs are also sensitive to magnetic fields through various mechanisms. For example, field fluctuations directly cause atom phase fluctuations, which could look like a ``fake”  dark matter or gravitational wave signal. For this effect, the magnetic field noise constraints are most critical in the detector sensitivity band of $\approx$ 20 mHz to 10 Hz. However, noise spikes at frequencies > 10 Hz could potentially alias down, e.g., those originating from the 16.7 Hz railway power line. AI detectors are also sensitive at any locations in which laser cooling is taking place – this will happen at regular intervals probably between 10 and 50 meters apart along the shaft. Regardless of the results of these field measurements, an AI experiment needs bias coils that can compensate the earth’s magnetic field and any other background fields at the $\approx$ 1 Gauss level.

The main sources of EM noise that need to be considered are the large HVAC plants distributed between the surface and the underground cavern. The standard telecommunication equipment installed and operated in the underground areas (3G/4G cellular networks) and the vertical lift / crane using industrial radio-controls also need to be considered. 

In order to reveal the most likely sources of potential problems, and to inform the detector design (e.g., the shielding), the following measurements were carried out, both at the base and a few meters below the top of the shaft: 
\begin{itemize}
%    \item At the base of the shaft:
    \item 3-axis fluxgate measurement in the band 1 mHz – 3 kHz with a noise floor in the 10 pT$/\sqrt{Hz}$ range;
    \item 3-axis pickup coil measurement in the band 1 kHz – 100 kHz with whatever noise floor is easy to achieve at low cost.
\end{itemize}

%\begin{itemize}
%    \item A few meters below the top of the shaft:
%    \item[] 3-axis fluxgate measurement in the band 1 mHz – 3 kHz with a noise floor in the 10 pT$/\sqrt{Hz}$ range;
%    \item[] 3-axis pickup coil measurement in the band 1 kHz – 100 kHz with whatever noise  floor is easy to achieve at low cost.
%\end{itemize}

In order to assess the field noise in all likely operating conditions of the detector and of the Gotthard Base Tunnel infrastructure, a set of separate measurements were taken over a period of weeks.

\section{Modus Operandi for the Environmental Measurement Campaign} %}
\label{sec:Intro}

\subsection{Vibration and seismic noise measurements}
\label{subsec:seismic-measurements_5-1}

% How to perform the measurements

The setup for ground motion investigation is based on a broadband seismometer consisting of a magnetic mass placed in the core of a coil. Oscillations of the mass due to ground motion create electro-magnetically-induced current changes in the coil. The voltage at the ends of the resistance is proportional to the ground velocity. High frequencies are filtered thanks to a low-pass filter. The frequency range of interest is higher than the resonance frequency of the mass-spring system and lower than the filter’s cut-off frequency. Such broadband seismometers are a typical sensitivity to ground velocities between 800 and 2500 V/(m/s).

\begin{figure}[h!]
    \centering
    \includegraphics[width=0.5\linewidth]{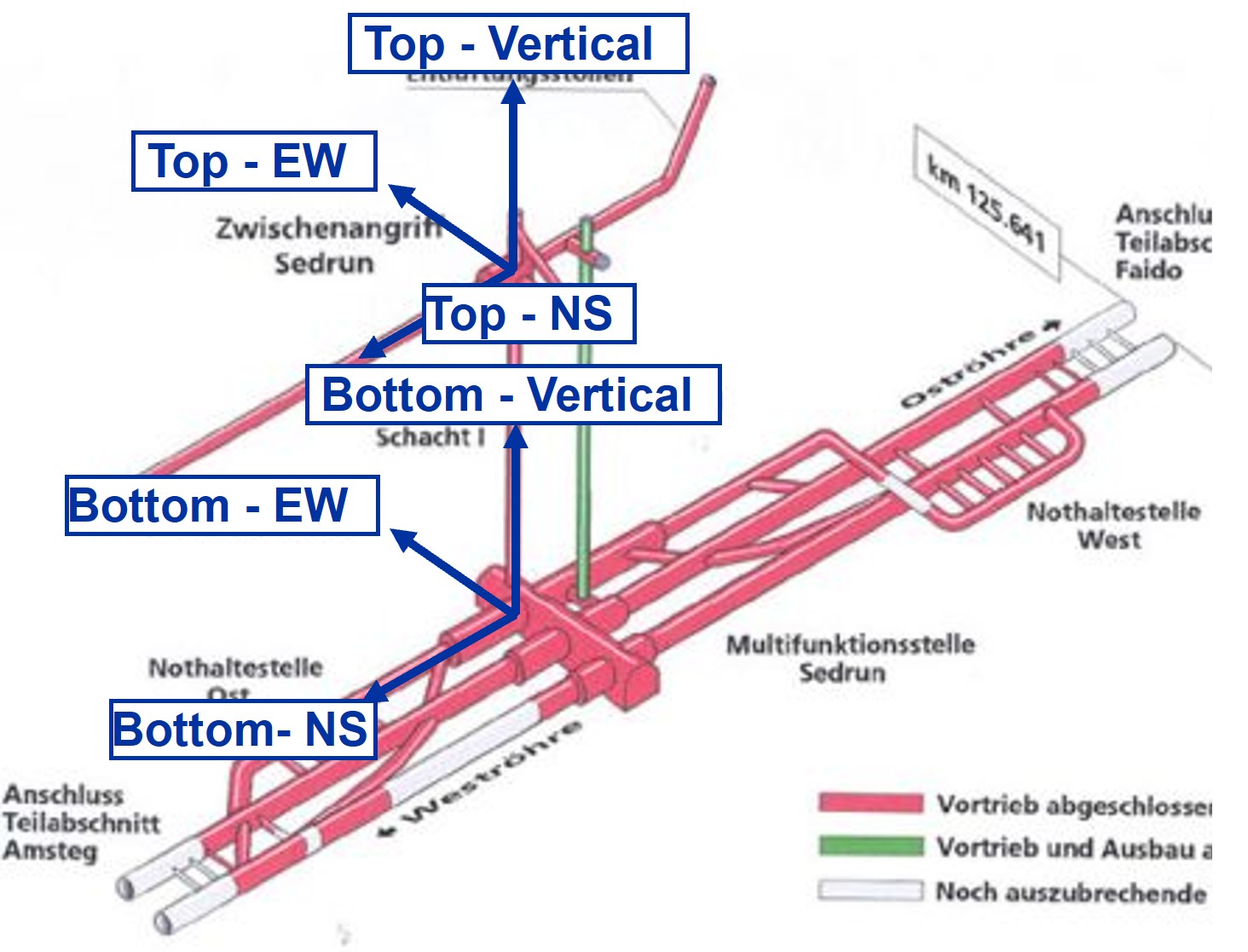}
    \caption{Orientation of the sensors in the infrastructure}
    \label{fig:Orientation of the sensors in the infrastructure}
\end{figure}

The broadband seismometers chosen for the investigation were the 6T series from \href{https://www.guralp.com/}{Güralp Systems Ltd} with a bandwidth from 30 s up to 100 Hz and a sensitivity of 2000 V/(m/s) in the three directions. They were supplied by a 230 V AC / 12 V DC switching mode power supply outputting 2 A with less than 50 mVpp ripple.
The sensors were installed on the floor and carefully levelled at both the upper and lower measurement sites. The North–South (NS) direction was aligned with the tunnel axis, while the East–West (EW) direction was oriented orthogonally to the tunnel axis, as shown in Figure~\ref{fig:Orientation of the sensors in the infrastructure}.

The data acquisition system (DAQ) controllers are 8-channel MICROQs from MullerBBM\texttrademark. The chosen modules allow simultaneous 24 bits sampling over a minimum +-100 mV dynamic range with less than 5 µVrms input noise. The DAQ controller is supplied by a 230 V power distribution unit and controlled by a laptop with signal treatment software permitting remote control via a 4G GSM connection and real-time FFT calculations.\\

\begin{figure}[h!]
    \centering
    \includegraphics[width=1\linewidth]{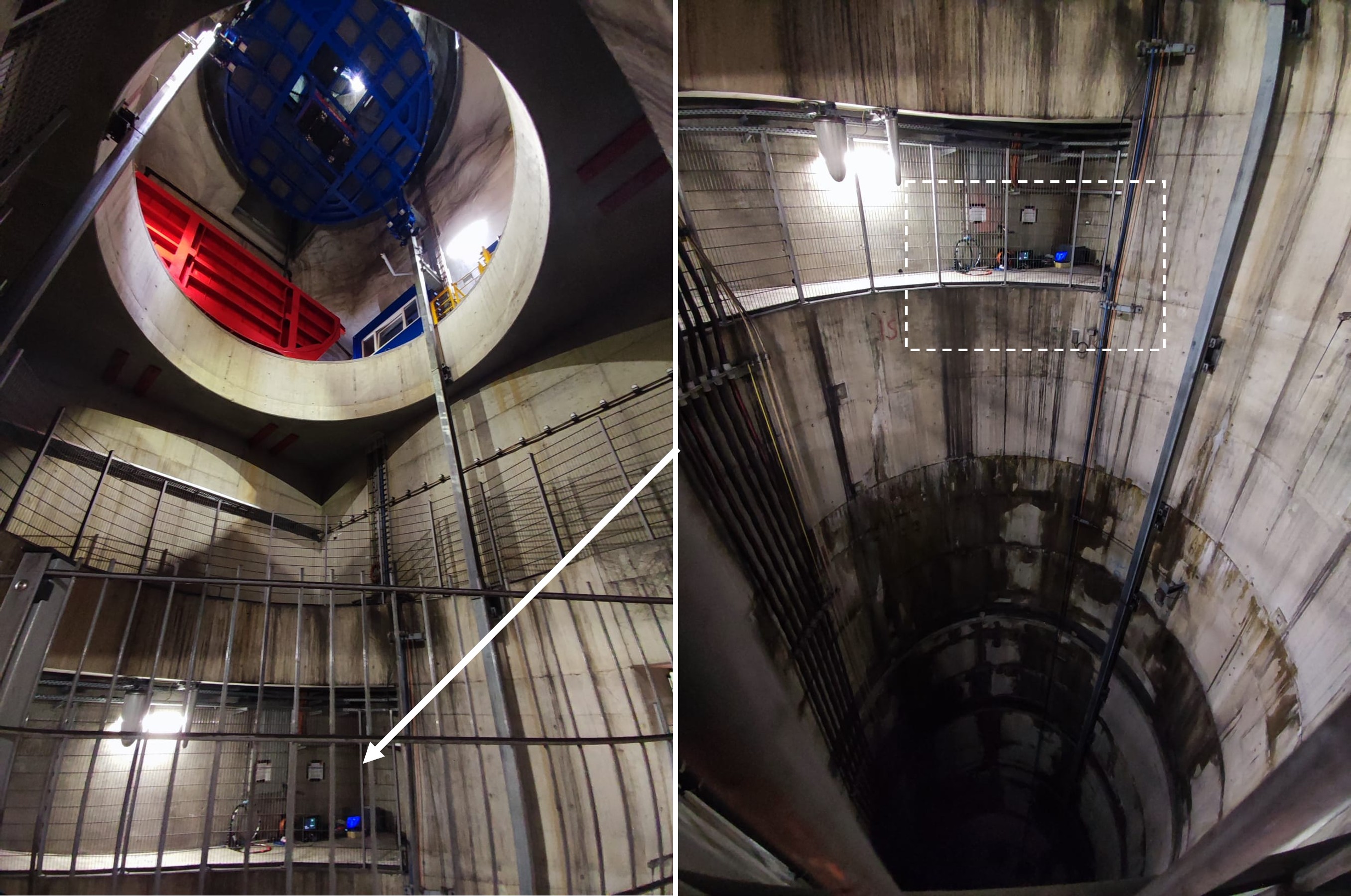}
    \caption{Equipment installed at the top of the lift shaft.}
    \label{fig:Equipments installed at the top of the lift shaft}
\end{figure}

\begin{itemize}
 \item \textit{Locations for measurements}\\
Following a first visit to the Sedrun MFS, two measurement sites were identified and validated in coordination with the relevant Swiss Federal Railway officials. 
 At the top of the lift shaft, the slab placed below the lift access slab, which has a rigid concrete connection to the shaft as visible in the left panel of Figure~\ref{fig:Equipments installed at the top of the lift shaft} was used as the TOP station.
 At the bottom of the lift shaft, the best location identified during the visit was close to the shaft metallic structure as visible in Figure~\ref{fig:MeasEquipmentUnderground}. This location was labeled the BOTTOM station for the data collected.
 
\end{itemize}

\begin{itemize}
 \item \textit{Measurement duration}\\
The ground motion measurements were performed continuously from 7th May 2025 to 10th July 2025, 24 hours per day, including weekends, in order to improve the statistical robustness of the dataset. Measurements during weekends were particularly important, as tunnel operations and cultural noise sources differ from those observed during weekdays. In total, more than 1,500 hours of data were collected, including the recorded passages of over 32,000 trains

\end{itemize}

\begin{itemize}
 \item \textit{Power supply}\\
The DAQ systems and the laptop computers were placed less than 2 meters away from the sensors and connected to 230 V power lines without backup solutions.
\end{itemize}

\begin{itemize}
 \item \textit{Equipment footprint}\\
As seen in Figure~\ref{fig:MeasEquipmentUnderground} showing the installation at the bottom of the lift shaft, the footprint of the ground motion equipment is limited to 1.5~m$^2$ at each measurement site.
\end{itemize}

\begin{itemize}
 \item \textit{Synchronization and communication}\\
In order to assess the ground-motion stability of the site and to evaluate potential phase shifts between the TOP and BOTTOM locations, synchronous measurements were carried out using the GSM network available in the MFS Sedrun infrastructure. The electronics were connected to external NTP servers, enabling second-level time synchronisation throughout the entire measurement period. In addition, a complete logbook of all train activities within the infrastructure was provided by SBB to be synchronized with data collected.
\end{itemize}

\begin{itemize}
 \item \textit{Data post-processing}

A total of \SI{1277.4}{\hour} were recorded at the TOP station and a total of \SI{1412.5}{\hour} at the BOTTOM station. The post-processing of this data is composed of three steps:

\begin{enumerate}
    \item Resampling from \SIrange{300}{256}{\hertz} with low-pass filtering after upsampling (FIR filter with Hanning window for anti-aliasing; cutoff is \SI{300}{\hertz}).
    \item Long-term analysis:
    \begin{enumerate}
        \item Computing PSDs using the Welch method using 16384 samples (= \SI{64}{\second}), Hanning window, and \SI{50}{\percent} overlap;
        \item In parallel, a histogram is updated with each PSD, which will later represent the probability distribution for the entire dataset (velocity resolution in log-scale with 300 bins, frequency resolution given by the PSD).
    \end{enumerate}
    \item Analysis of transitory aspects:
    \begin{enumerate}
        \item Computing PSDs with the Welch method using 1024 samples (= \SI{4}{\second}) and Hanning window; no overlap applied;
        \item Worst case events were identified by the highest integrated power in velocity (integral of the velocity PSD).
    \end{enumerate}
\end{enumerate}

In addition, the velocity channels are stored as .MSEED files with the absolute time based on the start time (+/- 1s).

\end{itemize}
\subsection{Electromagnetic noise measurements} 
\label{subsec:Site}

The environmental magnetic field is measured by combining the data from two probes: a three-axis fluxgate magnetometer to cover the low-frequency band and a three-axis passive loop antenna to cover the high-frequency band. Signals from both probes are digitized using a high-resolution digital oscilloscope.

The Bartington Mag-13 fluxgate magnetometer~\cite{Bartington:2024} is capable of measuring the magnetic field over the range from DC to 3~kHz. It is connected to a power supply and signal processing unit by a flexible cable with a maximum length of 100~m. Analogue outputs of the signal processing unit is ±10~V, corresponding to a full scale of $\pm70~\mu T$ (or $\pm100~\mu T$ depending on the electronics used).

A three-axis loop antenna (or H-field probe, similar to Schwarzbeck FMZB 1519-60 C \cite{Schwarzbeck:2024}) is capable of measuring the magnetic field in the range from a few hundred Hz to a few hundred kHz. The two antennae were custom designed and made for this measurement campaign by the Institute of Electrical Engineering of the Faculty of Electrical Engineering and Information Technology in Bratislava. 

Two high-resolution, 8-channel, remotely controlled oscilloscopes model LeCroy 8108HD \cite{LeCroy:2023} were used to acquire signals from the fluxgate magnetometer and loop antenna. Data were locally stored to a SSD drive and retrieved only at the end of the measurement campaign. Remote connection to the instruments was provided by an industrial cellular router Teltonika RUT956 using Swisscom Mobile Private Network (MPN) to have the remote device appear like a local device on the CERN network. The router was shared with the seismic measurement equipment. 

Prior to installation at the MFS Sedrun site, the noise floor of the measurement set-up was determined in an electromagnetically  shielded chamber in the CERN EMC laboratory, where the flux-gate magnetometer was additionally inserted into a zero Gauss chamber. 

\begin{figure}
\begin{center}
    \subfloat[]{\includegraphics[width=0.36\linewidth]{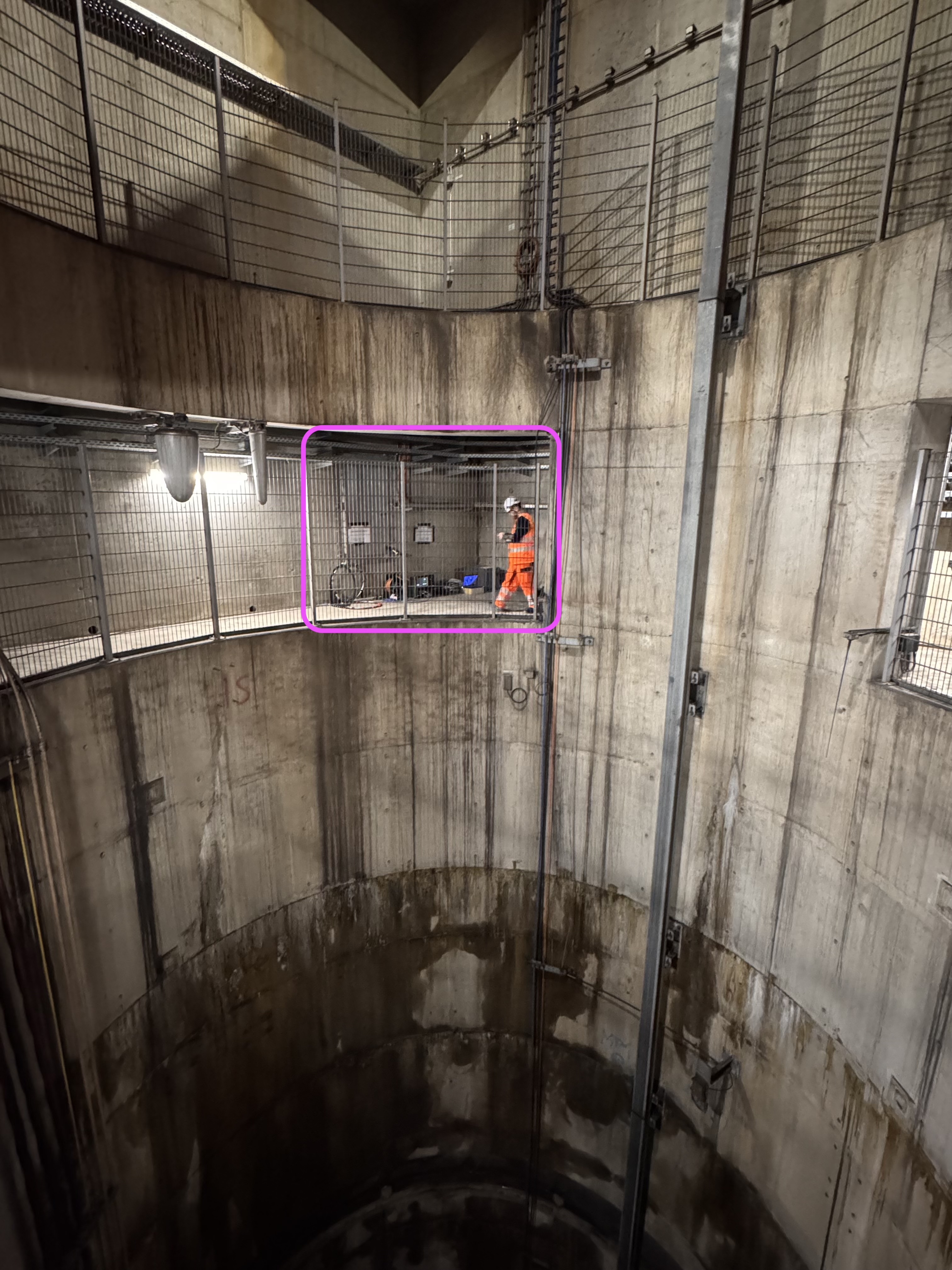}
    \label{fig:MeasEquipmentSurface_A}}
%    \quad
    \subfloat[]{\includegraphics[width=0.64\linewidth]{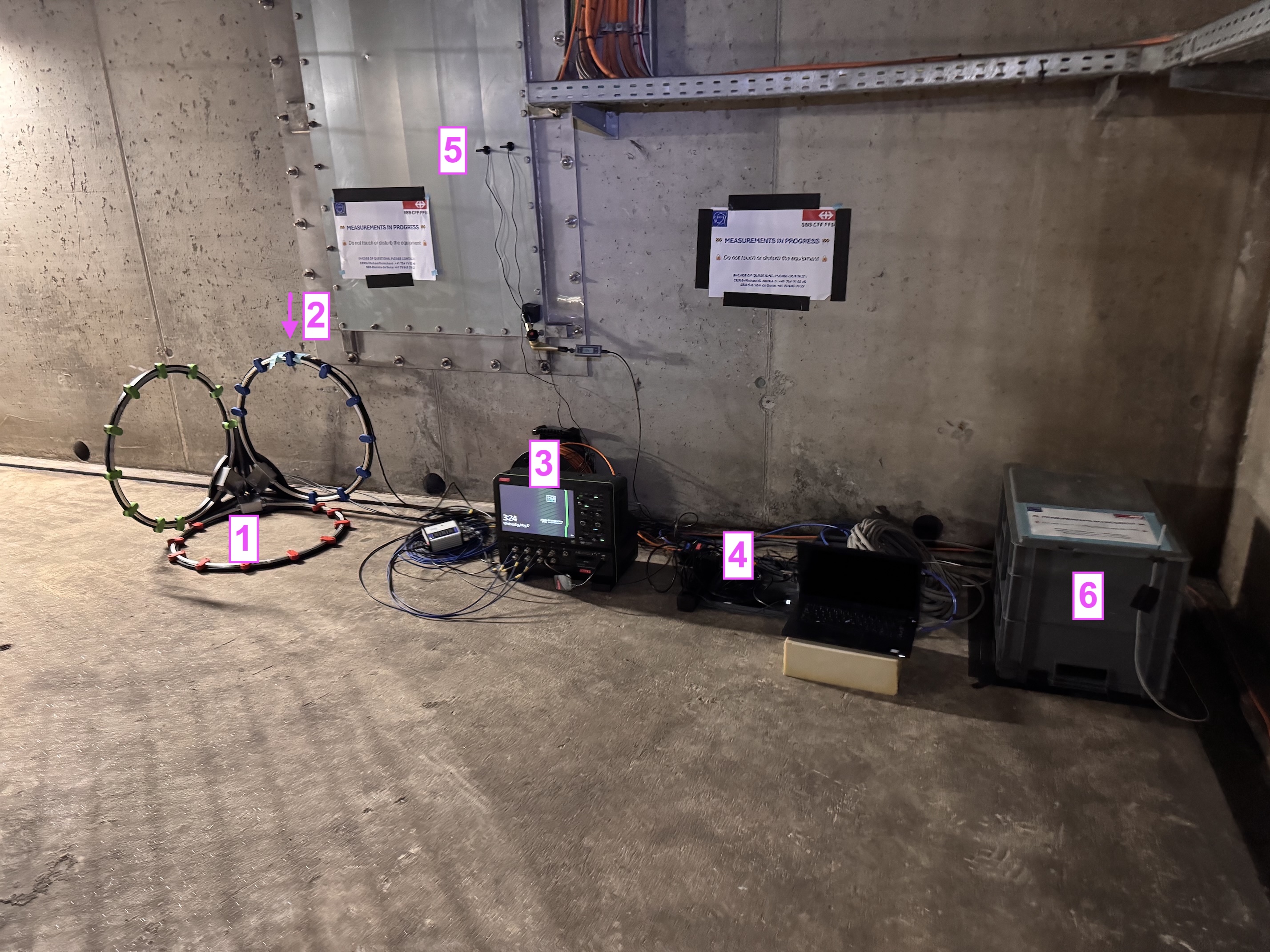}
    \label{fig:MeasEquipmentSurface_B}}
%    \quad
\caption{The electromagnetic field measurement station at the top of the elevator shaft: (a) general view, (b) detail of instruments: 1) 3-axis loop antenna, 2) fluxgate magnetometer, 3) oscilloscope, 4) cellular modem, 5) GSM antennas, 6) seismic measurement equipment.}   
\label{fig:MeasEquipmentSurface}
\end{center}
\end{figure}

\begin{figure}
\begin{center}
    \subfloat[]{\includegraphics[width=0.36\linewidth]{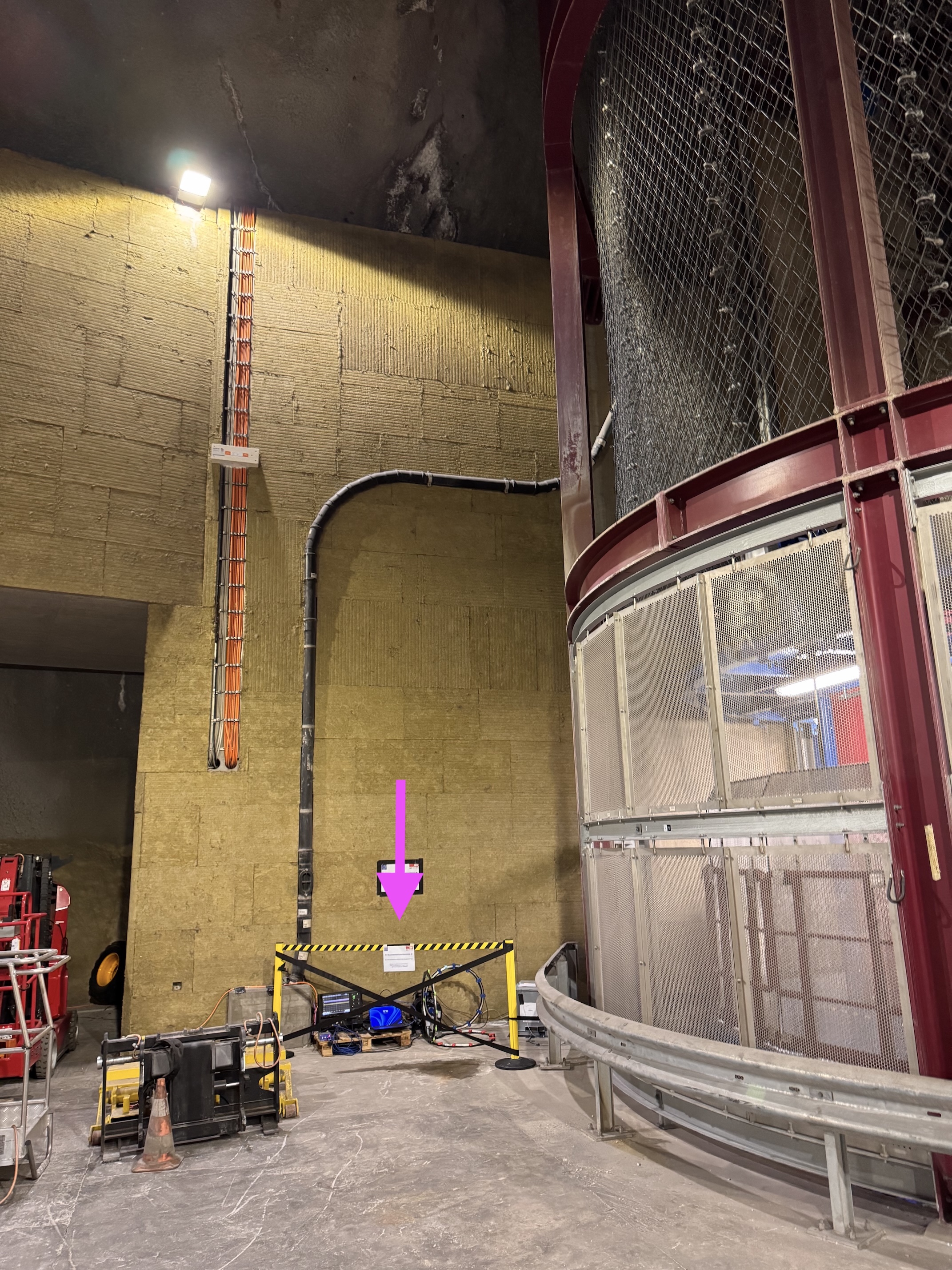}
    \label{fig:MeasEquipmentUnderground_A}}
%    \quad
    \subfloat[]{\includegraphics[width=0.64\linewidth]{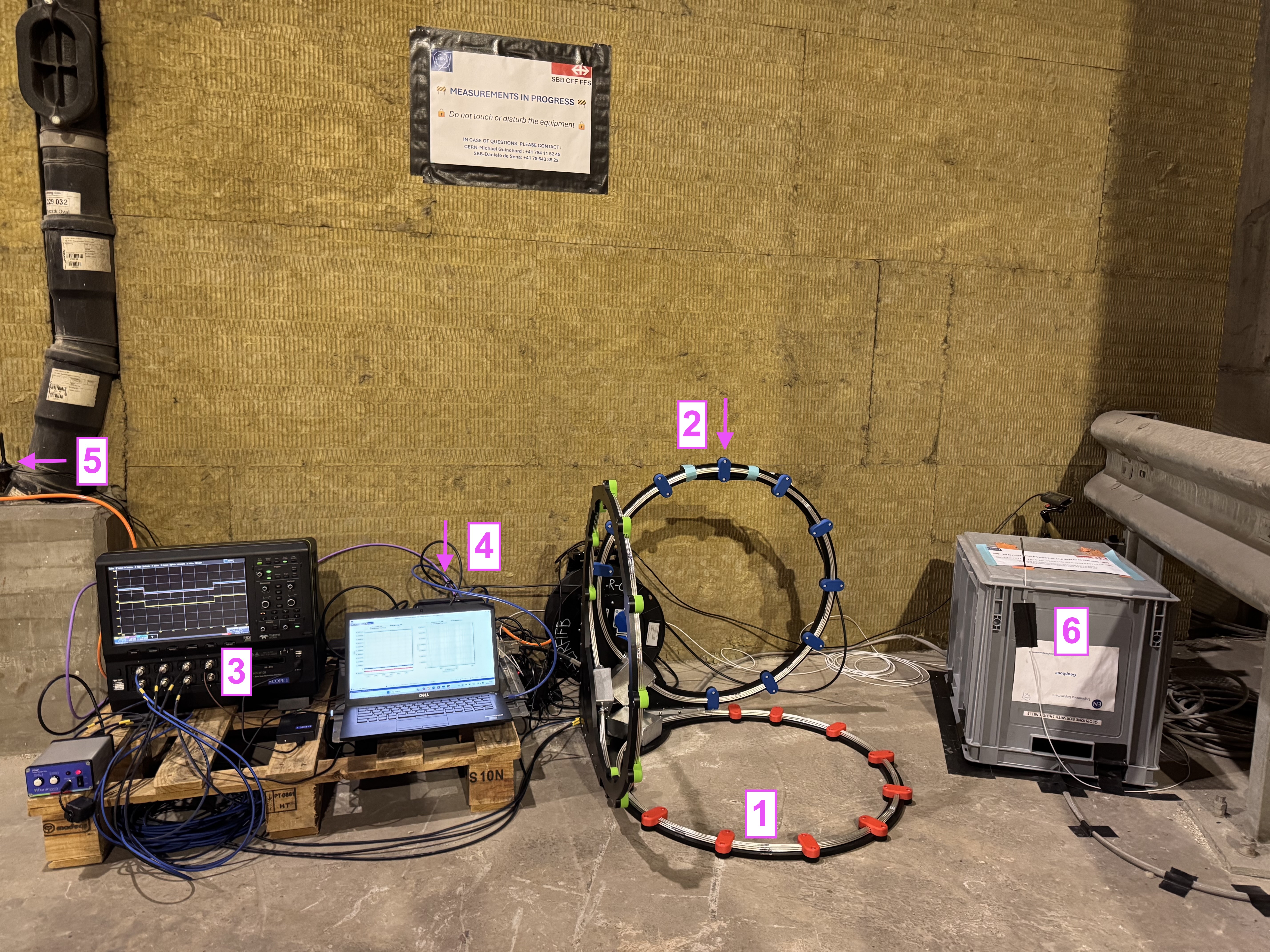}
    \label{fig:MeasEquipmentUnderground_B}}
%    \quad
\caption{The electromagnetic field measurement station at the bottom of the elevator shaft: (a) general view, (b) detail of instruments: 1) 3-axis loop antenna, 2) fluxgate magnetometer, 3) oscilloscope, 4) cellular modem, 5) GSM antennas, 6) seismic measurement equipment.}   
\label{fig:MeasEquipmentUnderground}
\end{center}
\end{figure}

The electromagnetic field measurement station at the top of the elevator shaft is shown in Fig.~\ref{fig:MeasEquipmentSurface} and at the bottom of the elevator shaft is shown in Fig.~\ref{fig:MeasEquipmentUnderground}. 

The low-frequency measurement ran with the oscilloscope in low-noise mode, sampling at 2500 Sa/s, record length 20000 seconds (i.e., 50 million points per channel/axis). The acquisition was triggered by the cellular modem every 6 hours (at 5:00, 11:00, 17:00, 23:00). Modem time was synchronized over the network. In total, 738 hours (top) and 1117 hours (bottom) of low-frequency data were collected from 13th May 2025 to 9th July 2025, including standard unmanned site operation and maintenance periods with personnel present on site and the elevator operating. 

The saved data from the oscilloscope (50 million points per axis) were scaled from voltage to B-field using the fluxgate magnetometer calibration factor. The power spectral density of each axis was calculated using the Welch method, 1 million samples and \SI{50}{\percent} overlap, Hanning window. The PSD modulus was finally calculated as $PSD = \sqrt{PSD_x^2+PSD_y^2+PSD_z^2}$.

The high-frequency (>1~kHz) measurements with 3-axial loop antennae were complementary and results are not presented in this report, as the AI environment specification in this frequency band is not critical. The oscilloscope was set to low-noise mode, sampling at 250~kSa/s, record length 200 seconds (i.e., 50 million points per channel/axis). 
The acquisition was triggered by the cellular modem every 5 minutes (at 0, 5, 10, 15 … 55 min). Modem time was synchronized over the network. High-frequency data were collected for 2 days from 9th July 2025 to 10th July 2025. 

\begin{figure}
\begin{center}
    \subfloat[]{\includegraphics[width=0.4\linewidth]{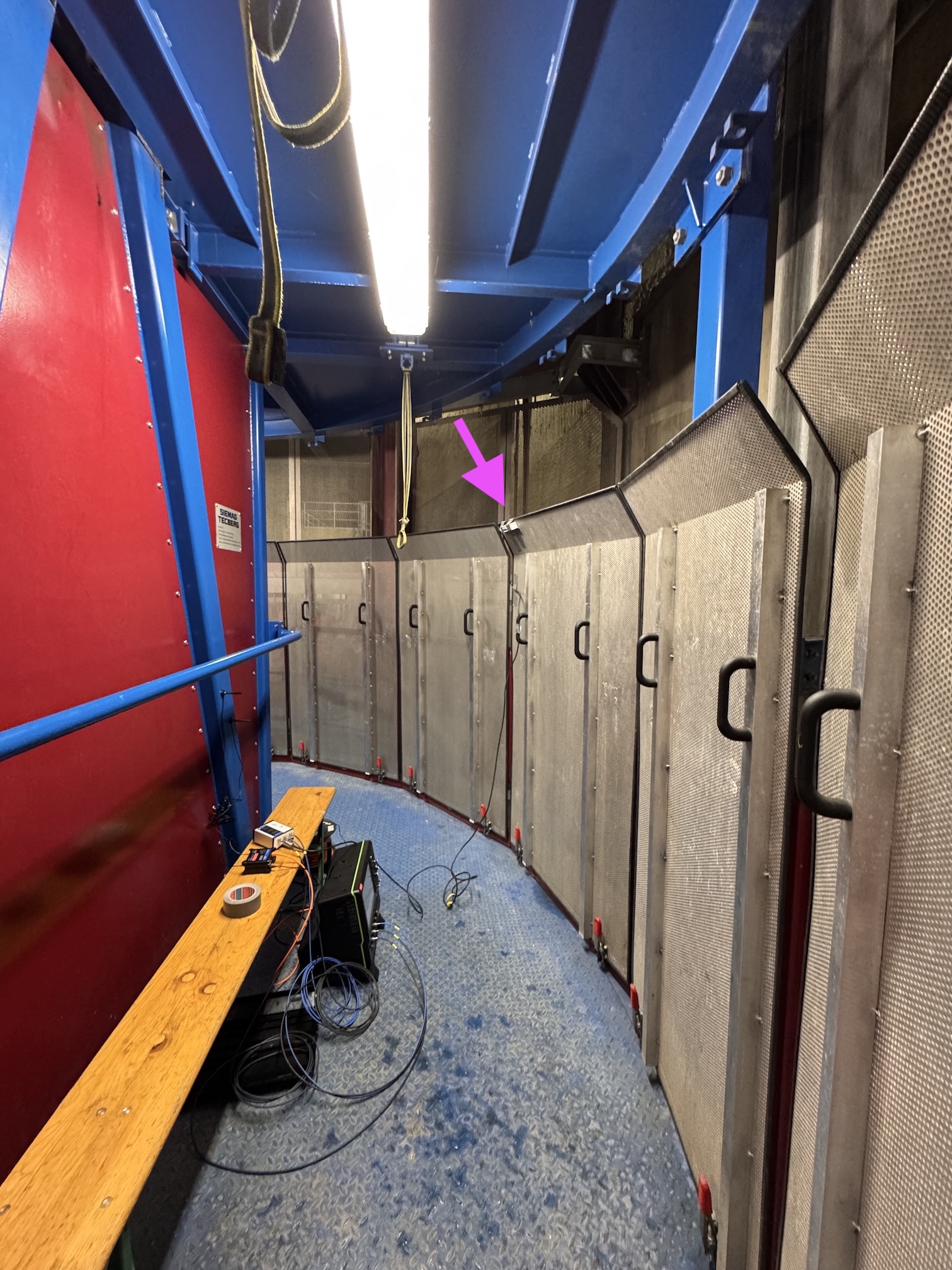}
    \label{fig:MeasEquipmentElevator_A}}
%    \quad
    \subfloat[]{\includegraphics[width=0.4\linewidth]{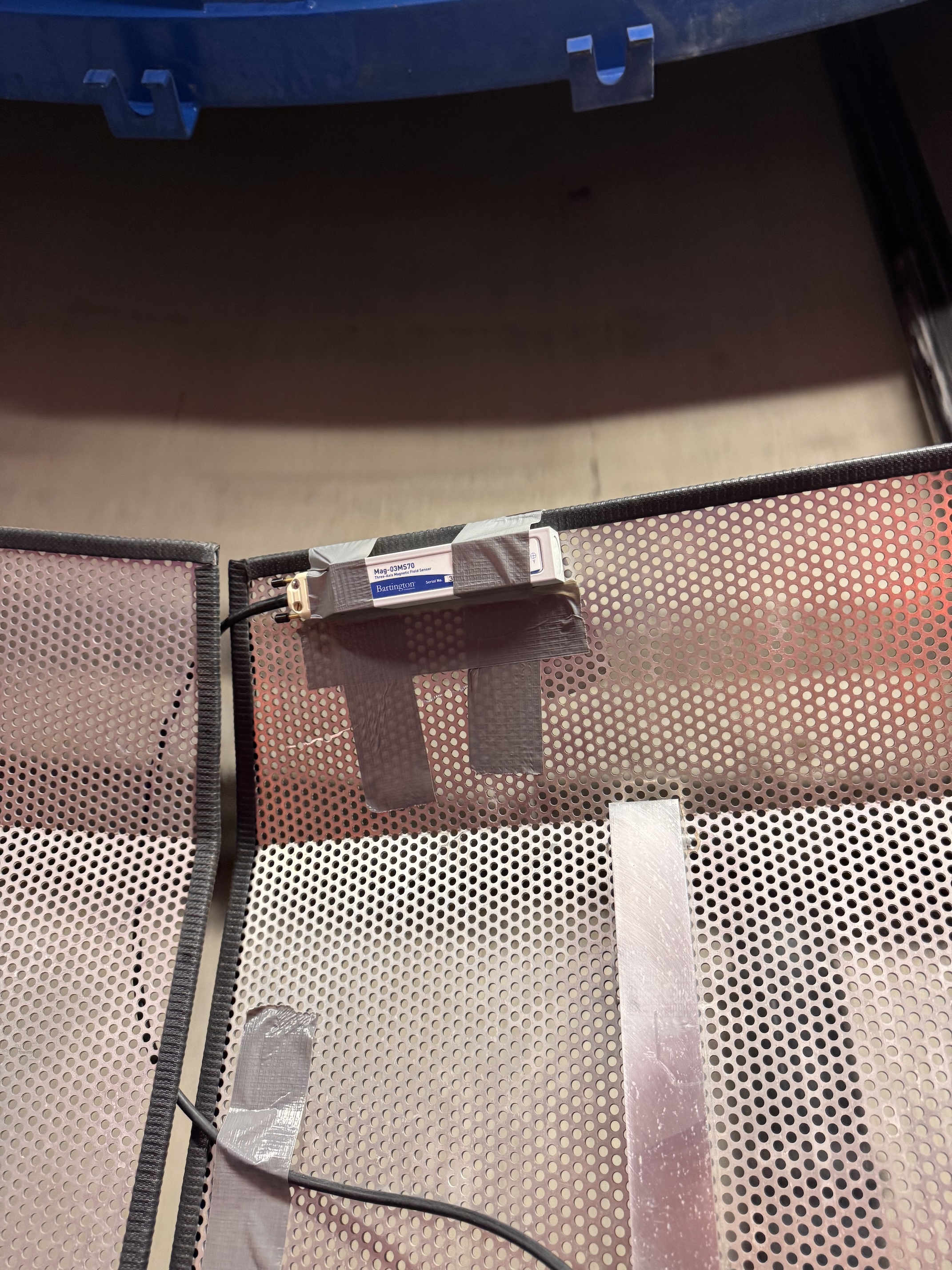}
    \label{fig:MeasEquipmentElevator_B}}
%    \quad
\caption{The low-frequency magnetic field measurement along the elevator shaft: (a) measurement station, (b) detail of the magnetic field probe location.}   
\label{fig:MeasEquipmentElevator}
\end{center}
\end{figure}

On the last day of the measurement campaign, it was possible to remove the electromagnetic measuring equipment from the underground station and install it in the elevator in order to map the low-frequency magnetic field along the 800~m elevator shaft. The installation of the measurement equipment in the elevator is shown in  Fig.~\ref{fig:MeasEquipmentElevator}.

\FloatBarrier

\section{Experimental Results} %}
\label{sec:Experimentalresults}

\subsection{Vibration and seismic noise results}
\label{subsec:seismic-results}
A maximum allowable acceleration amplitude of $\delta a \leq \SI{e-4}{(m/s^2)/\sqrt{Hz}}$ has been specified as a requirement for the AION-100 project (see Chapter~\ref{subsubsec:Vibrations}), which we use as a reference value for the viability of the MFS as a potential location for the proposed long-baseline AI. Although the focus of this analysis is on the long-term ground motion results, two transitory effects are also evaluated: the impact on the ground motion of passing trains and the impact of the local infrastructure.

\paragraph*{Long-term random vibration analysis}
The data processing for the long-term analysis includes the entire dataset for the TOP and BOTTOM stations, from which the distributions of vibration (Probability Power Spectral Density, PPSD) and the resulting mode curves (highest probability) are computed. The parameters for the underlying Power Spectral Densities (PSDs) are blocks of 16384 samples, with \SI{50}{\percent} overlap, and the application of the Hanning window (more details on the postprocessing are given in Chapter~\ref{subsec:seismic-measurements_5-1}). \newline

\begin{figure}[b!]
    \centering
    \includegraphics[width=1\linewidth]{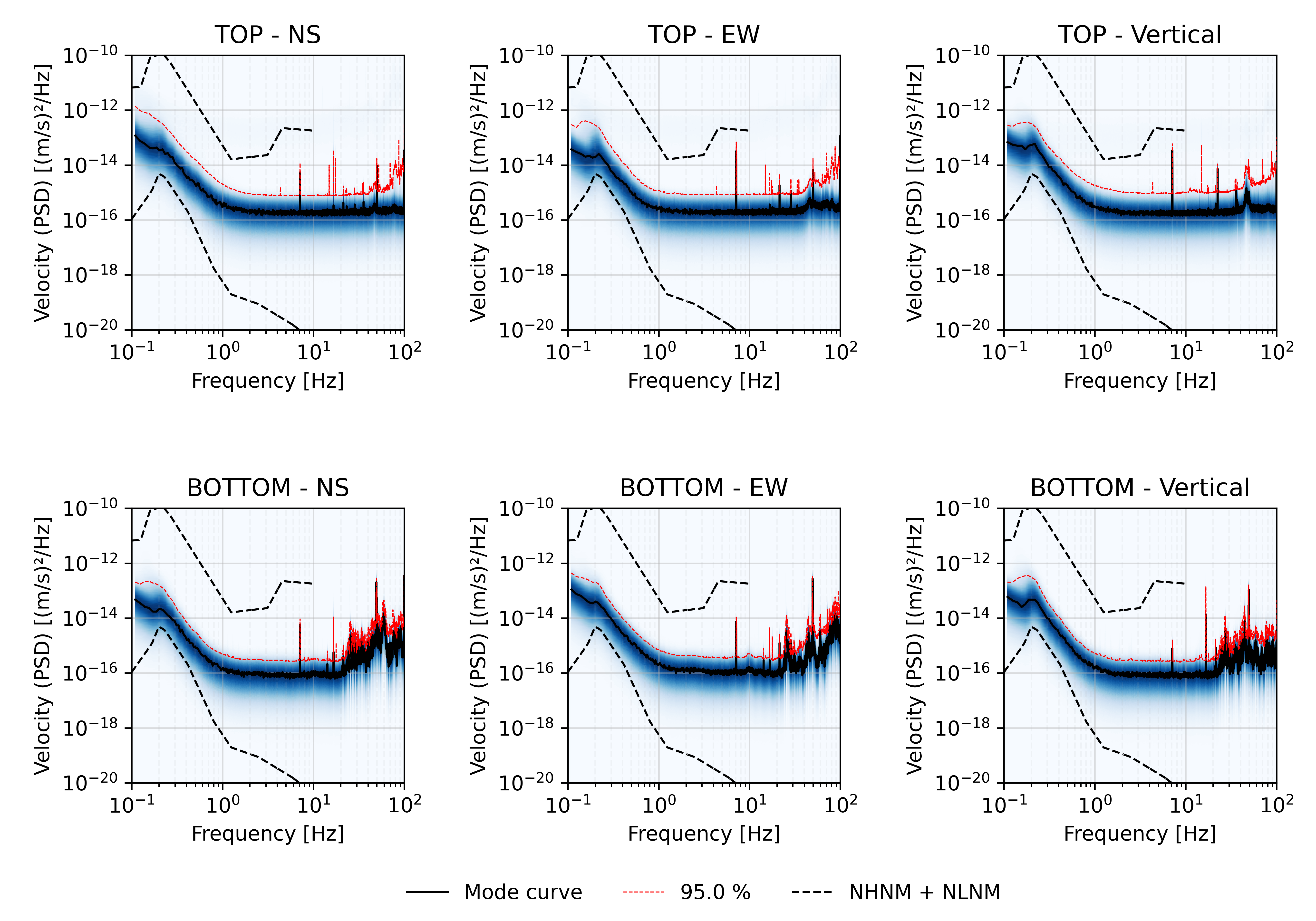}
    \caption{The PPSDs in velocity for each station and direction.}
    \label{fig:seismic-results_longterm}
\end{figure}

A global summary of the long-term ground motion is given in Figure~\ref{fig:seismic-results_longterm}, 
showing the PPSDs in velocity for each station and direction. Each graph includes the corresponding mode curve, a \SI{95}{\percent} curve (\SI{95}{\percent} of the data falls below this power), as well as the New Low Noise Model (NLNM) and New High Noise Model (NHNM)~\cite{peterson1993observations}. All graphs display similar behavior for frequencies below \SI{20}{\hertz}, above which the probability distributions of the BOTTOM stations shift towards higher values with increased noise. In contrast, the TOP station yielded a relatively constant power in velocity with less noise but a few peaks to higher values. Generally, three distinctive peaks are observable at approximately \SI{7}{\hertz}, \SI{16.7}{\hertz}, and \SI{50}{\hertz}. The \SI{7}{\hertz} peak has been identified from the local infrastructure (Cooling and ventilation equipments) while the \SI{16.7}{\hertz} peak corresponds to the frequency of the electrical power supply for the trains, and so is stronger at the BOTTOM station. The \SI{50}{\hertz} peak corresponds to the frequency of the power grid, and is also more pronounced at the BOTTOM station.\newline

\begin{figure}[b!]
    \centering    \includegraphics[width=1\linewidth]{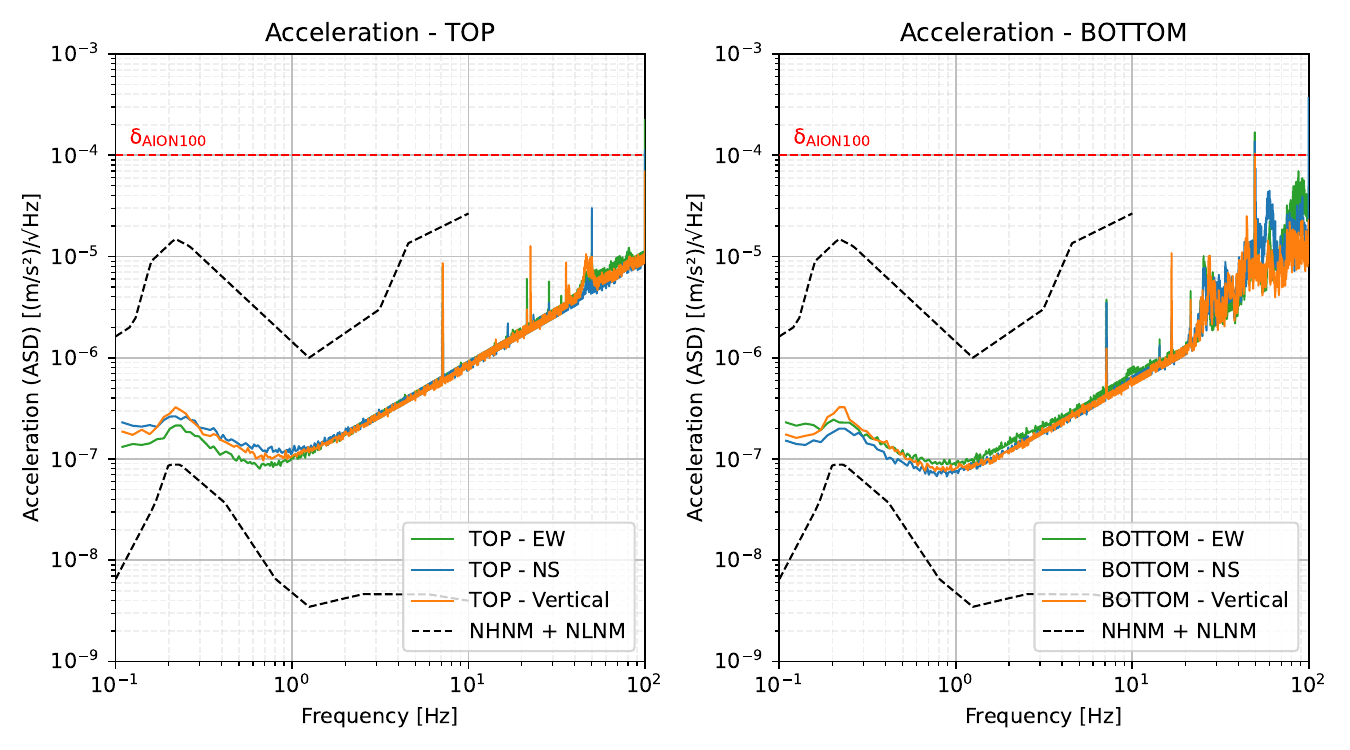}
    \caption{Acceleration Spectral Density mode curves at the TOP and BOTTOM stations compared to the NHNM and NLNM and to the AION-100 requirement.}
    \label{fig:seismic-results_mode-curves}
\end{figure}

By converting the mode curves from velocity power to acceleration amplitude, Figure~\ref{fig:seismic-results_mode-curves} places these results in relation to the reference requirement of AION-100. We see that the ground motion at both stations is within this limit of \SI{e-4}{(m/s^2)/\sqrt{Hz}} for frequencies from \SIrange{0.1}{100}{\hertz}. The acceleration amplitude exceeds the requirement only at the BOTTOM station at \SI{50}{\hertz}, where it reaches approximately \SI{1.7e-4}{(m/s^2)/\sqrt{Hz}}.\newline 

\begin{figure}[t!]
    \centering \hspace{0.2\linewidth}    \includegraphics[width=1\linewidth]{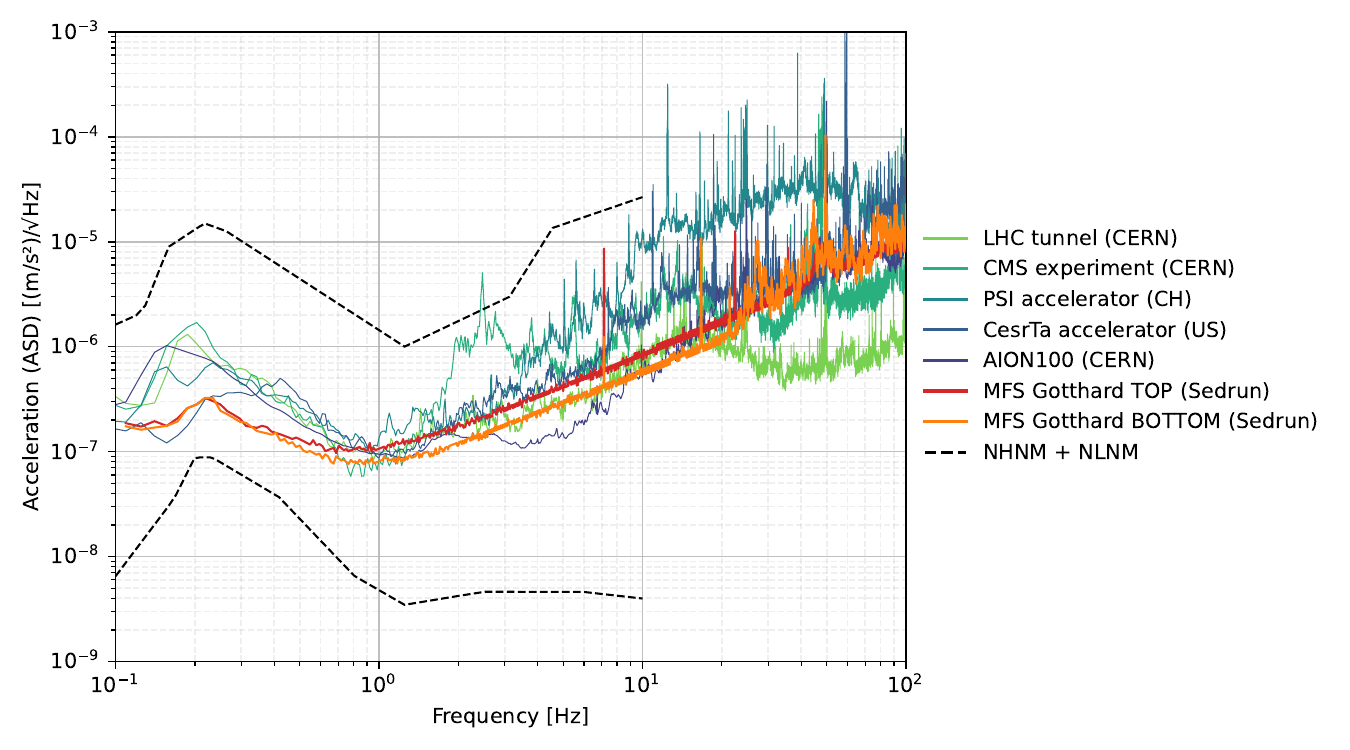}
    \caption{The acceleration ASD compared to other sites.}
    \label{fig:seismic-results_comparison}
\end{figure}

Figure~\ref{fig:seismic-results_comparison} illustrates the ground motion stability of the MFS site compared to data available from other particle accelerators and experiments. We see that below \SI{20}{\hertz} the amplitudes for both stations of the MFS site are generally lower compared to the ground motion observed at other sites.

\clearpage
\paragraph*{Impact on ground motion of passing trains}

 Based on a database provided by SBB, the train momentum was estimated for the full dataset using the maximum train speed and the train mass. To evaluate the impact of passing trains, a representative time window from \SIrange{8}{9}{\hour} (CEST) on 20 June 2025 was selected. During this interval, a total of five trains passed the MFS site, covering low, intermediate, and high momentum regimes. The characteristics of these trains are summarised in Table~\ref{tab:seismic_train-impact}. It should be noted that the reported train momentum values are indicative rather than exact.

\begin{table}[h!]
    \centering
    \caption{Trains passing in the time window from \SIrange{8}{9}{\hour} (CEST) on the 20/06/2025.}
    \begin{tabular}{c c c c c c c}
         \hline 
         Train & Time & Train & Max speed & Weight & Length & Momentum \\
         number & of passage & type & [\SI{}{\kilo\meter\per\hour}] & [t] & [\SI{}{\meter}] & [\SI{}{\kilo\gram\meter\per\second}] \\
         \hline
         1 & 08:12:58 & InterCity & 200 & 381 & 404 & \SI{42.3e6}{} \\
         2 & 08:18:13 & InterCity & 230 & 762 & 202 & \SI{24.3e6}{} \\
         3 & 08:26:47 & Goods train & 120 & 2310 & 389 & \SI{77.0e6}{} \\
         4 & 08:42:21 & InterCity & 230 & 381 & 202 & \SI{24.3e6}{} \\
         5 & 08:46:41 & EuroCity & 230 & 762 & 404 & \SI{48.7e6}{} \\
         \hline
    \end{tabular}
    \label{tab:seismic_train-impact}
\end{table}
\begin{figure}[h!]
    \centering
    % \hspace{0.2\linewidth}
    \includegraphics[width=1\linewidth]{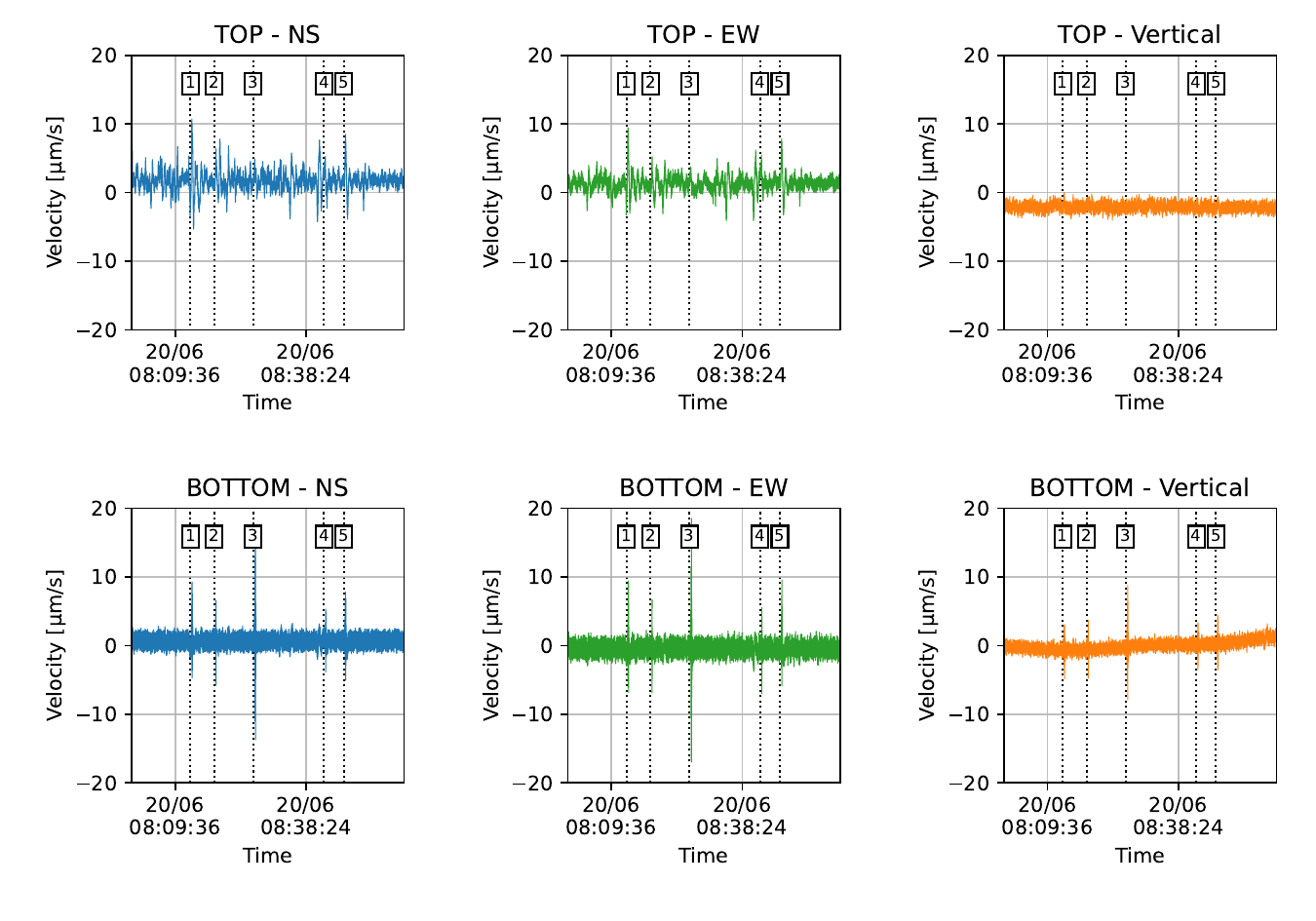}
    \caption{Velocity over time for both stations in all directions, recorded on the 20/06/2025 in the morning between 8h and 9h (CEST). The passages of five trains (see Table~\ref{tab:seismic_train-impact}) are indicated.}
    \label{fig:seismic-results_6-1b_vel-over-time}
\end{figure}

Figure~\ref{fig:seismic-results_6-1b_vel-over-time} displays the measured velocity as a function of  time for the two stations and all measurement directions. The ground motion excitation due to the five trains is clearly visible for the BOTTOM station. The graphs show a visual correlation between the momentum of each train and the maximum velocity excitation. However, we have not made a thorough analysis of the entire dataset to confirm this correlation, which would involve the integration of the velocity amplitude in the time domain rather than the maximum excitation. \newline %This is reasoned with the fact that the information on the trains are rather indicative and can deviate non-negligibly from the actual values.

\begin{figure}[t!]
    \centering
    % \hspace{0.2\linewidth}
    \includegraphics[width=1\linewidth]{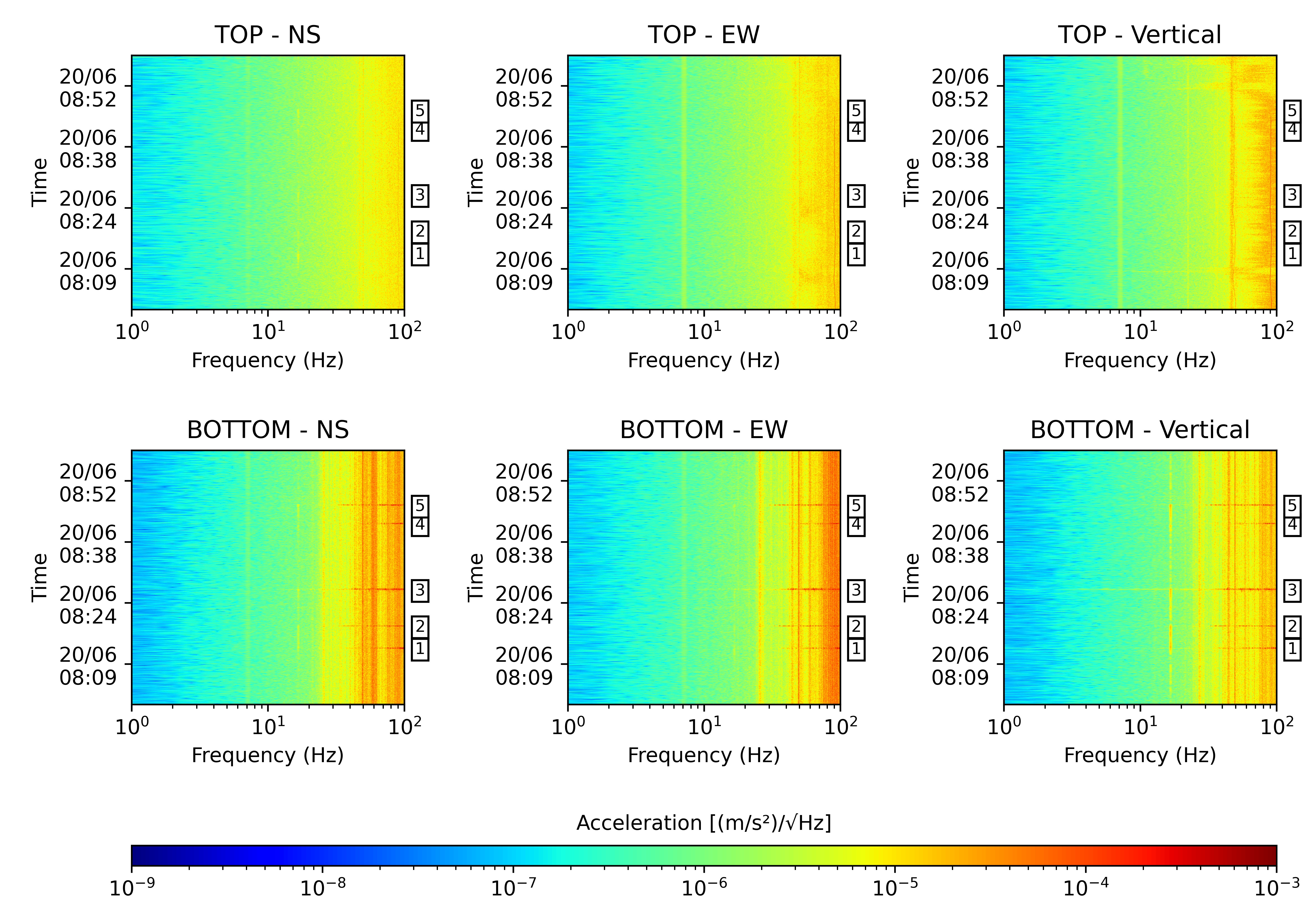}
    \caption{Spectrograms of the acceleration amplitudes for the data displayed in Figure~\ref{fig:seismic-results_6-1b_vel-over-time}.}
    \label{fig:seismic-results_6-1b_ASD-over-time}
\end{figure}

Figure~\ref{fig:seismic-results_6-1b_ASD-over-time} displays the same dataset but as a spectrogram of the amplitude of the acceleration. The PSDs were computed using blocks of 1024 samples and the Hanning window, but without any overlap. Due to the reduced frequency resolution, the graphs show the acceleration amplitude only in the range from \SIrange{1}{100}{\hertz}. 
Analogously to the time domain, all five trains can be identified in the frequency domain, and train 3 at around 8:27 shows the highest amplitudes, as already observed in the time domain. We see that the graphs indicate that the trains mainly affect the ground motion at higher frequencies. \newline

\clearpage

\begin{figure}[h!]
    \centering
    % \hspace{0.2\linewidth}
    \includegraphics[width=1\linewidth]{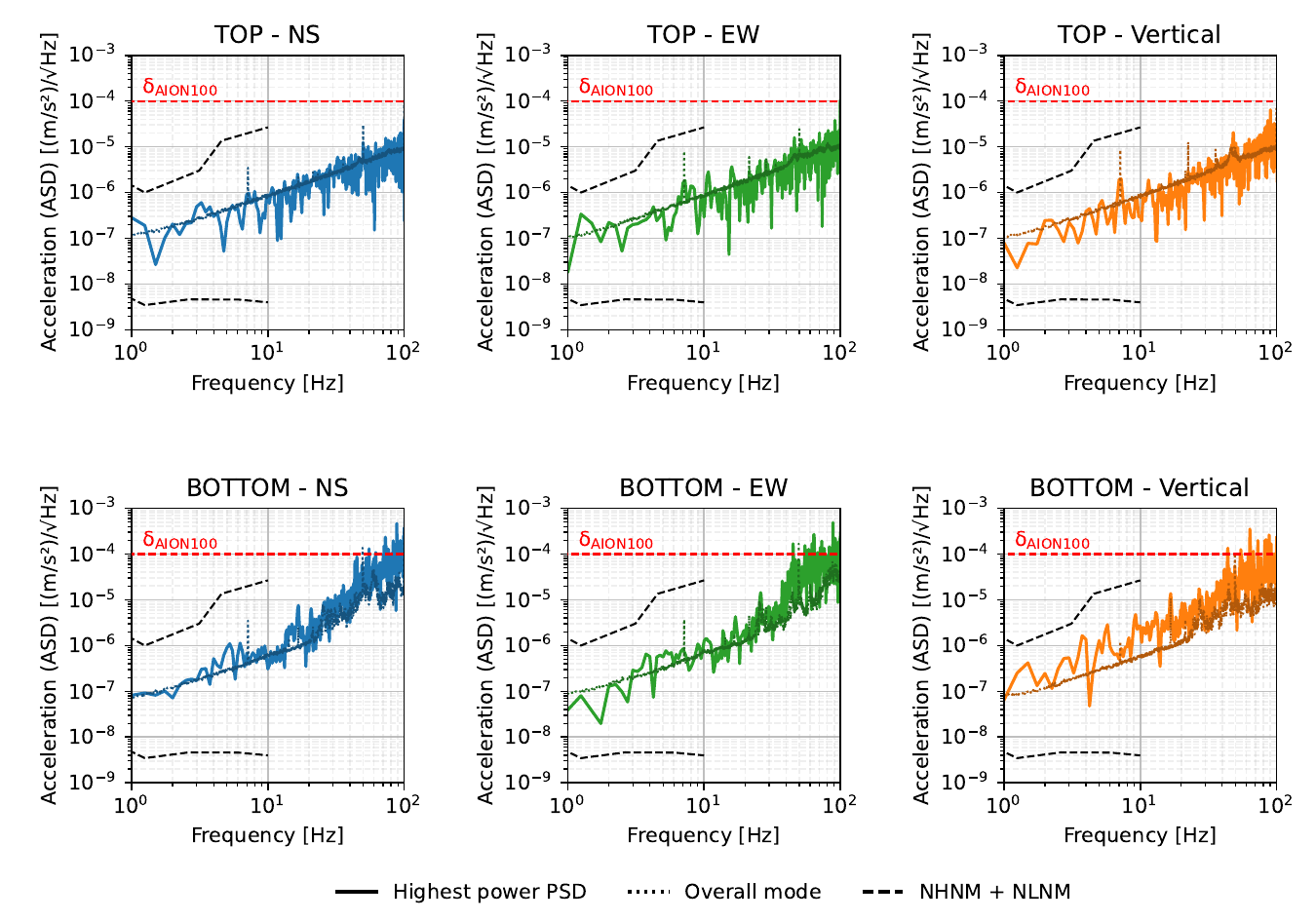}
    \caption{The strongest event compared in velocity amplitude to the mode curves of the long-term analysis, the NLNM and the NHNM  for the data displayed in Figure~\ref{fig:seismic-results_6-1b_vel-over-time}.}
    \label{fig:seismic-results_6-1b_worst-events}
\end{figure}

In Figure~\ref{fig:seismic-results_6-1b_worst-events} the strongest event of this time range (a \SI{4}{\second} window during the passage of train 3) is placed in relation to the overall mode curves, the AION-100 requirement, and the  NHNM and NLNM curves. The quantity displayed is the amplitude in acceleration.
At the BOTTOM station this event caused amplitude spikes in the high-frequency range above \SI{40}{\hertz} which exceed the AION-100 requirement with a maximum of about \SI{5e-4}{(\meter\per\square\second)/\sqrt{\hertz}}. As mentioned previously, the lower frequencies do not show any substantial impact. Only the vertical direction shows significantly increased ground motion.\newline 

Given these results, it can be concluded that individual trains lead to temporarily increased ground motion. However, the time window of significant excitation is limited to a range of \SIrange{10}{20}{\second}, which is small compared to the typical time between trains of several minutes.

\begin{figure}[h!]
    \centering
    % \hspace{0.2\linewidth}
    \includegraphics[width=1\linewidth]{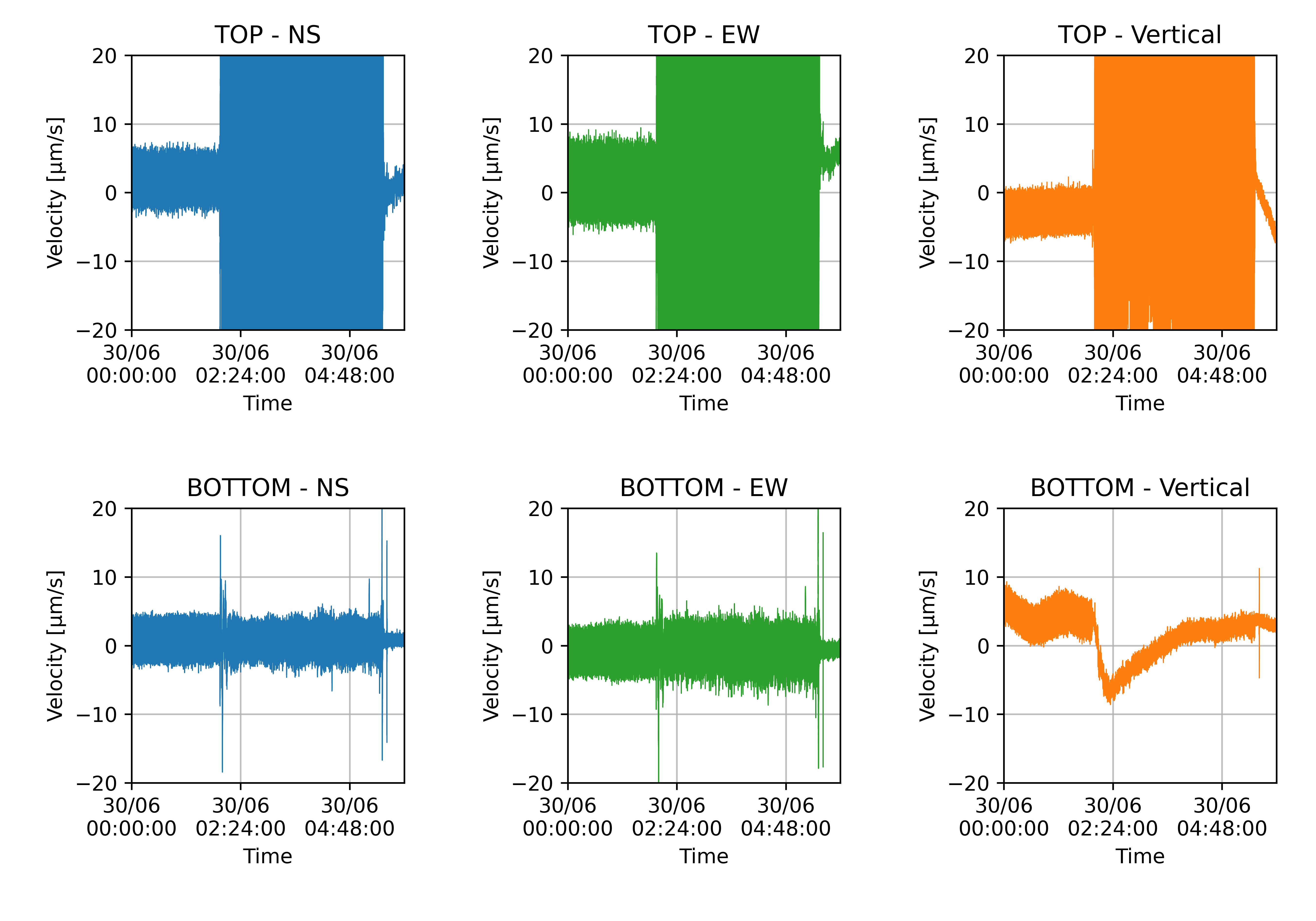}
    \caption{Velocity over time for both stations in all directions recorded on the 30/06/2025 in the morning between 0h and 6h (CEST).}
    \label{fig:seismic-results_6-1c_vel-over-time}
\end{figure}

\clearpage
\paragraph*{Impact on ground motion due to local infrastructure}
Over the full measurement duration, seven events with strong velocity amplitudes for multiple hours have been registered at the TOP station. All these events are related to maintenance activities, during which the ventilation is turned on. An example of such an event is given in Figure~\ref{fig:seismic-results_6-1c_vel-over-time} as velocity over time, which was recorded on the 30/06/2025 from \SIrange{0}{6}{\hour} (CEST). The plots indicate that the impact of the maintenance and ventilation affects mainly the TOP station, where the velocities reach far above $\pm\SI{20}{\micro\meter\per\second}$ for several hours.\newline

\begin{figure}
    \centering
    % \hspace{0.2\linewidth}
    \includegraphics[width=1\linewidth]{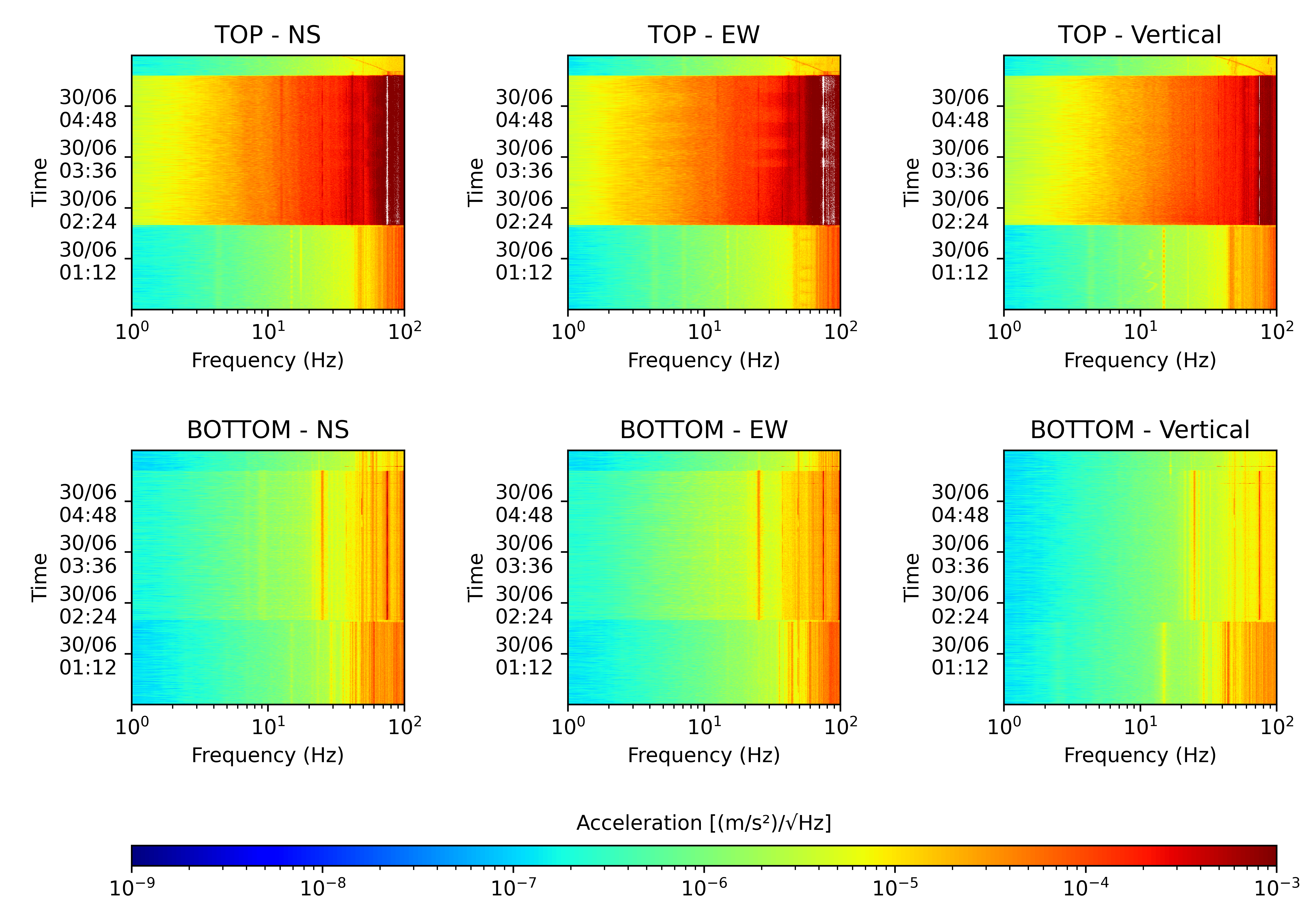}
    \caption{Spectrograms in acceleration amplitude for the data displayed in Figure~\ref{fig:seismic-results_6-1c_vel-over-time}.}
    \label{fig:seismic-results_6-1c_ASD-over-time}
\end{figure}

Spectograms of acceleration amplitude in the same time window are displayed in Figure~\ref{fig:seismic-results_6-1c_ASD-over-time}. The underlying PSDs were calculated with the same parameters as for the previous transient case: 1024 samples (= \SI{4}{\second}), no overlap and the Hanning window. As in the previous discussion, the frequency range is reduced to \SIrange{1}{100}{\hertz} due to the limited frequency resolution provided by the Fourier transform.
The graphs indicate that the acceleration amplitude at the TOP station is significantly elevated over all frequencies in the three directions (large excitation in all the directions). The BOTTOM station displays as well a sudden change in the acceleration amplitude with a vibration signature different from the standard operation on the three directions.
\newline

\begin{figure}
    \centering
    % \hspace{0.2\linewidth}
    \includegraphics[width=1\linewidth]{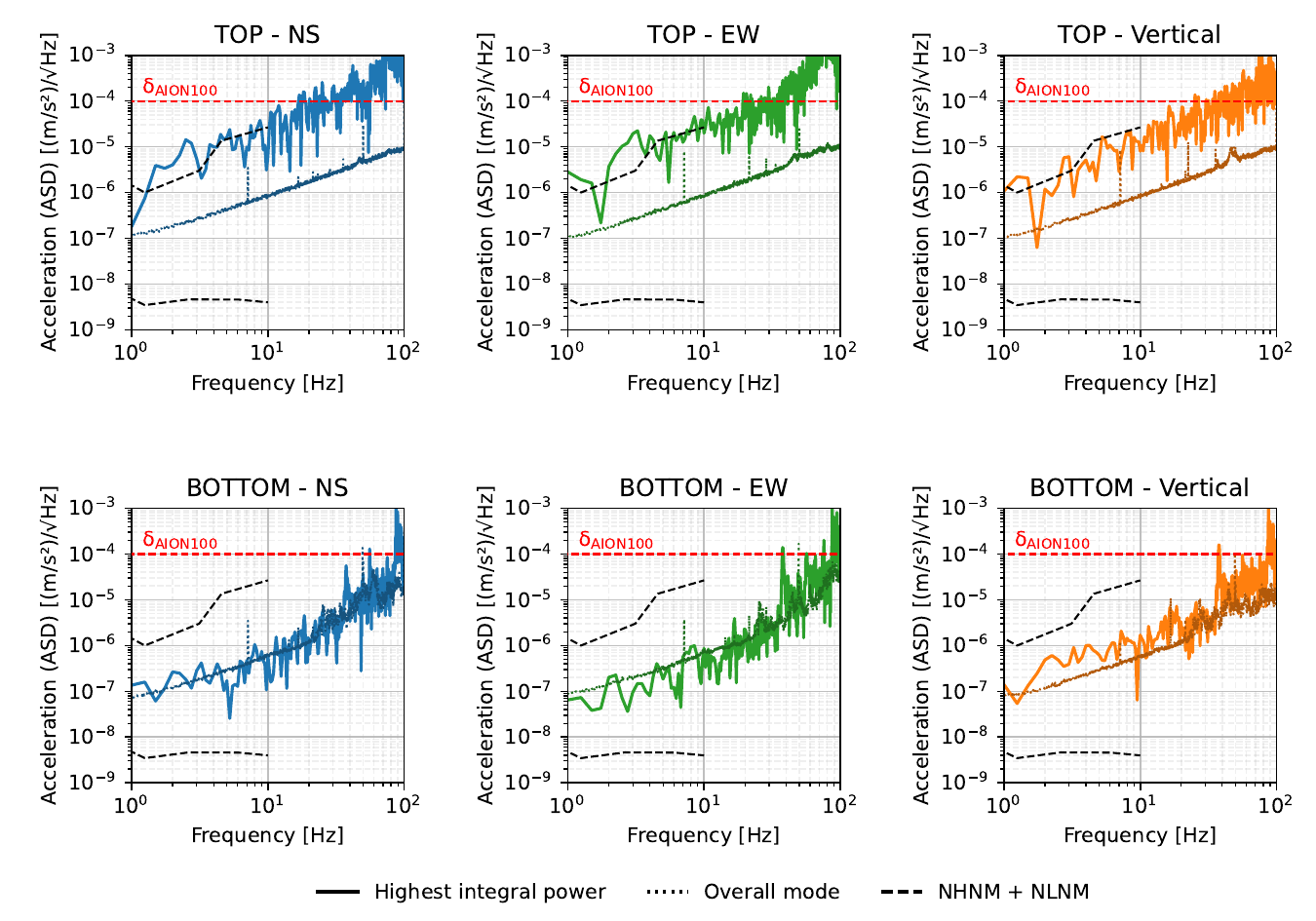}
    \caption{The highest integral-power in acceleration amplitude for the data displayed in Figure~\ref{fig:seismic-results_6-1c_vel-over-time} compared to the mode curves of the long-term analysis, the NLNM and the NHNM.}
    \label{fig:seismic-results_6-1c_worst-events}
\end{figure}

\clearpage
Figure~\ref{fig:seismic-results_6-1c_worst-events} presents the amplitude in acceleration for the window with the highest integral power at each station and direction. Additionally, the mode curves of the long-term data are given as comparison. The graphs show that the impact due to the maintenance event on the acceleration amplitude can reach up to 1 to 2 magnitudes over all frequencies, with the higher frequencies being more strongly affected.

\paragraph*{Data availability}
For data sharing, the resampled velocity channels are stored as .MSEED files, containing synchronized data for the TOP and BOTTOM stations with an uncertainty in synchronization of $\pm\SI{1}{\second}$. In order to maintain consistency, only the data available simultaneously at both stations is provided (more than \SI{73}{\percent} of total data recorded) and made available here:\newline

\href{https://edms.cern.ch/document/3195246/1
}{https://edms.cern.ch/document/3195246/1}

\FloatBarrier
\clearpage
\subsection{Electromagnetic noise results}
\label{subsec:ElectromagneticNoise}
AIs are sensitive to magnetic fields through various mechanisms as described in section \ref{subsubsec:EMnoise}. The magnetic field noise constraints are most critical in the detector sensitivity band of $\approx$ 20 mHz to 10 Hz. However, two prominent peaks are expected at the train traction frequency 16.7 Hz and at the mains frequency 50 Hz. The long term measurements and analysis focuses on this frequency band. Input to all analyses were datasets, containing x, y and z data from the fluxgate magnetometers and 3-axial coils, 50 million points for each axis. The power spectral density of each axis was calculated using the Welch method, using 1 million samples with \SI{50}{\percent} overlap and Hanning window. 

Inspecting the combined data of 1855 hours collected from 13.5.2025 to 9.7.2025 we have identified three typical scenarios relevant for the electromagnetic background:
\begin{enumerate}
  \item Standard remote MFS operation and an electromagnetically quiet environment, when the traction frequency 16.7~Hz is dominant;
  \item Remote operation, but MFS technical services are being operated/tested remotely - the train  traction frequency present as in 1) but 50~Hz mains and other components become visible;
  \item Maintenance periods with personnel present on site and the elevator operating. Electromagnetic field components as in 1) and 2) still present, but major perturbation by elevator operation is visible at each passage.
\end{enumerate}

For better readability, the following sections contain only the main measurement results. Detailed results, including all plots (Figures \ref{fig:Top_quiet_A} through \ref{fig:Bottom_elevator_sigma}), are attached in Appendix.

%John : What fractions of the time in which scenarios?
\newpage

\paragraph*{Magnetic field at the TOP station}
\label{subsec:ElectromagneticNoise_Top}

\begin{figure}[h!]
    \centering
    \includegraphics[width=1\linewidth]{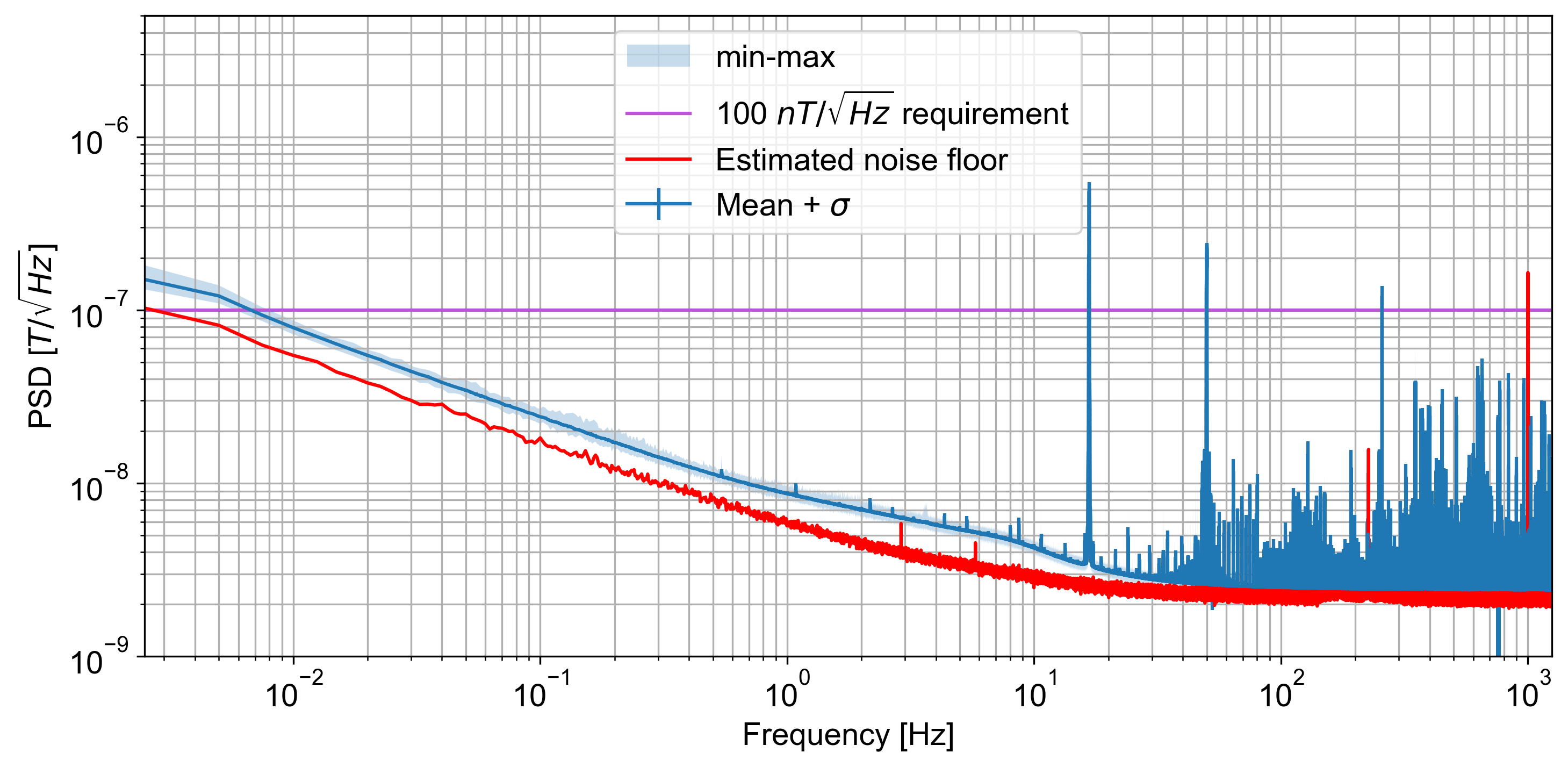}
    \caption{Distribution of PSDs of all 101 "quiet" datasets recorded at the TOP station.}
    \label{fig:Top_quiet_sigma_loc}
\end{figure}

Measurements in a typical "quiet" magnetic field background (case 1) are shown in Figures \ref{fig:Top_quiet_A}, \ref{fig:Top_quiet_B} and \ref{fig:Top_quiet_C}. The traction current component at 16.7~Hz is visible at varying amplitude, corresponding to the train traffic. A certain level of 50~Hz mains (and its harmonics) is present, otherwise no magnetic activity is observed in the spectrum of interest. We note that the fluxgate magnetometer probe was installed in the vicinity of a steel-reinforced concrete wall, so the DC component of the magnetic field (the Earth's magnetic field) is distorted, as discussed later. 
%in \autoref{subsec:ElectromagneticNoise_Elevator}, \nameref{subsec:ElectromagneticNoise_Elevator}.
A summary of all "quiet" datasets (case 1), where no external perturbation was observed is shown in Figures \ref{fig:Top_quiet_sigma_loc}, \ref{fig:Top_quiet_minmax} and \ref{fig:Top_quiet_sigma}. This represents 101 datasets out of the total of 133, i.e., 561 hours out of 738 hours. 

%John : ????

\FloatBarrier
Electromagnetic signatures of remote operation, but with MFS technical services operated/tested remotely (case 2) are shown in Figures \ref{fig:Top_case2_A}, \ref{fig:Top_case2_B}, \ref{fig:Top_case2_B_detail} and \ref{fig:Top_case2_C}. The traction frequency and 50~Hz mains are still present, with additional components becoming visible, Fig. \ref{fig:Top_case2_B_detail} shows in detail the spectrum from DC to 60~Hz.  Significant activity is measured below 10~Hz. It is thought that this is not a real low-frequency magnetic field, but rather a measurement artifact due to vibration of the fluxgate probe in the static terrestrial magnetic field when exposed to a strong airflow at the top of the elevator/ventilation shaft. This activity is visible also in the PSD plot. 

\begin{figure}[h!]
    \centering
    \includegraphics[width=1\linewidth]{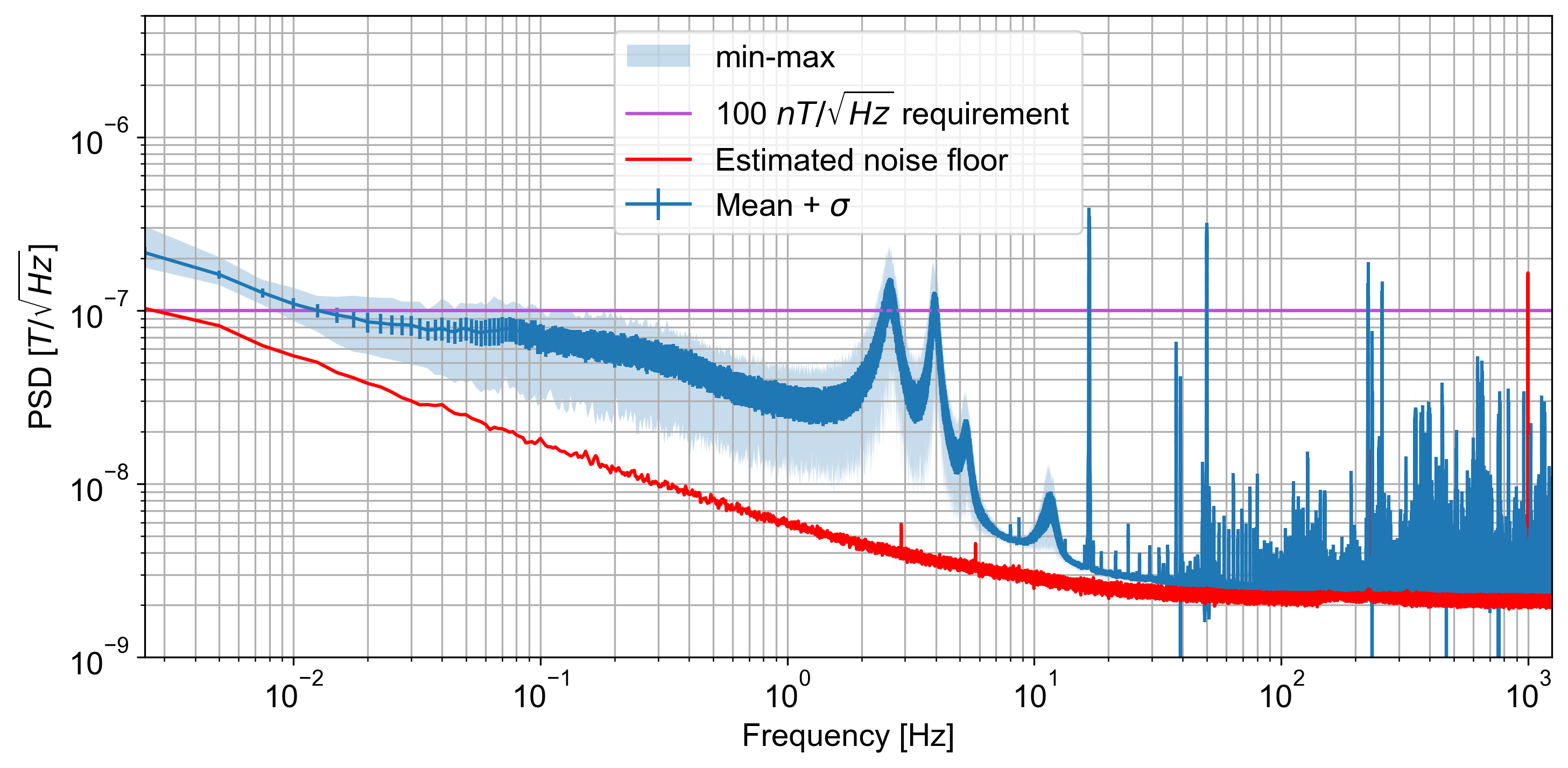}
    \caption{The PSD distributions of all 11 datasets with technical services operated recorded at the TOP station.}
    \label{fig:Top_services_sigma_loc}
\end{figure}

A summary of all acquisitions with MFS technical services running (case 2) is shown in Figures \ref{fig:Top_services_sigma_loc}, \ref{fig:Top_services_minmax} and \ref{fig:Top_services_sigma}. This represents 11 datasets out of total 133, i.e. 61 hours out of 738 hours. The services are typically being tested/operated in the night, for a period of about 6 hours. The low frequency components are believed to be a measurement artifact introduced by vibration of the fluxgate sensor due to a strong air-flow.

\FloatBarrier

\begin{figure}[h!]
    \centering
    \includegraphics[width=1\linewidth]{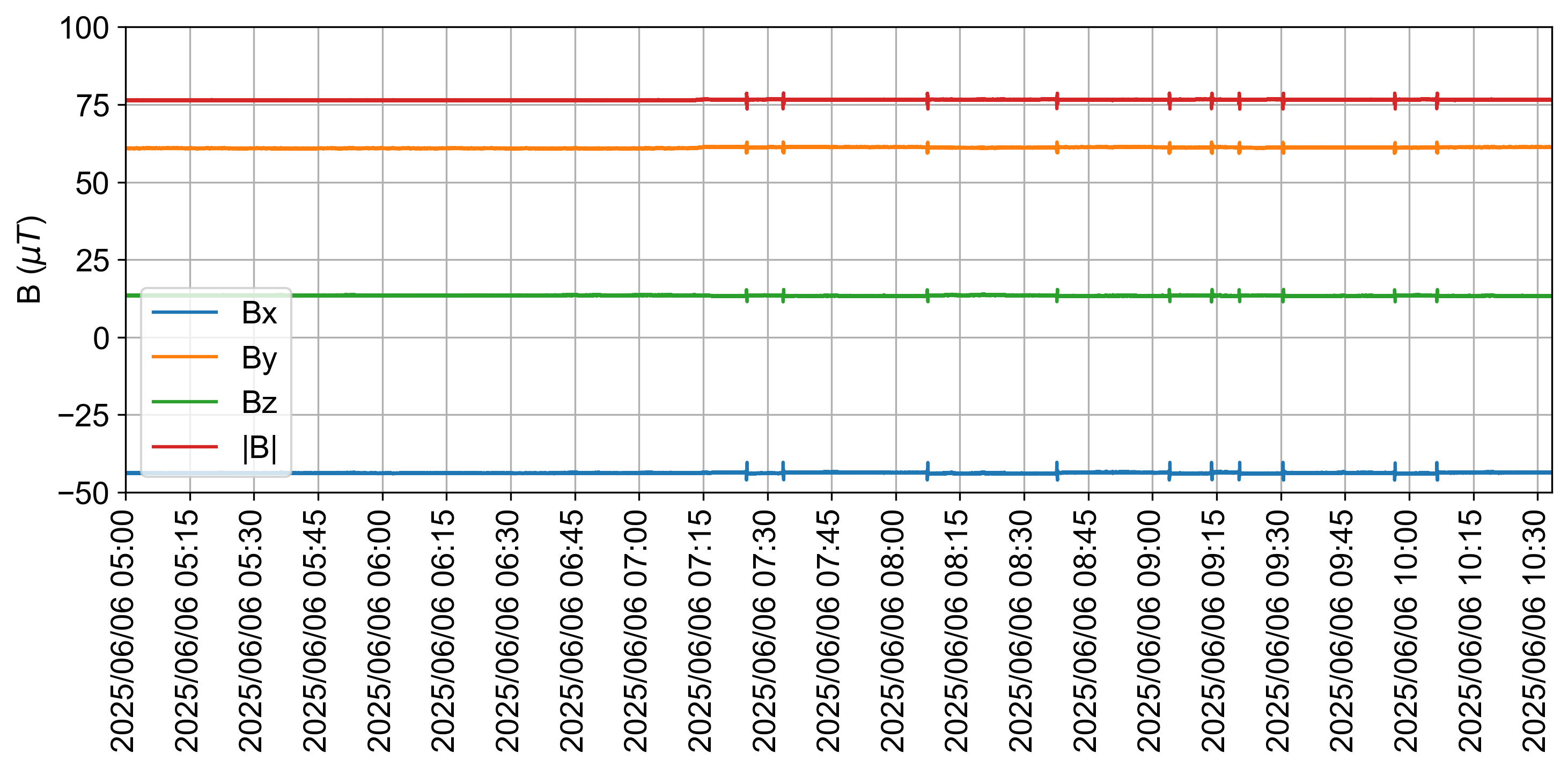}
    \caption{The magnetic field background with the elevator passing through the shaft: the B-field as a function of time. The peaks indicate the elevator passage (up or down).}
    \label{fig:Top_case3_A_loc}
\end{figure}

The third electromagnetic signature we have identified was of maintenance periods with personnel present on site and the elevator operating. The electromagnetic field components of regular operation (cases 1 and 2) are still present. A magnetic perturbation is introduced by each elevator passage as shown in Figures \ref{fig:Top_case3_A_loc}, \ref{fig:Top_case3_A}, \ref{fig:Top_case3_B} and \ref{fig:Top_case3_C}. The elevator passage generates additional activity in the very low frequency part of the PSD. 

\begin{figure}[h!]
    \centering
    \includegraphics[width=1\linewidth]{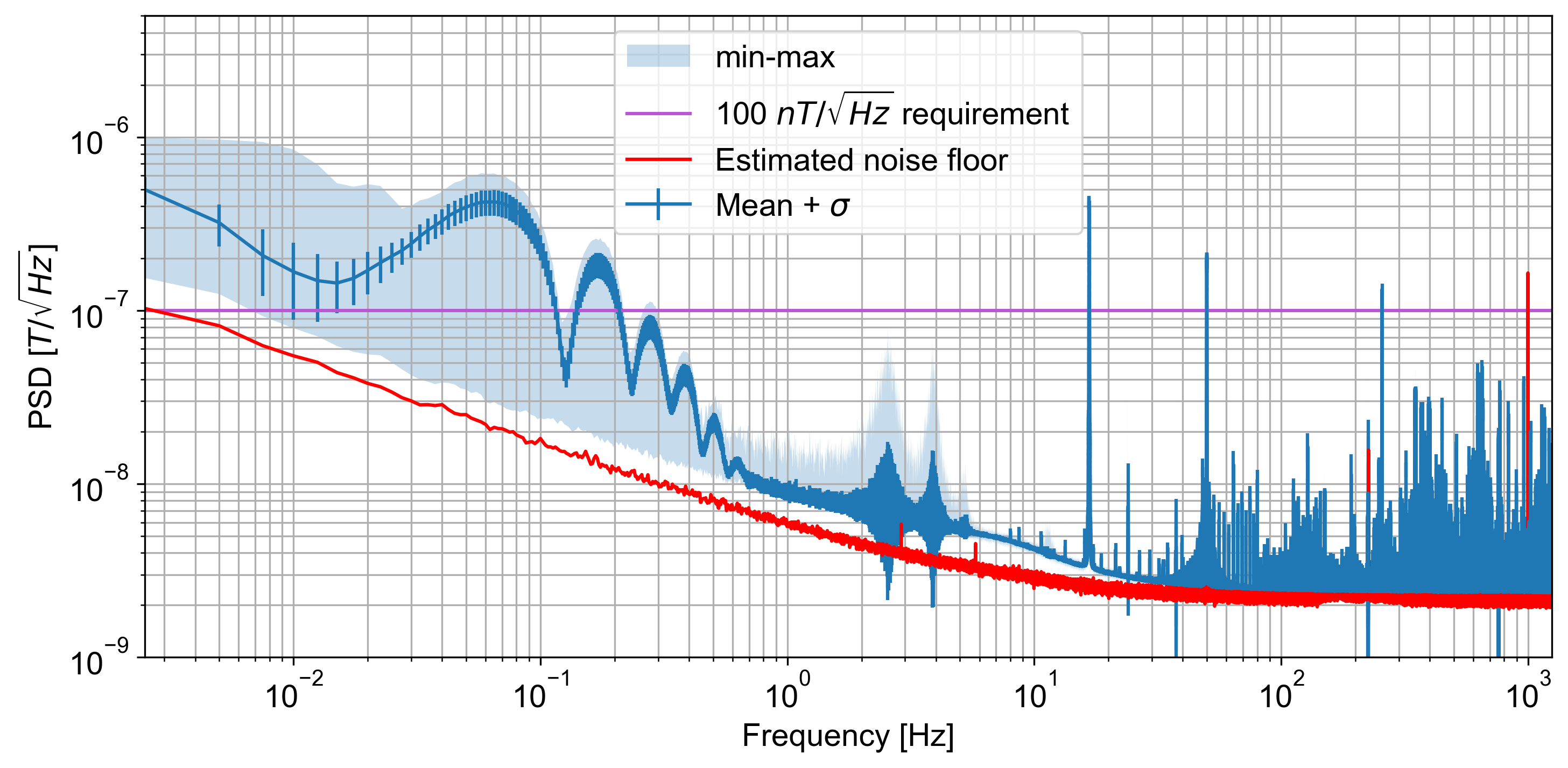}
    \caption{The PSD distributions of all 22 datasets recorded at the TOP station with the MFS elevator being operated.}
    \label{fig:Top_elevator_sigma_loc}
\end{figure}

A summary of all datasets with the elevator operated (case 3) is shown in Figures \ref{fig:Top_elevator_sigma_loc}, \ref{fig:Top_elevator_minmax} and \ref{fig:Top_elevator_sigma}. This represents 22 datasets out of the total of 133, i.e., 122 hours out of 738 hours. The elevator is typically operated within normal working hours during MFS maintenance periods. 

\FloatBarrier

% --------------------------------------------------------------------------
\paragraph*{Magnetic field at the BOTTOM station}
\label{subsec:ElectromagneticNoise_Bottom}
The magnetic field at the BOTTOM station is dominated by effects of the train traction current and operation of the elevator. The 16.7~Hz traction component is considerably higher than at the TOP station. The BOTTOM location is also an end station for the elevator, which is often parked there. Therefore, unlike the TOP station where the elevator is only briefly present, the magnetic field perturbation at the BOTTOM station is persistent, lasting as long as the elevator cage is parked there.  

\begin{figure}[h!]
    \centering
    \includegraphics[width=1\linewidth]{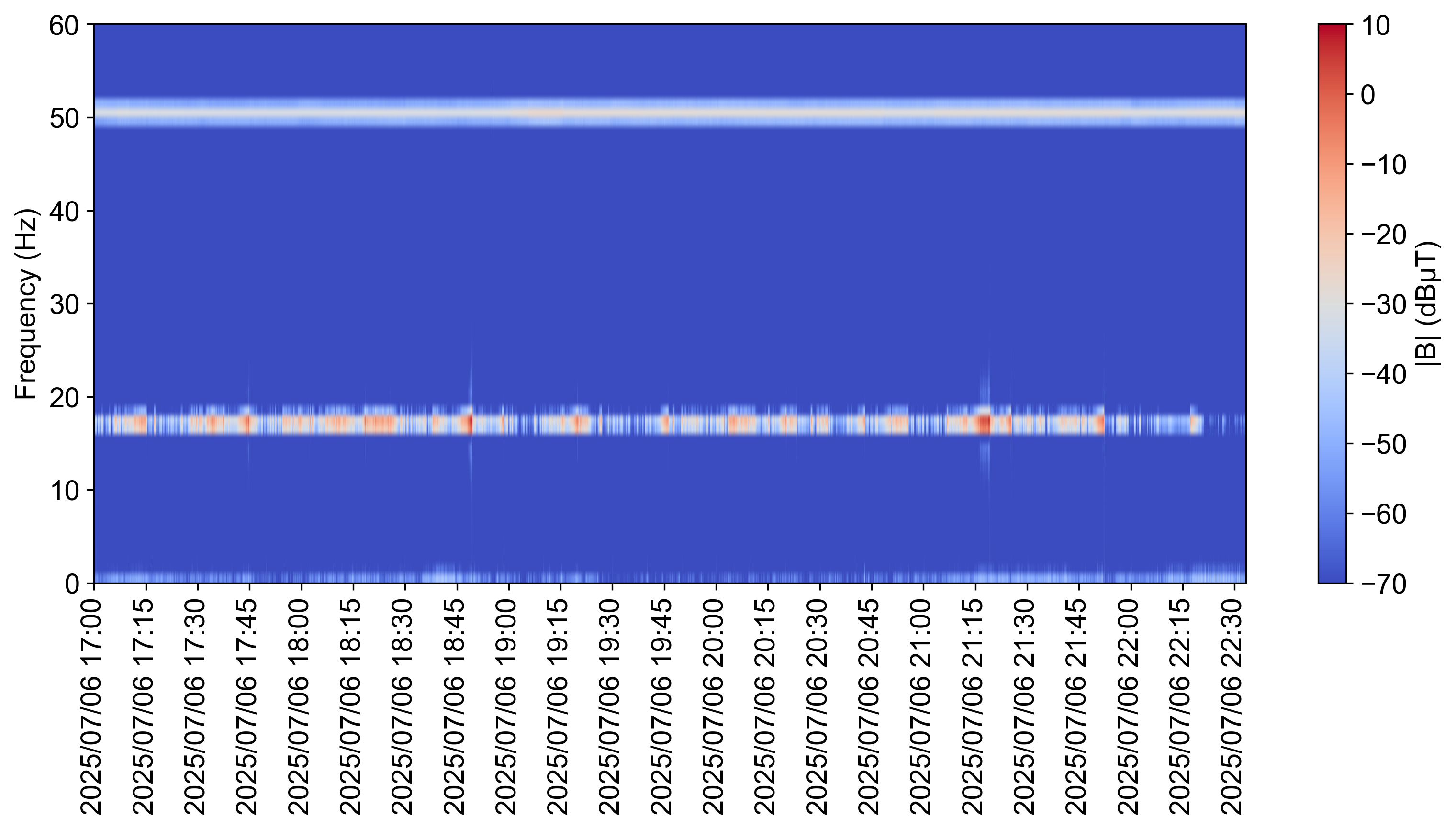}
    \caption{Heavy traffic example - spectrogram.}
    \label{fig:Bottom_spectrogram_traffic_loc}
\end{figure}

\begin{figure}[h!]
    \centering
    \includegraphics[width=1\linewidth]{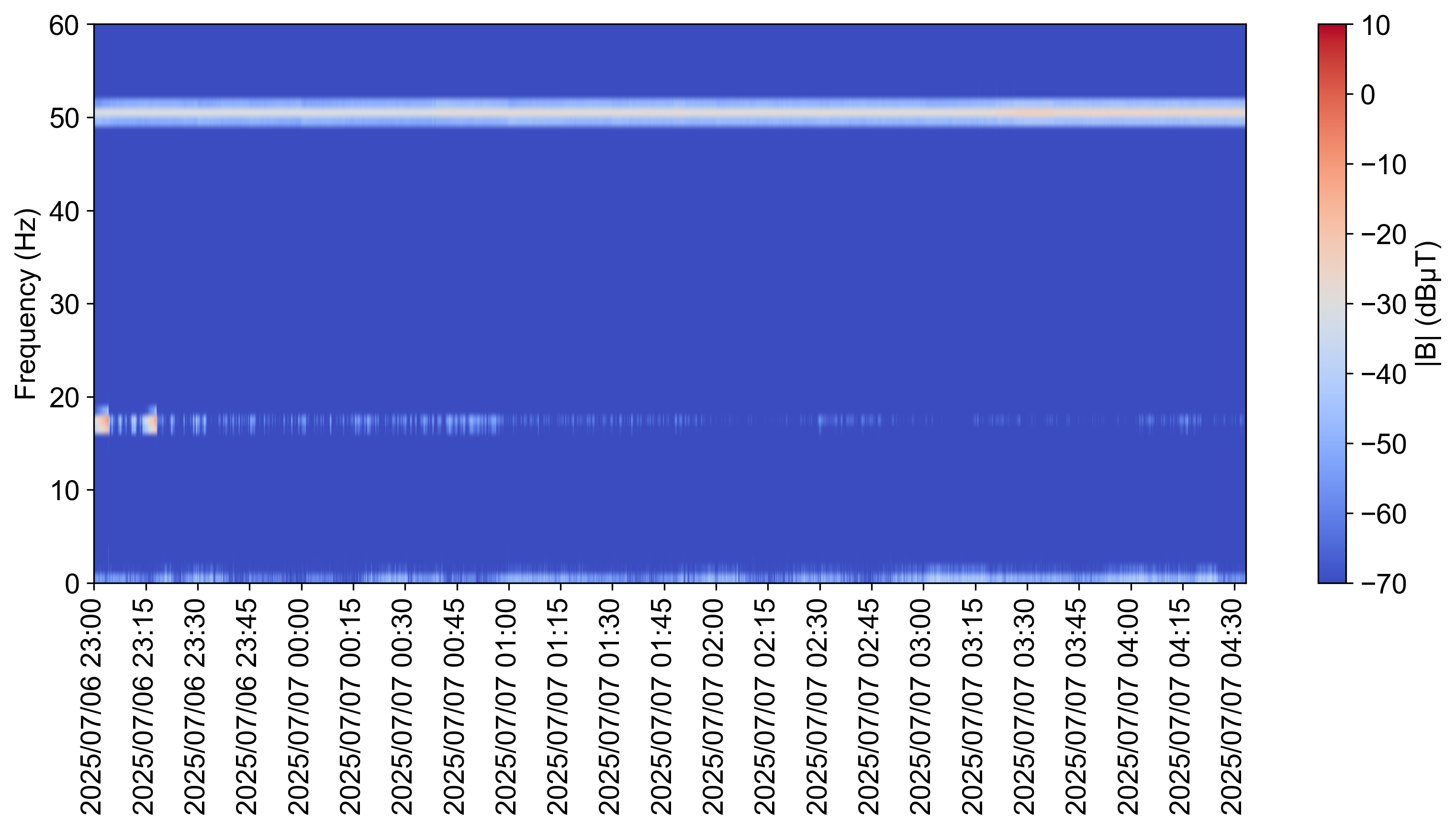}
    \caption{No traffic example - spectrogram.}
    \label{fig:Bottom_spectrogram_notraffic_loc}
\end{figure}

The magnetic field due to the traction current is clearly visible in the measurement and can be correlated with the traffic in the tunnel. Figure \ref{fig:Bottom_time_traffic} shows a time domain measurement in heavy traffic condition, whereas Fig. \ref{fig:Bottom_time_notraffic} covers the following night with very little traffic. The difference is even more visible in the spectrograms: Figures \ref{fig:Bottom_spectrogram_traffic_loc} and \ref{fig:Bottom_spectrogram_traffic} show the case with traffic, and Figures \ref{fig:Bottom_spectrogram_notraffic_loc} and \ref{fig:Bottom_spectrogram_notraffic} with no traffic. 

\begin{figure}[h!]
    \centering
    \includegraphics[width=1\linewidth]{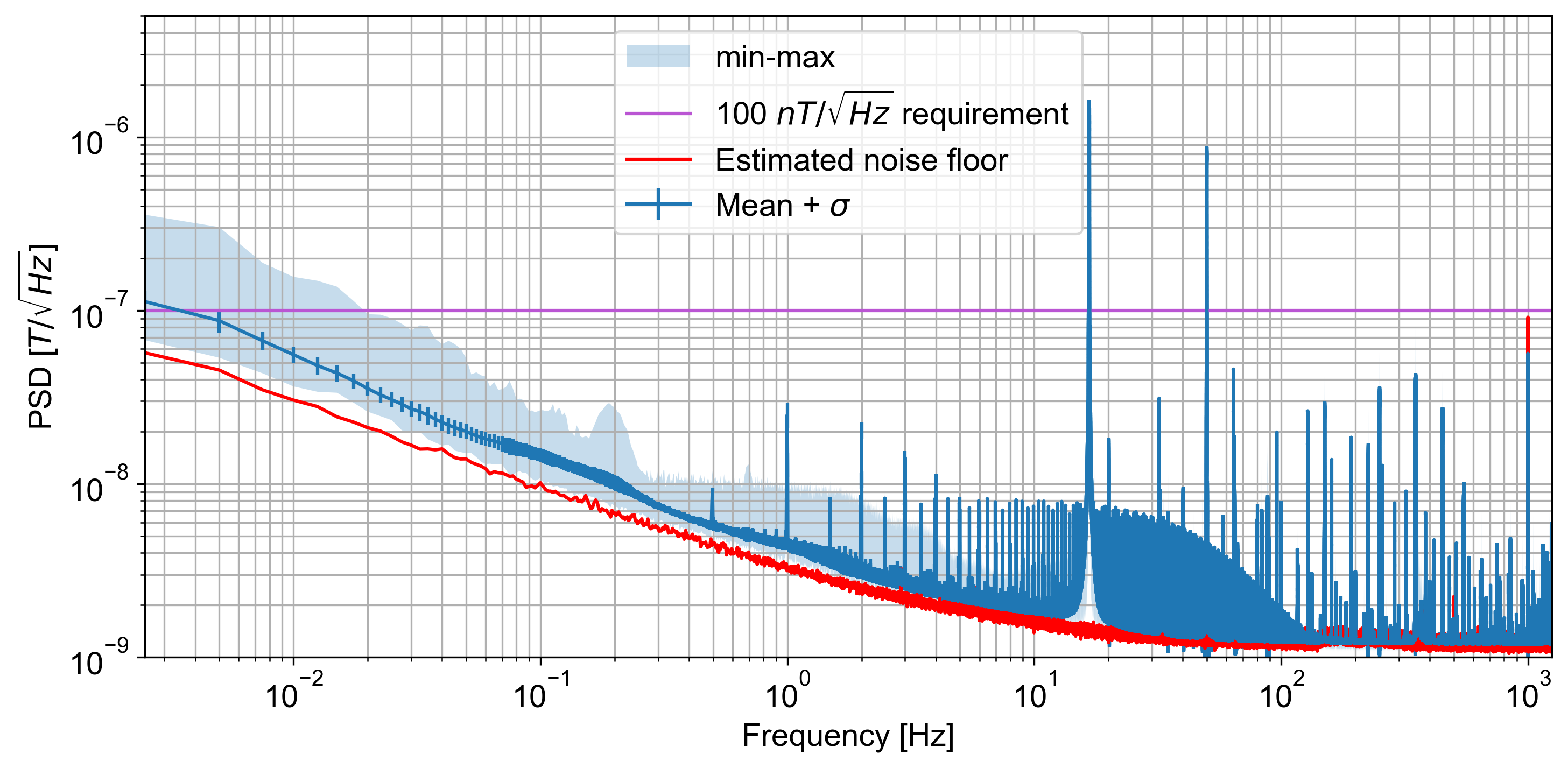}
    \caption{PSD distributions of all 165 "quiet" datasets measured at the BOTTOM station.}
    \label{fig:Bottom_quiet_sigma_loc}
\end{figure}

\FloatBarrier
Similarly to the TOP station, we observe "quiet" periods, in other words regular operation, with no additional perturbation (case 1). Summary plots of the magnetic noise are shown in Figures \ref{fig:Bottom_quiet_minmax}, \ref{fig:Bottom_quiet_sigma_loc} and \ref{fig:Bottom_quiet_sigma}. They are based on 165 datasets out of the total of 201, i.e.,  917 hours out of 1117 observed hours for the BOTTOM station. 

\FloatBarrier
No perturbation from the technical services operation was observed at the BOTTOM station. The measurement equipment was installed next to the elevator cage (see Figure \ref{fig:MeasEquipmentUnderground_A}), and not subject to air stream of the powerful ventilation system. This supports our hypothesis that the low-frequency perturbation observed at the top station was caused by vibration of the fluxgate sensor, and not an actual electromagnetic disturbance. In all of the BOTTOM station data (1117 hours), we have not identified signals of this origin (no case 2). 

\begin{figure}[h!]
    \centering
    \includegraphics[width=1\linewidth]{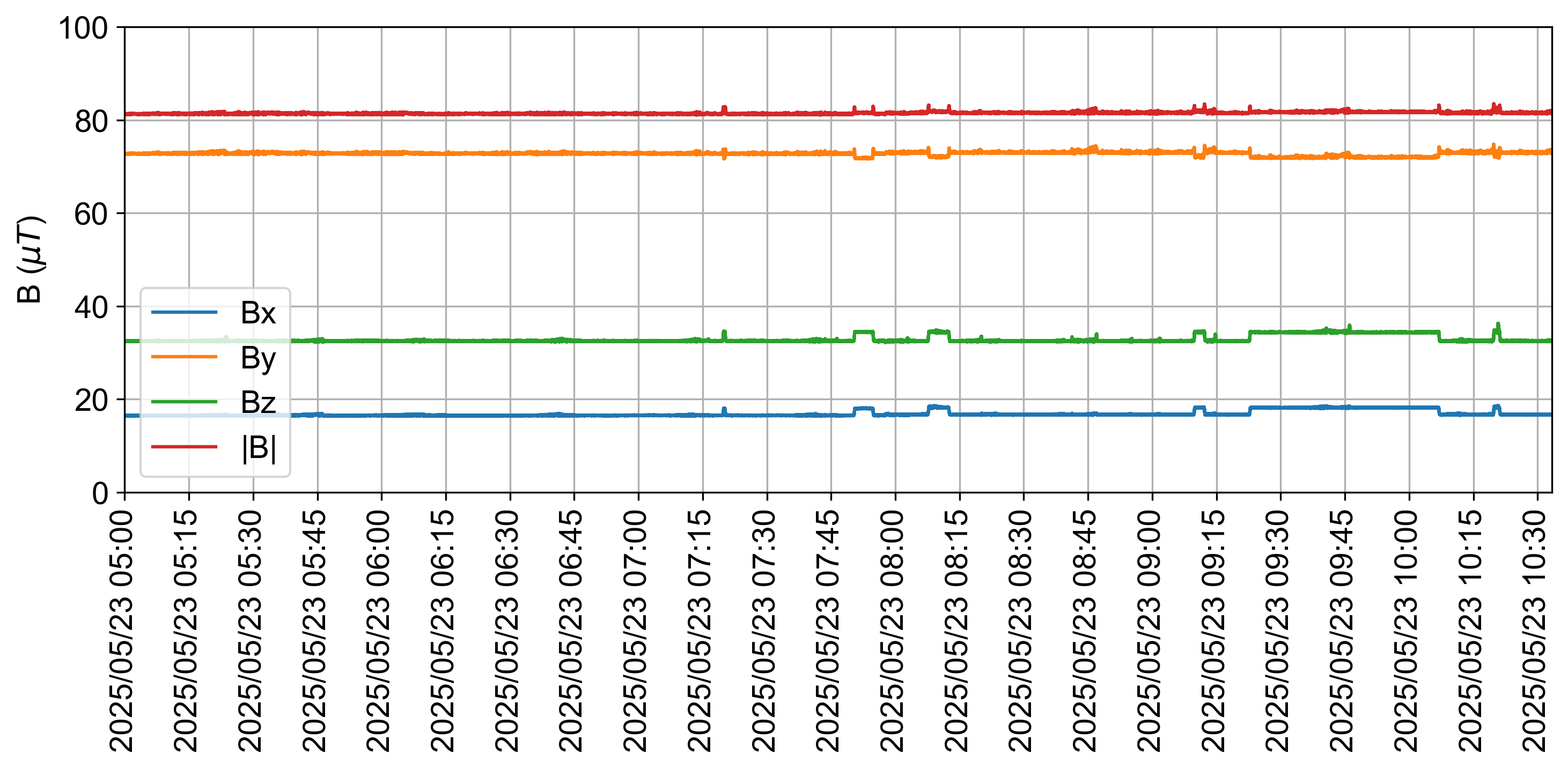}
    \caption{Perturbations of the magnetic field at the BOTTOM station when the elevator is stationary there.}
    \label{fig:Bottom_elevator_perturbation_loc}
\end{figure}

The elevator operation has a very different signature at the BOTTOM station that of the TOP station. When the elevator cage is parked there, a permanent magnetic field perturbation is present until the elevator is removed as can be seen in Figures \ref{fig:Bottom_elevator_perturbation_loc} and \ref{fig:Bottom_elevator_perturbation}.

\begin{figure}[h!]
    \centering
    \includegraphics[width=1\linewidth]{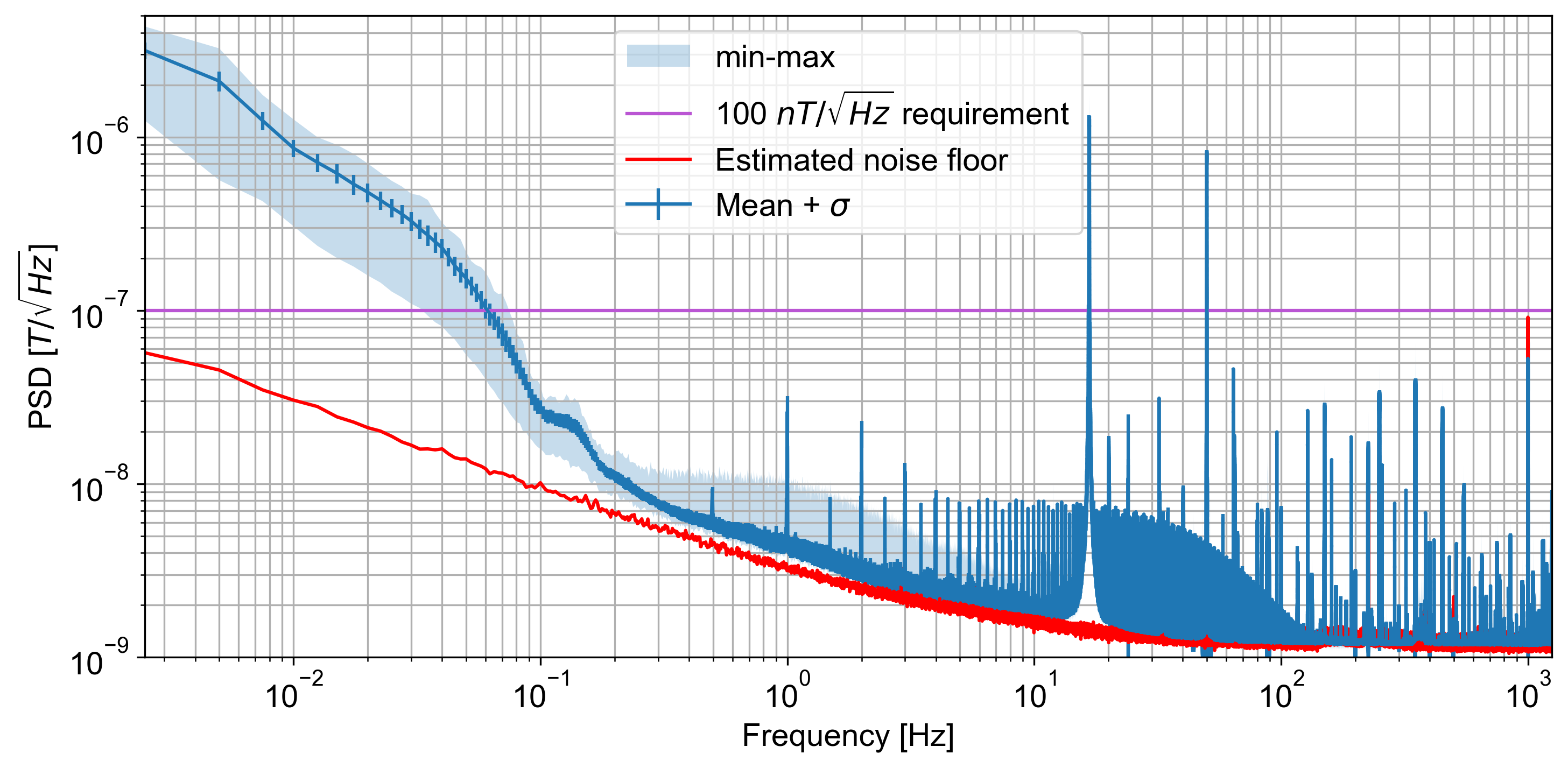}
    \caption{PSD distributions of all 34 datasets recorded at the BOTTOM station while the MFS elevator was being operated.}
    \label{fig:Bottom_elevator_sigma_loc}
\end{figure}

A summary of all datasets during elevator operation (case 3) is shown in Figures \ref{fig:Bottom_elevator_minmax}, \ref{fig:Bottom_elevator_sigma_loc} and \ref{fig:Bottom_elevator_sigma}. These represent 34 datasets out of the total of 201, i.e., 189 hours out of 1117 hours. The elevator is typically operated within normal working hours during the MFS maintenance periods. 

\FloatBarrier

% --------------------------------------------------------------------------
\newpage
\paragraph*{Magnetic field along the vertical elevator shaft}
\label{subsec:ElectromagneticNoise_Elevator}

The 800~m vertical elevator shaft of MFS is enclosed by steel-reinforced concrete rings. Ferromagnetic material contributes to the electromagnetic background in the shaft, and structural steel is known to be prone to remanent magnetization when exposed to mechanical work (e.g., bending), or welding. This effect was observed also at the MFS Sedrun. Measurements during the setting-up phase showed that the fluxgate magnetometer with full scale $\pm100~\mu T$ was easily getting fully saturated when located in the vicinity of a concrete wall or the elevator shaft cage. 

\begin{figure}[h!]
    \centering
    \includegraphics[width=1\linewidth]{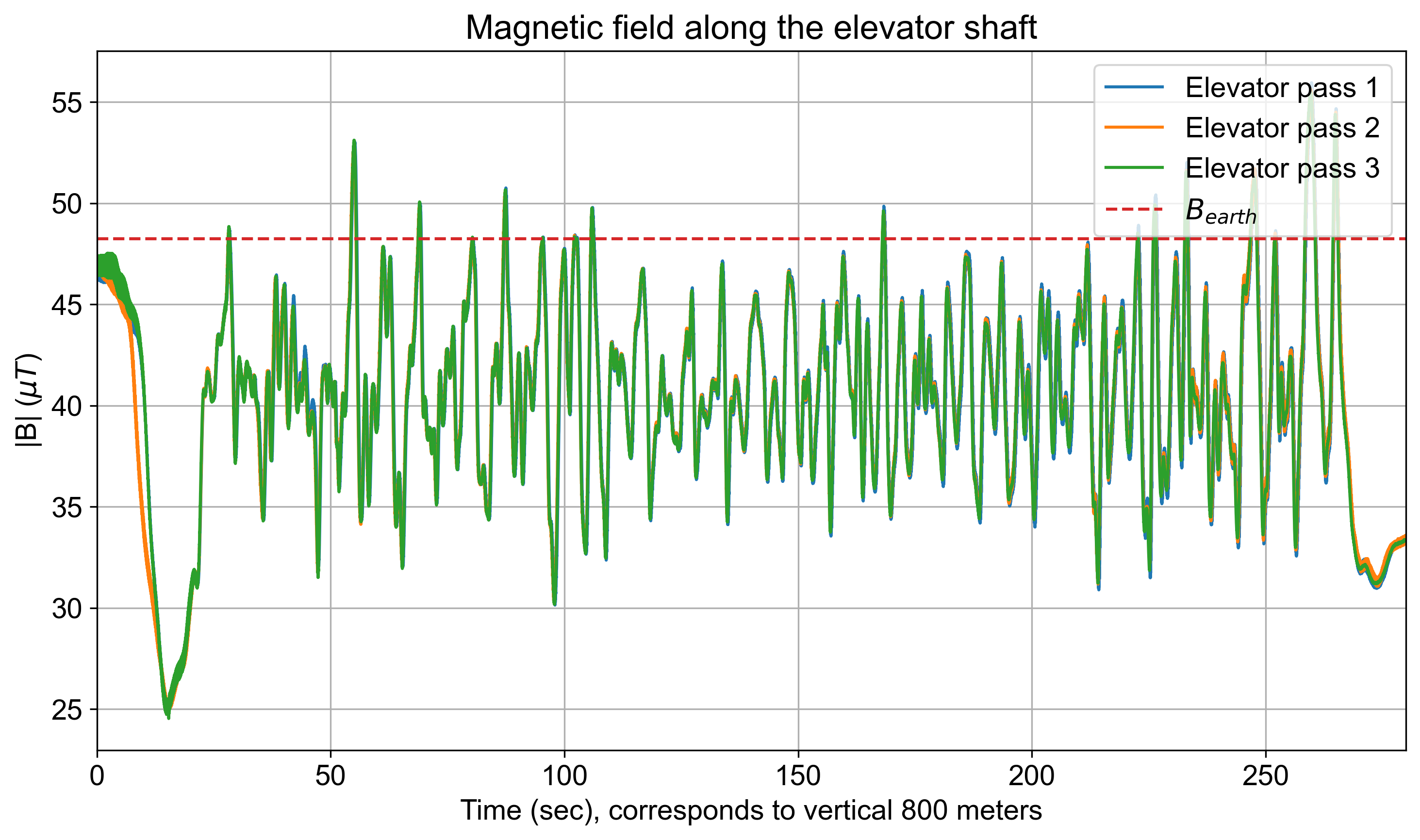}
    \caption{The static magnetic field along the elevator shaft.}
    \label{fig:Static_field_elevator}
\end{figure}

\begin{figure}[h!]
    \centering
    \includegraphics[width=1\linewidth]{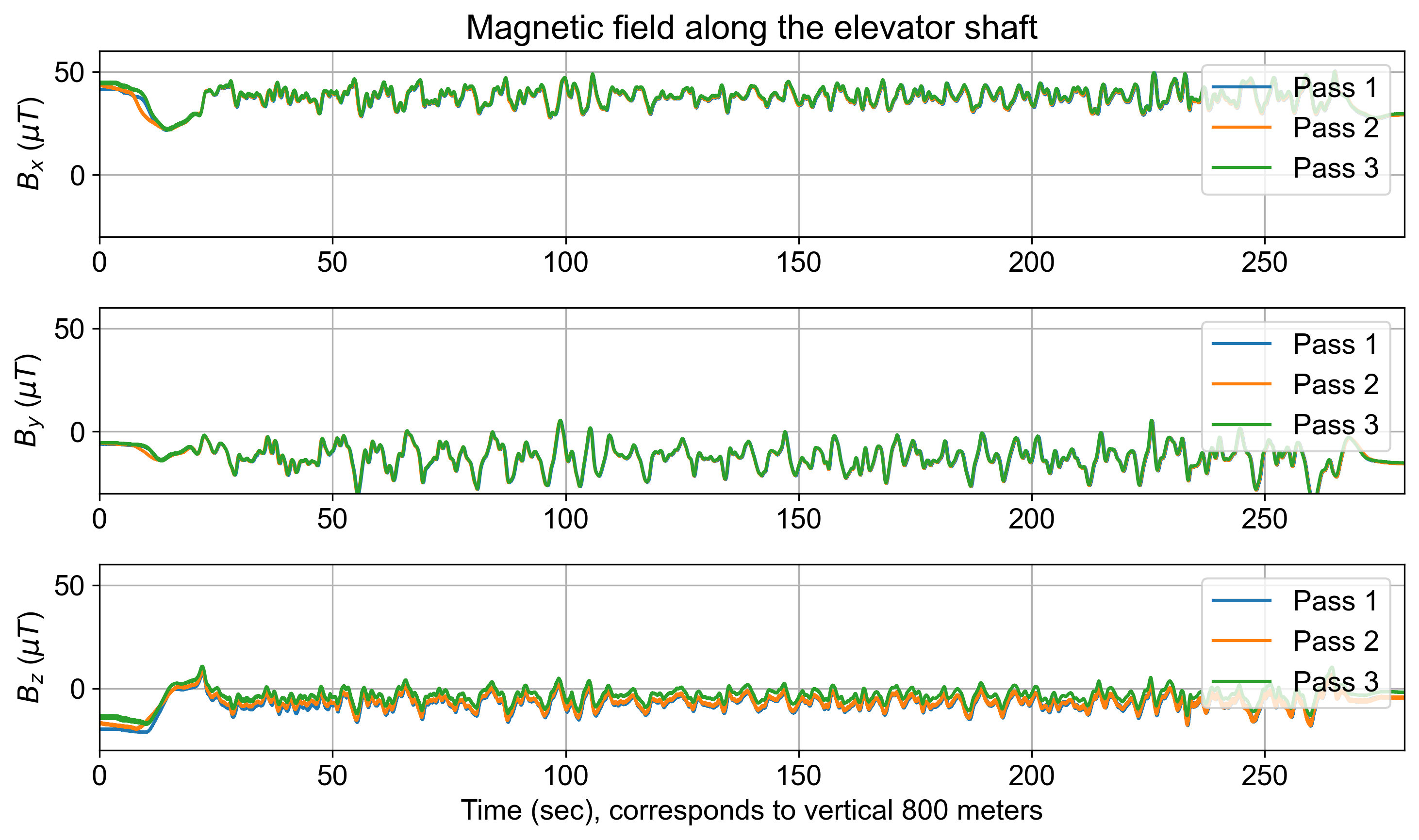}
    \caption{The x,y and z components of the static magnetic field along the elevator shaft.}
    \label{fig:Static_field_elevator3ax}
\end{figure}

Using three consecutive elevator passes, we could map variability of the static magnetic field along the full depth of the 800~m elevator shaft. Figure~\ref{fig:Static_field_elevator} shows the modulus |B|, with the Earth's magnetic field shown for comparison. Figure~\ref{fig:Static_field_elevator3ax} shows in detail the three x,y,z components. As the elevator started to move and reached the nominal speed, we could use time to align measurement data from the three consecutive passes. The x-scale on these plots corresponds to the depth, except the first and the last 15-20 seconds, when the elevator is in manual control and accelerating towards the nominal speed and decelerating to stop. 

The proposed AI would be installed in this shaft, enclosed by a densely interconnected conductive and ferromagnetic mesh cage formed by the reinforced concrete rings. Return/ground currents should be expected in this structure that can contribute to variability of the magnetic background. 

\FloatBarrier

\section{Conclusions} 
\label{sec:Conclusions}

This environmental measurement study has not identified any showstoppers for the installation and operation of an $\sim 800$~m atom interferometer experiment in the MFS of the Gotthard Base Tunnel at Sedrun. This report indicates that the experimental requirements for the spectra of vibrations, seismic and electromagnetic noise could all be met in the TOP and BOTTOM positions of the MFS infrastructure, confirming its high quality and suitability for an 800-m atom interferometer.

As discussed in Section~3, such a detector would have unique capabilities to search for ultralight dark matter, and would open the way to searches for gravitational waves and other interesting phenomena in fundamental physics.

If the international community interested in future large-scale atom interferometer experiments wishes to pursue the possibility of siting an $\sim 800$~m detector in the MFS of the Gotthard Base Tunnel at Sedrun, a programme of relevant future studies could be discussed with the local authorithies and SBB to evaluate the technical feasibility of the project.

\vspace{0.5cm}
\clearpage
\paragraph{Acknowledgements}

\textit{The authors would like to thank Fabrizia Toletti, Regional Head RSD at SBB, and her colleagues for the extremely valuable discussions held during the visit to the Sedrun MFS site. The authors also gratefully acknowledge SBB for agreeing to host and facilitate this study, for their support during the initial site visit, and for the continuous assistance provided throughout the environmental measurement campaign, from management level, to SBB engineers who helped with the local coordination during measurements and on-site operators. In addition, the authors acknowledge the availability and accessibility of operational logbooks and train traffic data, which were essential for the interpretation and correlation of the MFS Sedrun measurements.}

% \textbf{\color{red} Others to be acknowledged?}

\bigskip

%This is an example of citation \cite{Gorzawski:1697022}
%\clearpage
%\section*{Acknowledgments}
%Place for acknowledgements.\\
%The work of J.E. was supported in part by the United Kingdom STFC Grants
%ST/T00679X/1 and ST/T000759/1.

%\begin{thebibliography}{90}

%\bibitem{review} A reference

%\end{thebibliography}
\newpage

\bibliography{PBCreport}

@misc{bib:PBC_public,
  title = {The {P}hysics {B}eyond {C}olliders {S}tudy {G}roup},
  year = {2024},
  note = {Last accessed 10 January 2025},
  url = {https://pbc.web.cern.ch}
}

@article{Badurina:2019hst,
    author = "Badurina, L. and others",
    title = "{AION: An Atom Interferometer Observatory and Network}",
    eprint = "1911.11755",
    archivePrefix = "arXiv",
    primaryClass = "astro-ph.CO",
    reportNumber = "AION-2019-001, CERN-TH-2019-199",
    doi = "10.1088/1475-7516/2020/05/011",
    journal = "JCAP",
    volume = "05",
    pages = "011",
    year = "2020"
}

@proceedings{Proceedings:2024foy,
    author = "Abdalla, Adam and others",
    title = "{Terrestrial Very-Long-Baseline Atom Interferometry: Summary of the Second Workshop}",
    eprint = "2412.14960",
    archivePrefix = "arXiv",
    primaryClass = "hep-ex",
    month = "arXiv:2412.14960,",
    year = "2024"
}

@misc{Cooper:2005,
  author       = {Cooper.ch},
  title        = {Scheme of the {Gotthard Base Tunnel}},
  year         = {2005},
  howpublished = {\url{https://upload.wikimedia.org/wikipedia/commons/5/5f/Nrla_scheme.png}},
}

@article{FABBRI2019379,
title = {Risk, Contract Management, and Financing of the {Gotthard Base Tunnel} in {Switzerland}},
journal = {Engineering},
volume = {5},
number = {3},
pages = {379-383},
year = {2019},
issn = {2095-8099},
doi = {https://doi.org/10.1016/j.eng.2019.04.001},
url = {https://www.sciencedirect.com/science/article/pii/S2095809919304102},
author = {Davide Fabbri}
}

@article{AION:2025igp,
    author = "Baynham, C. F. A. and others",
    collaboration = "AION",
    title = "{A Prototype Atom Interferometer to Detect Dark Matter and Gravitational Waves}",
    eprint = "2504.09158",
    archivePrefix = "arXiv",
    primaryClass = "hep-ex",
    reportNumber = "AION-REPORT/2025-02",
    month = "4",
    year = "2025"
}

@article{MAGIS-100:2021etm,
    author = "Abe, Mahiro and others",
    collaboration = "MAGIS-100",
    title = "{Matter-wave Atomic Gradiometer Interferometric Sensor (MAGIS-100)}",
    eprint = "2104.02835",
    archivePrefix = "arXiv",
    primaryClass = "physics.atom-ph",
    reportNumber = "FERMILAB-PUB-21-031-AD-DI-FESS-QIS-T",
    doi = "10.1088/2058-9565/abf719",
    journal = "Quantum Sci. Technol.",
    volume = "6",
    number = "4",
    pages = "044003",
    year = "2021"
}

@book{peterson1993observations,
  title={Observations and modeling of seismic background noise},
  author={Peterson, Jon and others},
  volume={93},
  year={1993},
  publisher={US Geological Survey Reston, VA, USA}
}

@article{LIGOScientific:2014pky,
    author = "Aasi, J. and others",
    collaboration = "LIGO Scientific",
    title = "{Advanced LIGO}",
    eprint = "1411.4547",
    archivePrefix = "arXiv",
    primaryClass = "gr-qc",
    doi = "10.1088/0264-9381/32/7/074001",
    journal = "Class. Quant. Grav.",
    volume = "32",
    pages = "074001",
    year = "2015"
}

@article{VIRGO:2014yos,
    author = "Acernese, F. and others",
    collaboration = "VIRGO",
    title = "{Advanced Virgo: a second-generation interferometric gravitational wave detector}",
    eprint = "1408.3978",
    archivePrefix = "arXiv",
    primaryClass = "gr-qc",
    doi = "10.1088/0264-9381/32/2/024001",
    journal = "Class. Quant. Grav.",
    volume = "32",
    number = "2",
    pages = "024001",
    year = "2015"
}

@article{Aso:2013eba,
    author = "Aso, Yoichi and others",
    collaboration = "KAGRA",
    title = "{Interferometer design of the KAGRA gravitational wave detector}",
    eprint = "1306.6747",
    archivePrefix = "arXiv",
    primaryClass = "gr-qc",
    doi = "10.1103/PhysRevD.88.043007",
    journal = "Phys. Rev. D",
    volume = "88",
    number = "4",
    pages = "043007",
    year = "2013"
}

@article{LISA:2017pwj,
    author = "Amaro-Seoane, Pau and others",
    collaboration = "LISA",
    title = "{Laser Interferometer Space Antenna}",
    eprint = "arXiv:1702.00786",
    archivePrefix = "arXiv",
    primaryClass = "astro-ph.IM",
    month = "arXiv:1702.00786,",
    year = "2017"
}

@misc{TVLBAI2023WorkshopSummary,
  title         = {Terrestrial Very-Long-Baseline Atom Interferometry: Workshop Summary},
  author        = {Abend, Sven and Allard, Baptiste and Alonso, Iv{\'a}n and Antoniadis, John and Araujo, Henrique and Arduini, Gianluigi and Arnold, Aidan and A{\ss}mann, Tobias and Augst, Nadja and Badurina, Leonardo and others},
  year          = {2023},
  eprint        = {2310.08183},
  archivePrefix = {arXiv},
  primaryClass  = {hep-ex},
  doi           = {10.48550/arXiv.2310.08183},
  url           = {https://arxiv.org/abs/2310.08183},
  note          = {arXiv:2310.08183}
}

@techreport{AION100feasibilitystudy,
  author       = {Arduini, Gianluigi and Badurina, Leonardo and Balazs, K. and
                  Baynham, C. and Buchmueller, O. and Buzio, M. and
                  Calatroni, Sergio and Corso, J.-P. and Ellis, John and
                  Gaignant, Ch. and Guinchard, Michael and Hakulinen, T. and
                  Hobson, R. and Infantino, A. and Lafarge, D. and
                  Langlois, R. and Marcel, C. and Mitchell, J. and
                  Parodi, M. and Pentella, M. and Valuch, D. and Vincke, H.},
  title        = {A Long-Baseline Atom Interferometer at CERN: Conceptual Feasibility Study},
  institution  = {CERN},
  address      = {Geneva},
  year         = {2023},
  reportNumber = {CERN-PBC-REPORT-2023-002},
  number       = {CERN-TH-2023-051},
  url          = {https://cds.cern.ch/record/2851946},
  note         = {51 pages, 39 figures}
}

@MISC{Bartington:2024,
  author         = "{Bartington instruments}",
  title          = "{MAG-13 three-axis magnetometer}",
  howpublished   = "\href{https://www.bartington.com/products/mag-13/}{link}",
  year           = " 2024",
}

@MISC{LeCroy:2023,
  author         = "{Teledyne LeCroy}",
  title          = "{WaveRunner 8108HD Oscilloscope}",
  howpublished   = "\href{https://www.teledynelecroy.com/oscilloscope/oscilloscopemodel.aspx?modelid=11299&capid=102&mid=506}{link}",
  year           = " 2023",
}

@article{Buchmueller:2023nll,
    author = "Buchmueller, Oliver and Ellis, John and Schneider, Ulrich",
    title = "{Large-scale atom interferometry for fundamental physics}",
    eprint = "2306.17726",
    archivePrefix = "arXiv",
    primaryClass = "astro-ph.CO",
    reportNumber = "AION-REPORT/2023-04, KCL-PH-TH/2023-33, CERN-TH-2023-105",
    doi = "10.1080/00107514.2023.2239008",
    journal = "Contemp. Phys.",
    volume = "64",
    number = "2",
    pages = "93--110",
    year = "2023"
}

@MISC{Schwarzbeck:2024,
  author         = "{Schwarzbeck Messtechnik}",
  title          = "{Active Receive Loop Antenna FMZB 1519-60 C}",
  howpublished   = "\href{https://schwarzbeck.de/Datenblatt/k1519-60C.pdf}{link}",
  year           = " 2024",
}

@article{Arduini:2023wce,
    author = "Arduini, G. and others",
    title = "{A Long-Baseline Atom Interferometer at CERN: Conceptual Feasibility Study}",
    eprint = "2304.00614",
    archivePrefix = "arXiv",
    primaryClass = "physics.atom-ph",
    reportNumber = "CERN-PBC-REPORT-2023-002",
    month = "arXiv:2304.00614,",
    year = "2023"
}

@article{EventHorizonTelescope:2022wkp,
    author = "Akiyama \ , Kazunori and others",
    collaboration = "Event Horizon Telescope",
    title = "{First Sagittarius A* Event Horizon Telescope Results. I. The Shadow of the Supermassive Black Hole in the Center of the Milky Way}",
    doi = "10.3847/2041-8213/ac6674",
    journal = "Astrophys. J. Lett.",
    volume = "930",
    number = "2",
    pages = "L12",
    year = "2022"
}

@article{EventHorizonTelescope:2019dse,
    author = "Akiyama, Kazunori and others",
    collaboration = "Event Horizon Telescope",
    title = "{First M87 Event Horizon Telescope Results. I. The Shadow of the Supermassive Black Hole}",
    eprint = "1906.11238",
    archivePrefix = "arXiv",
    primaryClass = "astro-ph.GA",
    doi = "10.3847/2041-8213/ab0ec7",
    journal = "Astrophys. J. Lett.",
    volume = "875",
    pages = "L1",
    year = "2019"
}

\newpage
\section*{Appendix - Top station: Detailed results of electromagnetic background measurements} 
%Here the appendix to place graphs 

% Measurements in a typical "quiet" magnetic field background
\begin{figure}[h!]
    \centering
    \includegraphics[width=1\linewidth]{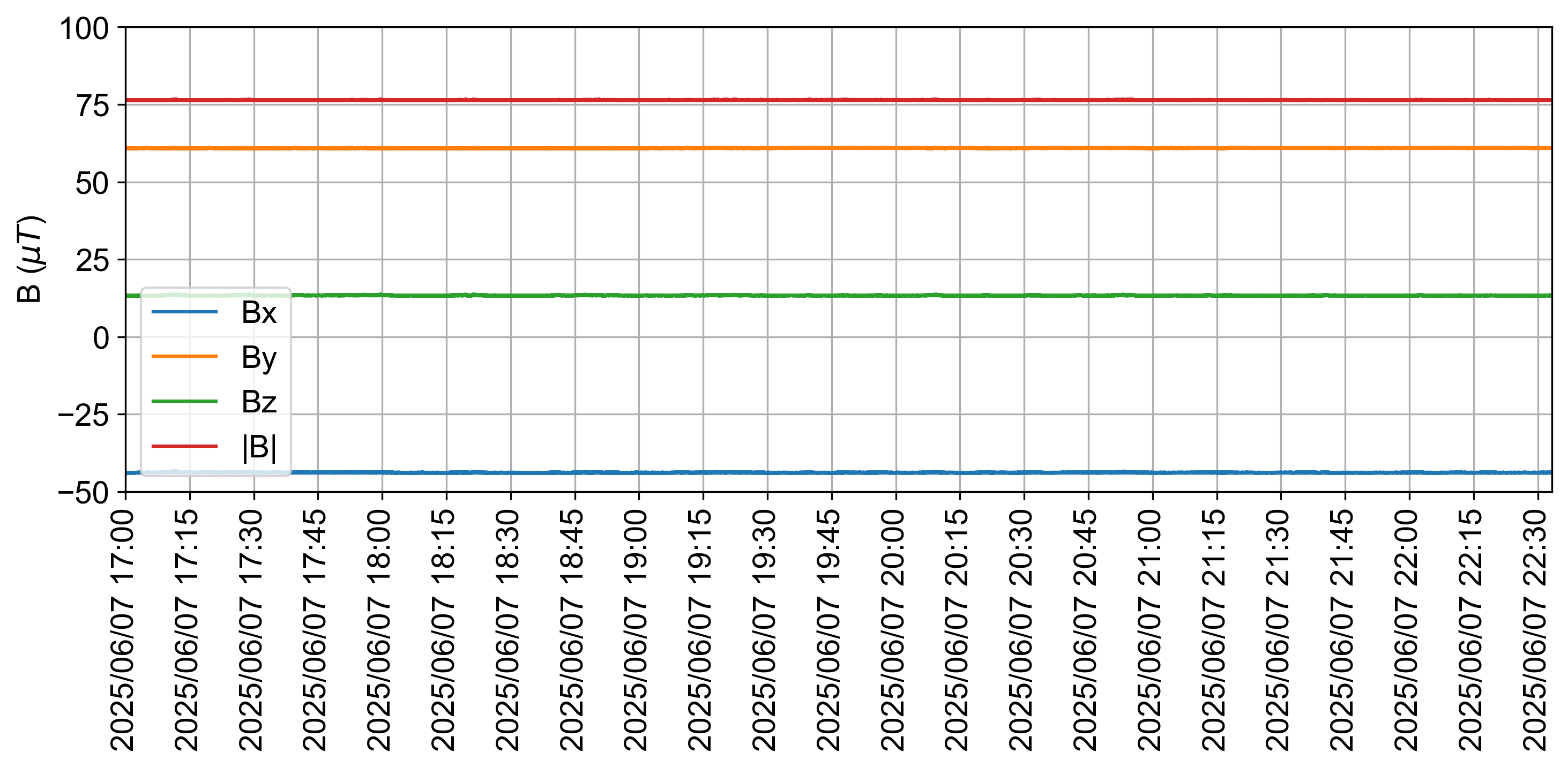}
    \caption{Typical magnetic field background in a "quiet" scenario: the B-field as a function of time.}
    \label{fig:Top_quiet_A}
\end{figure}
\begin{figure}[h!]
    \centering
    \includegraphics[width=1\linewidth]{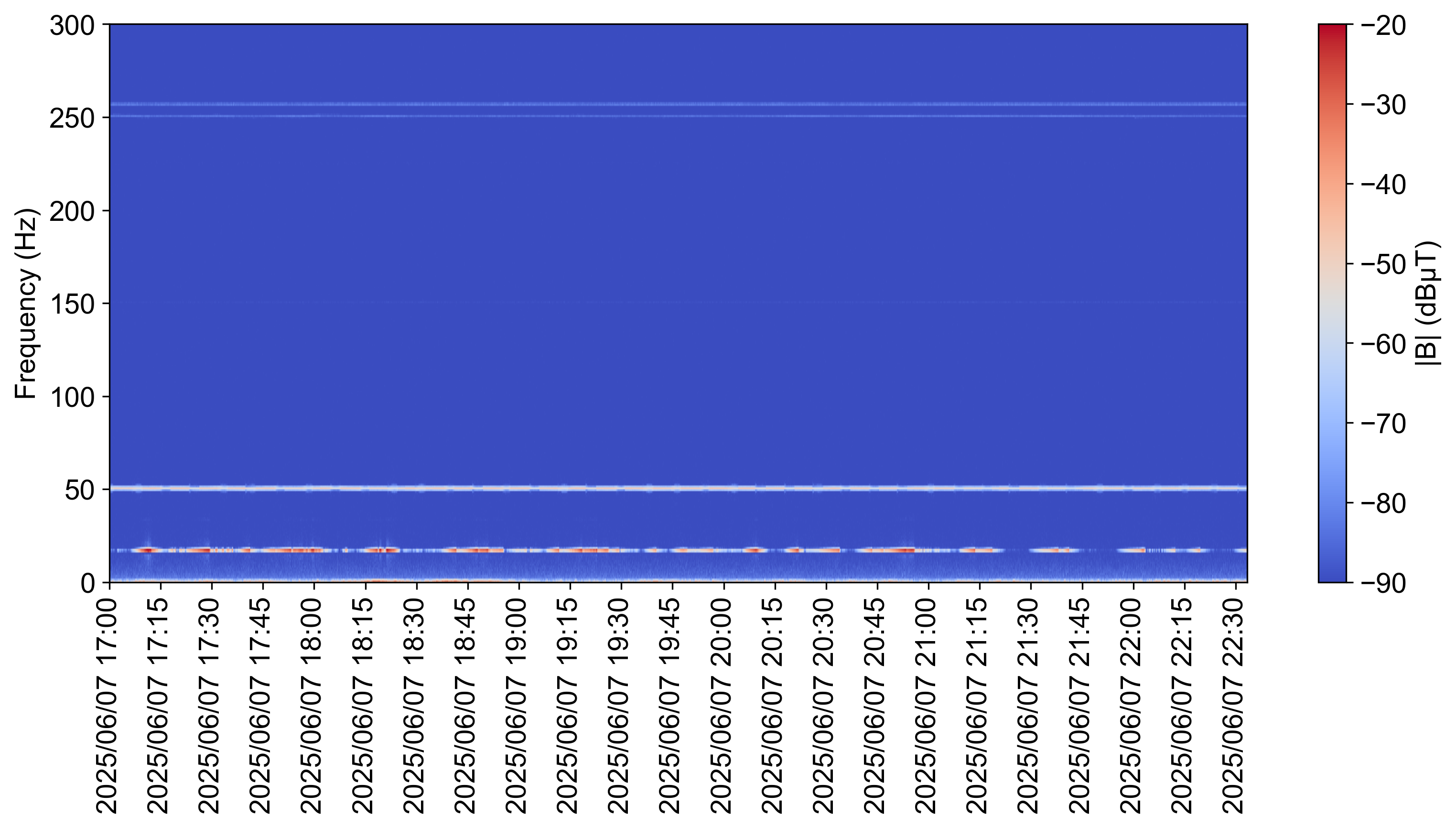}
    \caption{Typical magnetic field background in a "quiet" scenario: spectrogram.}
    \label{fig:Top_quiet_B}
\end{figure}
\begin{figure}[h!]
    \centering
    \includegraphics[width=1\linewidth]{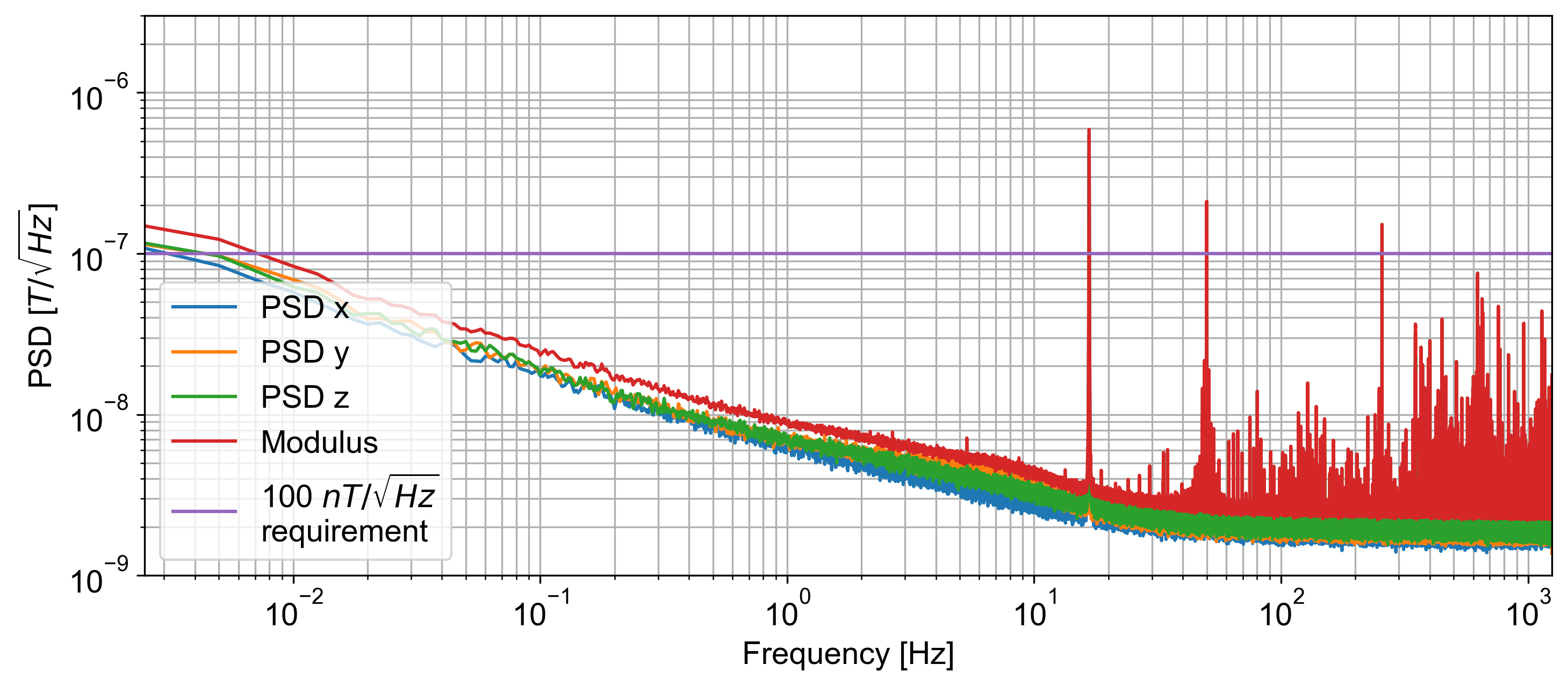}
    \caption{Typical magnetic field background in a "quiet" scenario: the corresponding PSD, record length 20000 seconds.}
    \label{fig:Top_quiet_C}
\end{figure}

\begin{figure}[h!]
    \centering
    \includegraphics[width=1\linewidth]{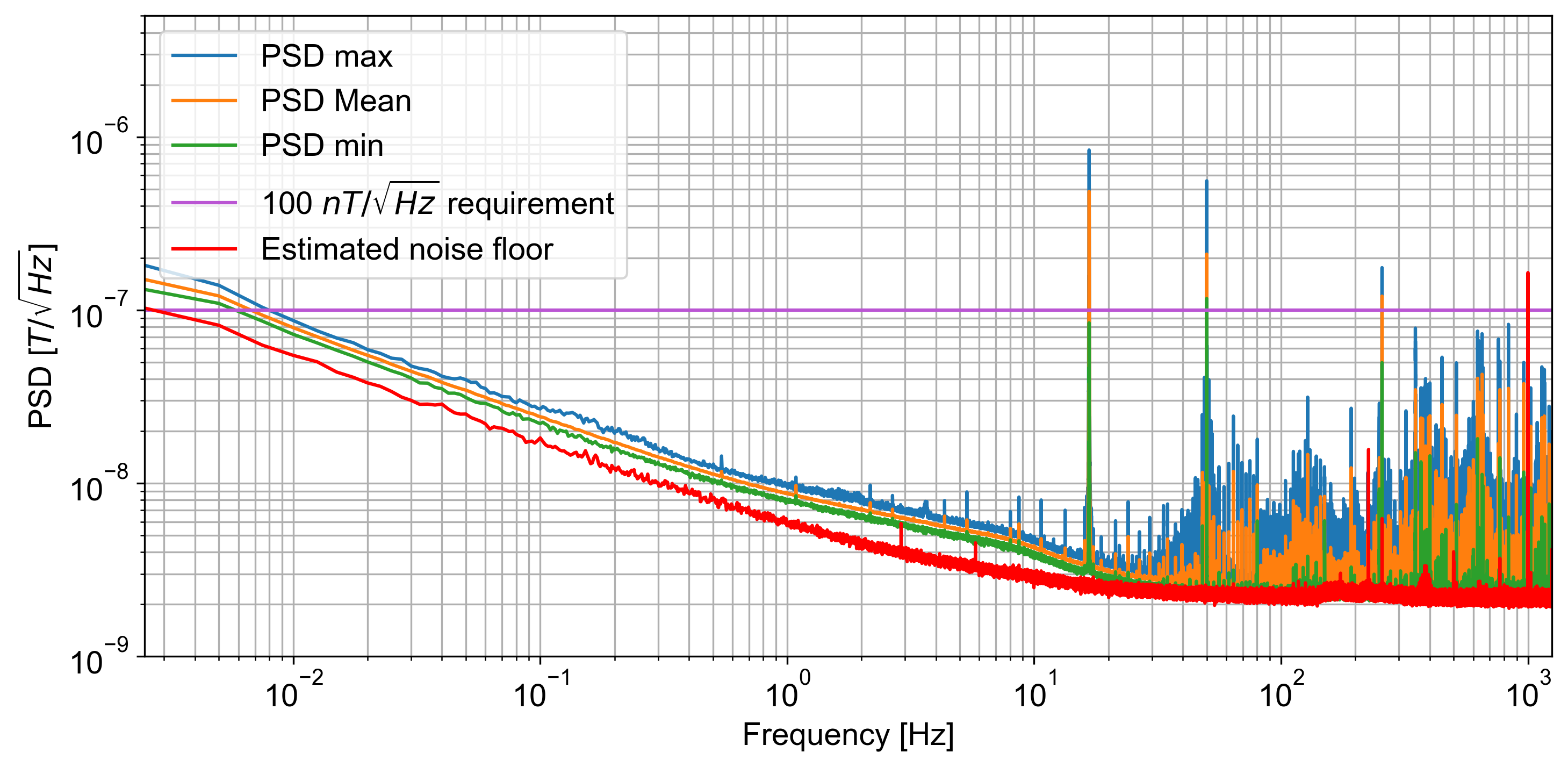}
    \caption{Minimum, mean and maximum PSD of all 101 "quiet" datasets recorded at the TOP station.}
    \label{fig:Top_quiet_minmax}
\end{figure}
\begin{figure}[h!]
    \centering
    \includegraphics[width=1\linewidth]{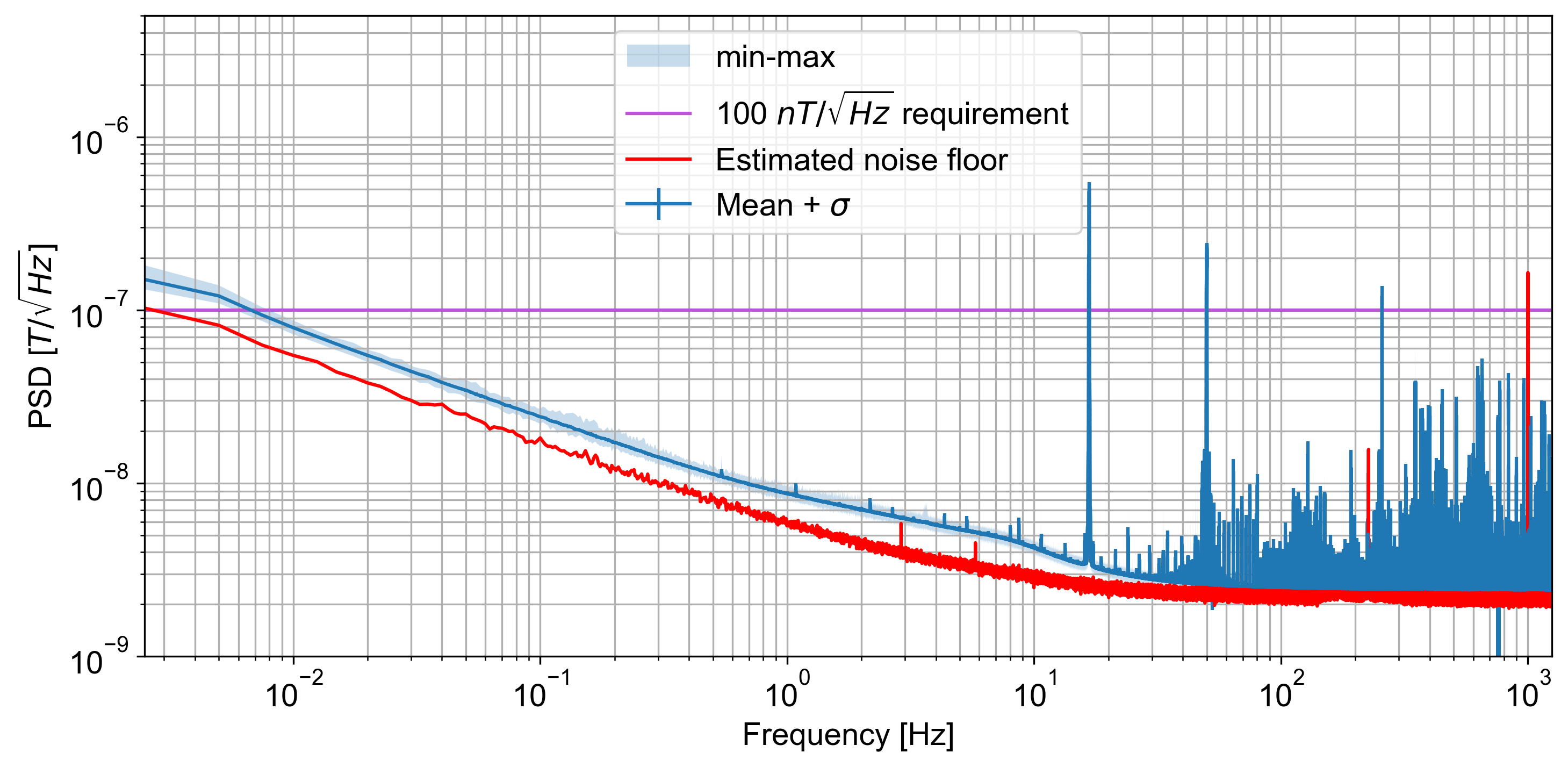}
    \caption{Distribution of PSDs of all 101 "quiet" datasets recorded at the TOP station.}
    \label{fig:Top_quiet_sigma}
\end{figure} 

% Electromagnetic signatures of remote operation
\begin{figure}[h!]
    \centering
    \includegraphics[width=1\linewidth]{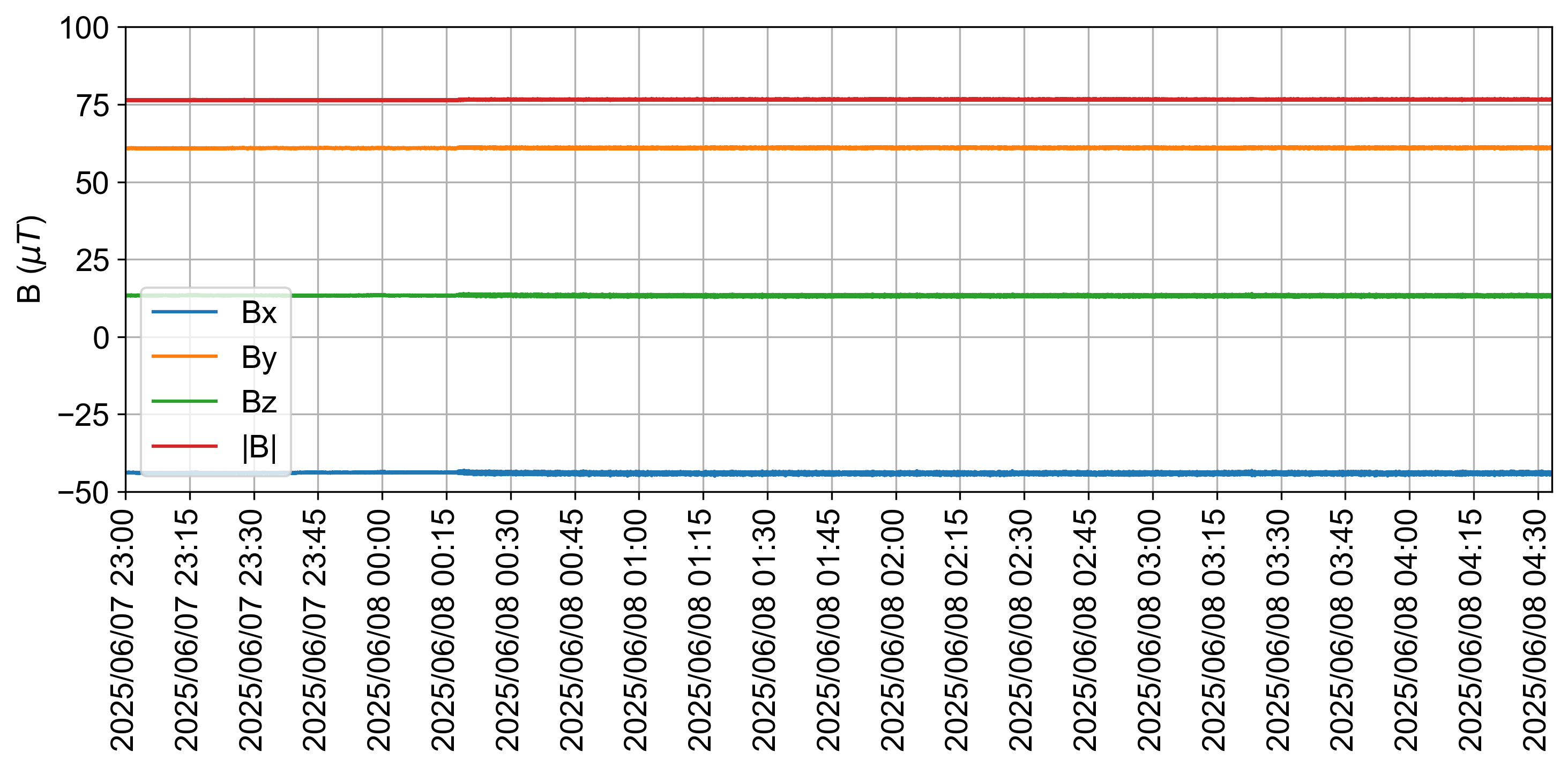}
    \caption{Magnetic field background with technical services operated: B-field as a function of time. The equipment was turned on shortly after 00:15.}
    \label{fig:Top_case2_A}
\end{figure}
\begin{figure}[h!]
    \centering
    \includegraphics[width=1\linewidth]{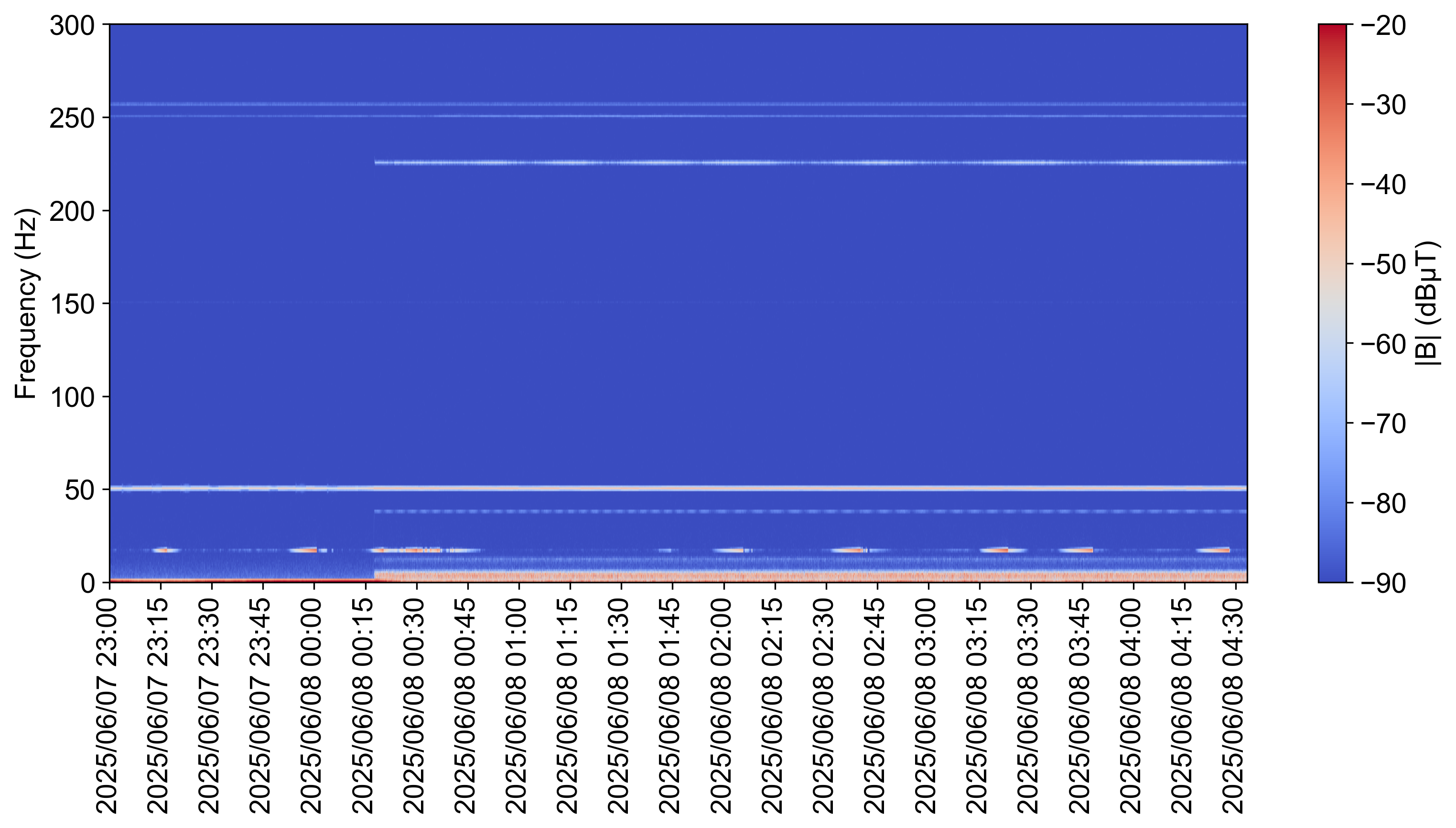}
    \caption{Magnetic field background with technical services operated: spectrogram.}
    \label{fig:Top_case2_B}
\end{figure}
\begin{figure}[h!]
    \centering
    \includegraphics[width=1\linewidth]{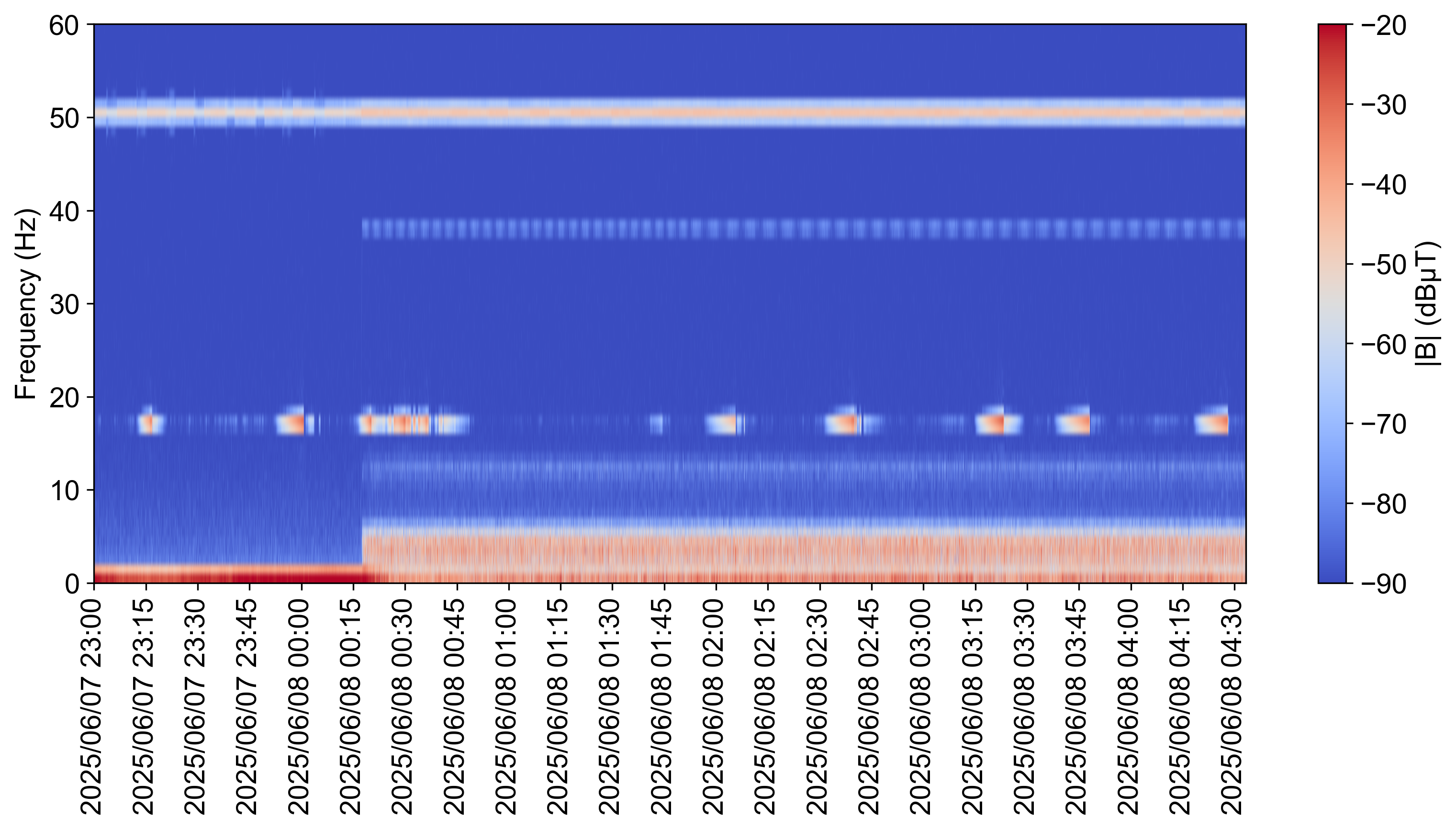}
    \caption{Expanded detail of low frequencies.}
    \label{fig:Top_case2_B_detail}
\end{figure}

\begin{figure}[h!]
    \centering
    \includegraphics[width=1\linewidth]{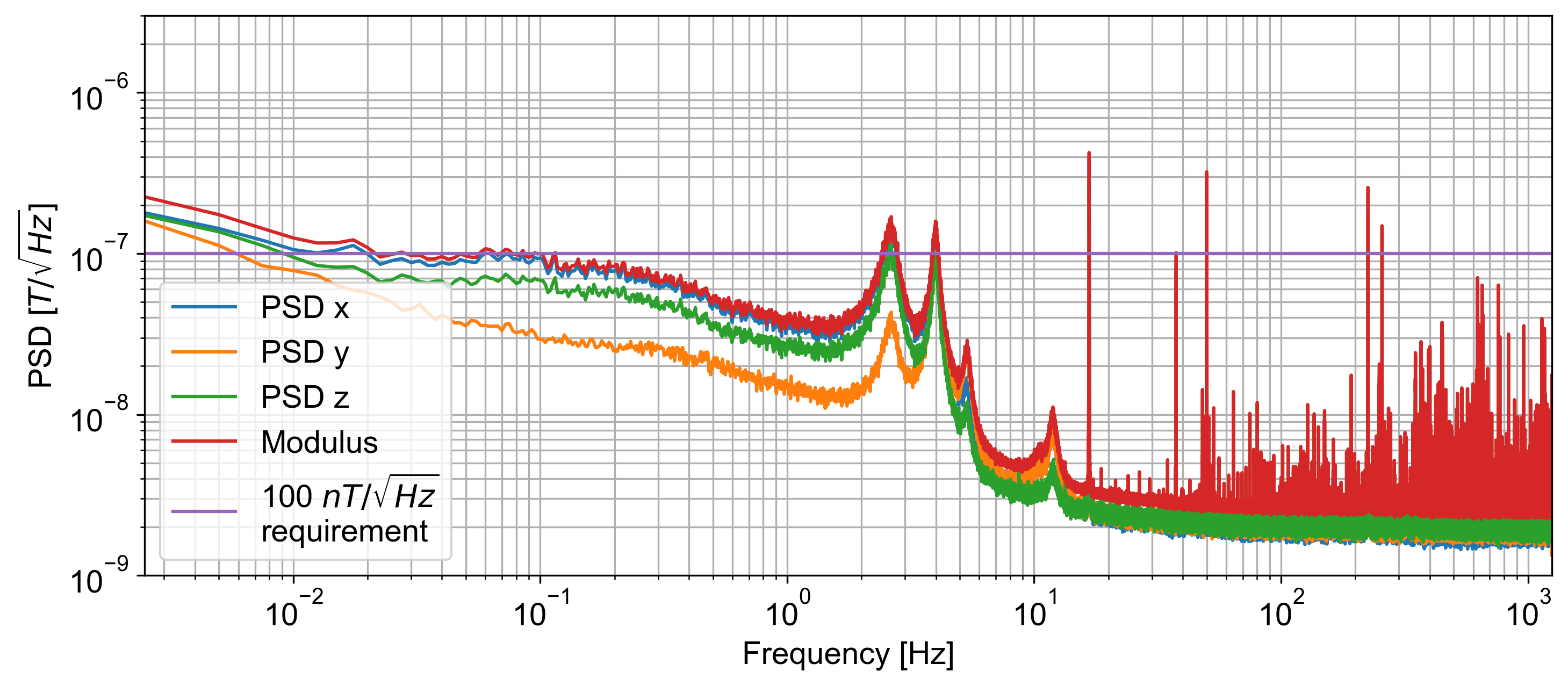}
    \caption{The PSDs of the magnetic field background with technical services operated. Record length 20000 seconds.}
    \label{fig:Top_case2_C}
\end{figure}

%A summary of all acquisitions with MFS technical services running (case 2)

\begin{figure}[h!]
    \centering
    \includegraphics[width=1\linewidth]{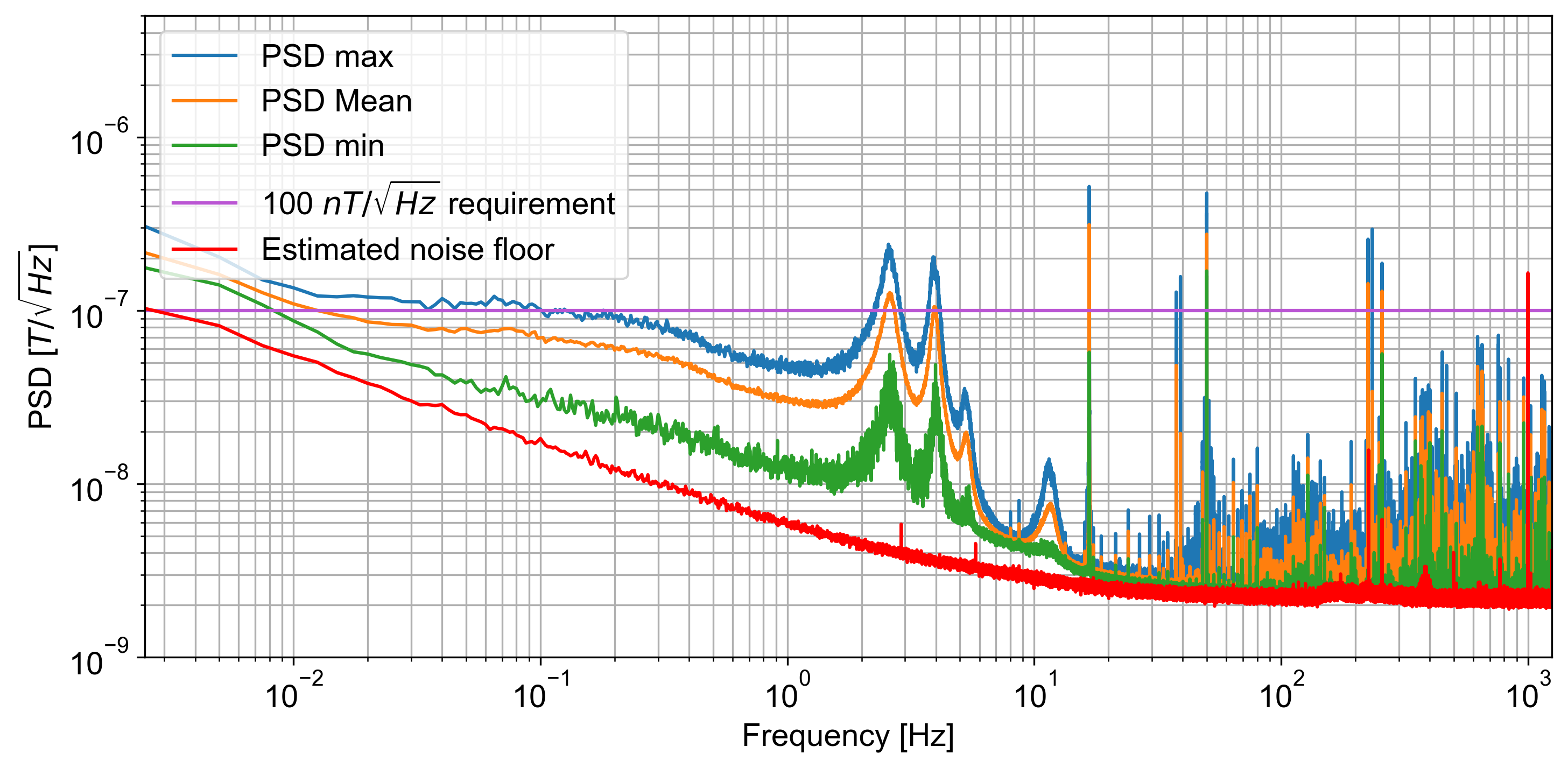}
    \caption{The minimum, mean and maximum PSDs of all 11 datasets with technical services operated recorded at the TOP station.}
    \label{fig:Top_services_minmax}
\end{figure}
\begin{figure}[h!]
    \centering
    \includegraphics[width=1\linewidth]{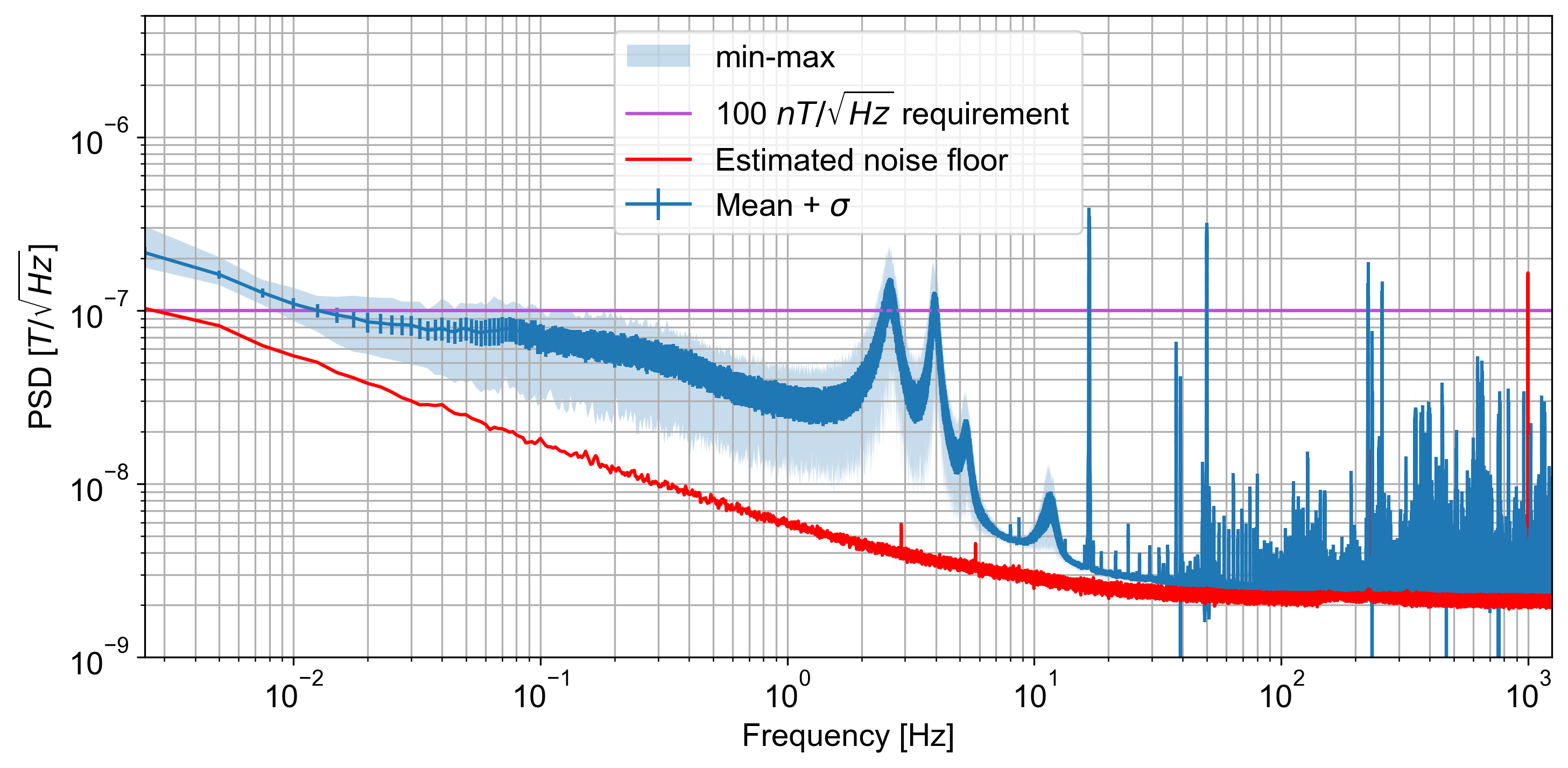}
    \caption{The PSD distributions of all 11 datasets with technical services operated recorded at the TOP station.}
    \label{fig:Top_services_sigma}
\end{figure}

% The third electromagnetic signature we have identified was of maintenance periods with personnel present on site and the elevator operating. 

\begin{figure}[h!]
    \centering
    \includegraphics[width=1\linewidth]{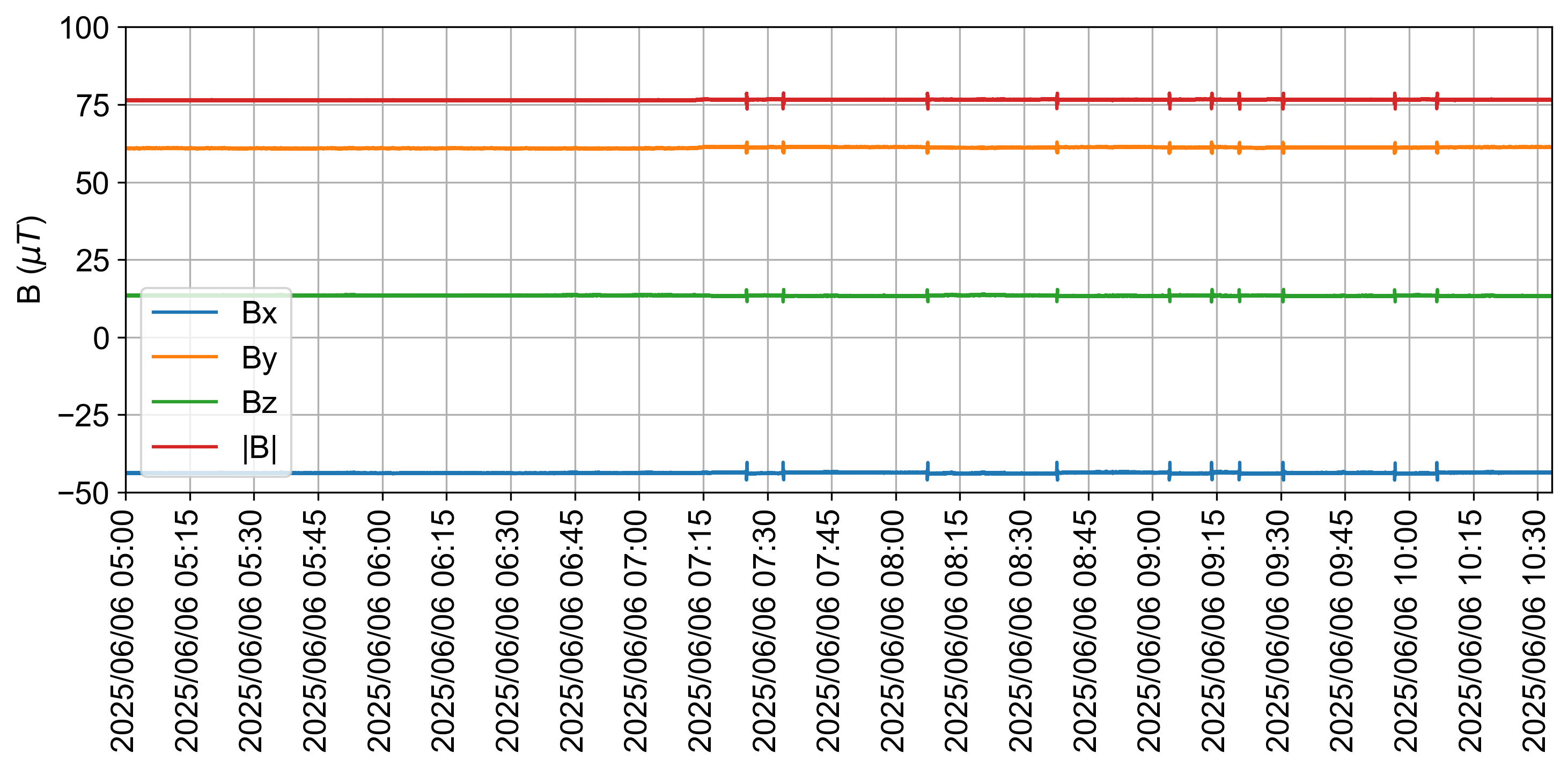}
    \caption{The magnetic field background with the elevator passing through the shaft: the B-field as a function of time. The peaks indicate the elevator passage (up or down).}
    \label{fig:Top_case3_A}
\end{figure}
\begin{figure}[h!]
    \centering
    \includegraphics[width=1\linewidth]{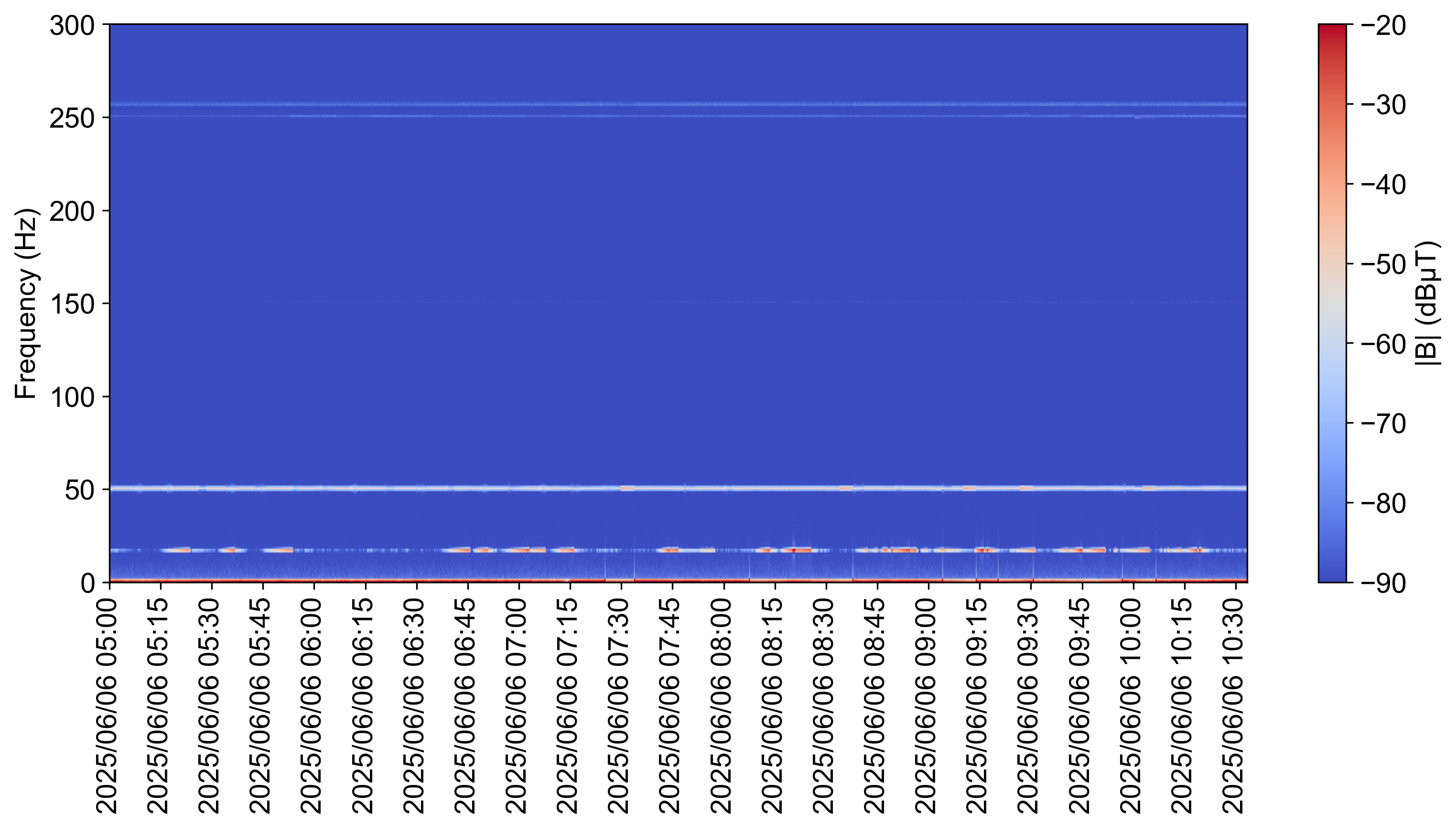}
    \caption{The magnetic field background with the elevator passing through the shaft: spectrogram.}
    \label{fig:Top_case3_B}
\end{figure}
\begin{figure}[h!]
    \centering
    \includegraphics[width=1\linewidth]{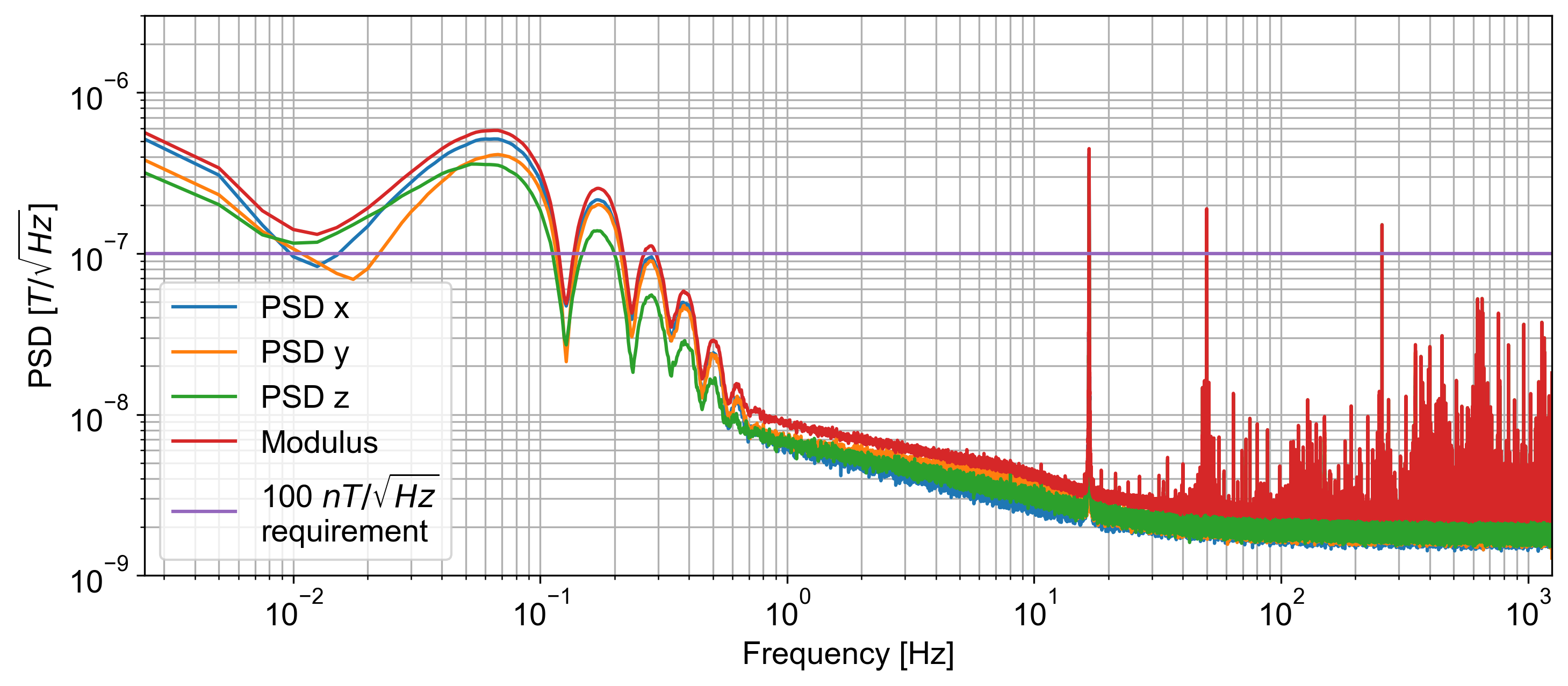}
    \caption{The magnetic field background with the elevator passing through the shaft: the corresponding PSDs. Record length 20000 seconds.}
    \label{fig:Top_case3_C}
\end{figure}

\begin{figure}[h!]
    \centering
    \includegraphics[width=1\linewidth]{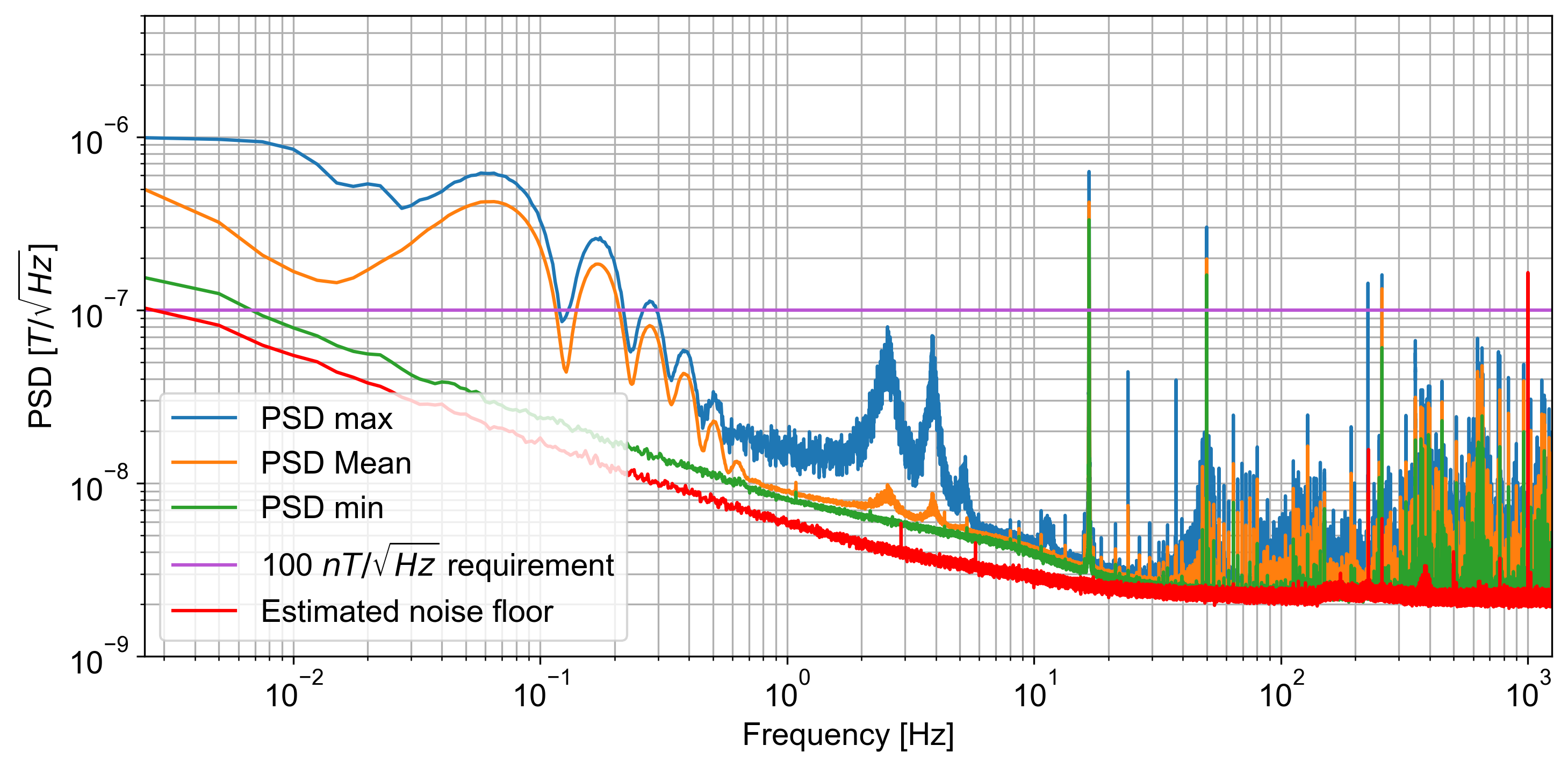}
    \caption{The minimum, mean and maximum PSDs of all 22 datasets recorded at the TOP station with the elevator being operated.}
    \label{fig:Top_elevator_minmax}
\end{figure}
\begin{figure}[h!]
    \centering
    \includegraphics[width=1\linewidth]{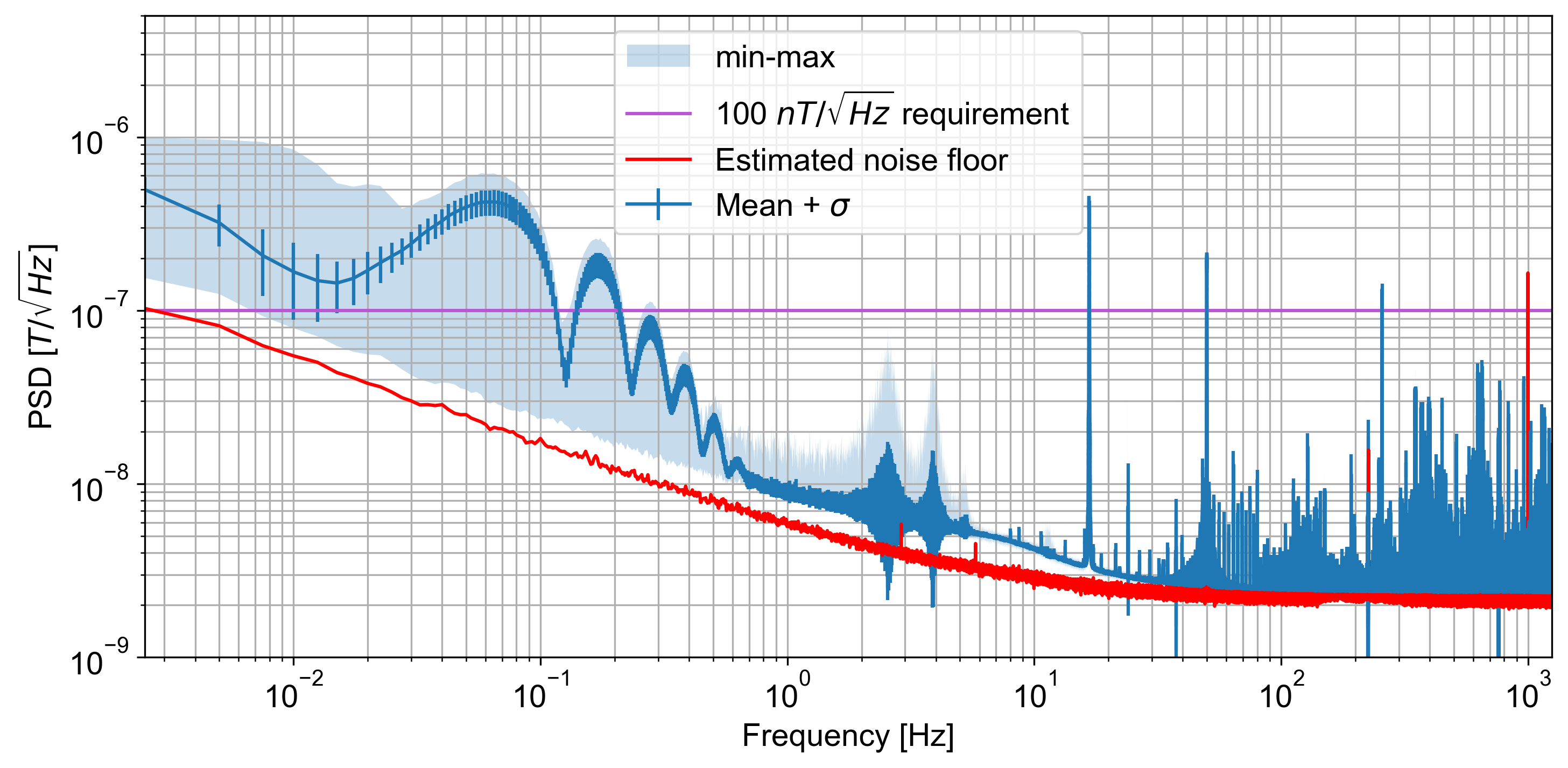}
    \caption{The PSD distributions of all 22 datasets recorded at the TOP station with the MFS elevator being operated.}
    \label{fig:Top_elevator_sigma}
\end{figure}

% The magnetic field at the BOTTOM station 
\newpage

\section*{Bottom station: Detailed results of electromagnetic background measurements} 

\begin{figure}[h!]
    \centering
    \includegraphics[width=1\linewidth]{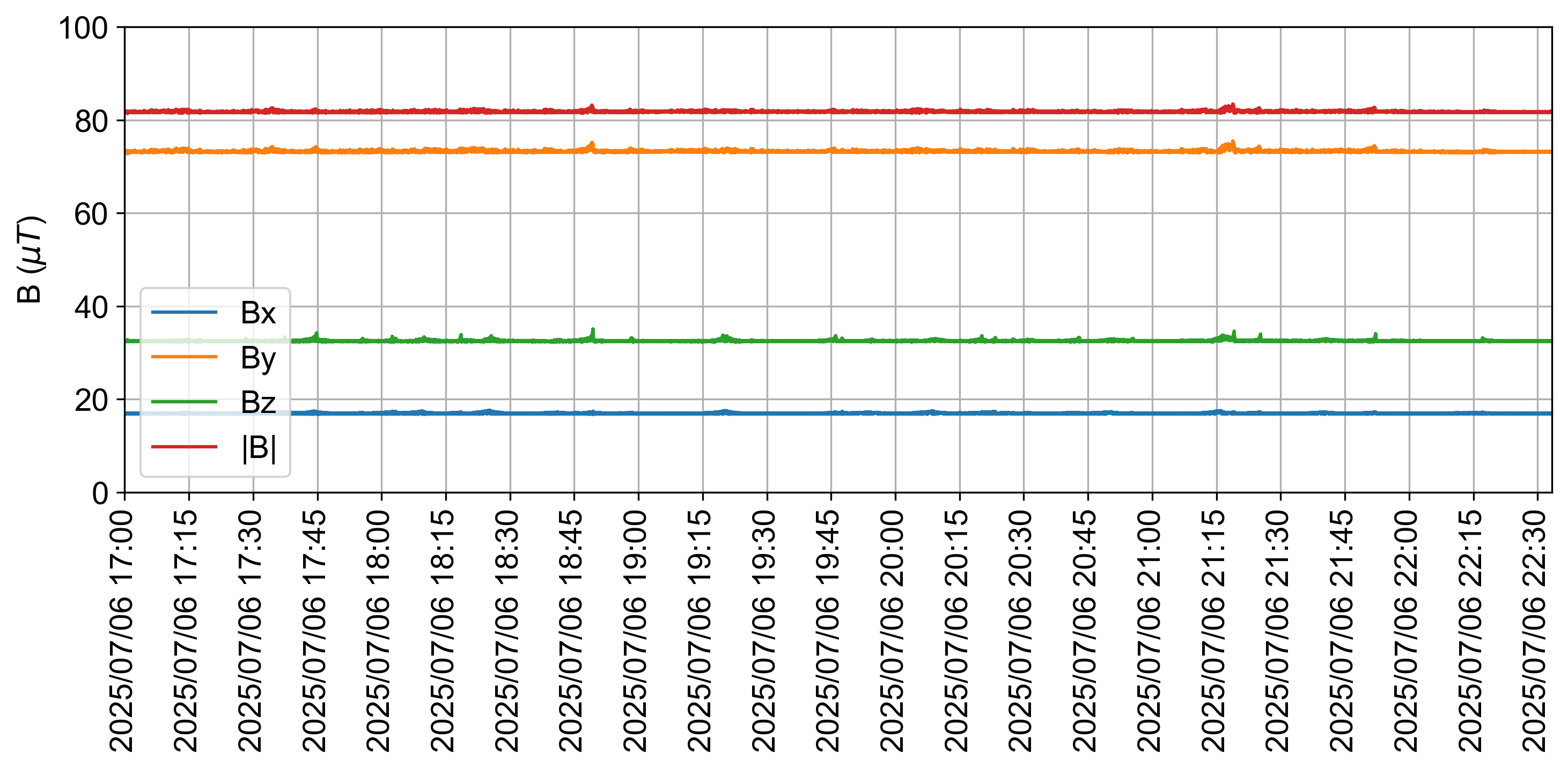}
    \caption{Heavy traffic example - time domain measurement.}
    \label{fig:Bottom_time_traffic}
\end{figure}
\begin{figure}[h!]
    \centering
    \includegraphics[width=1\linewidth]{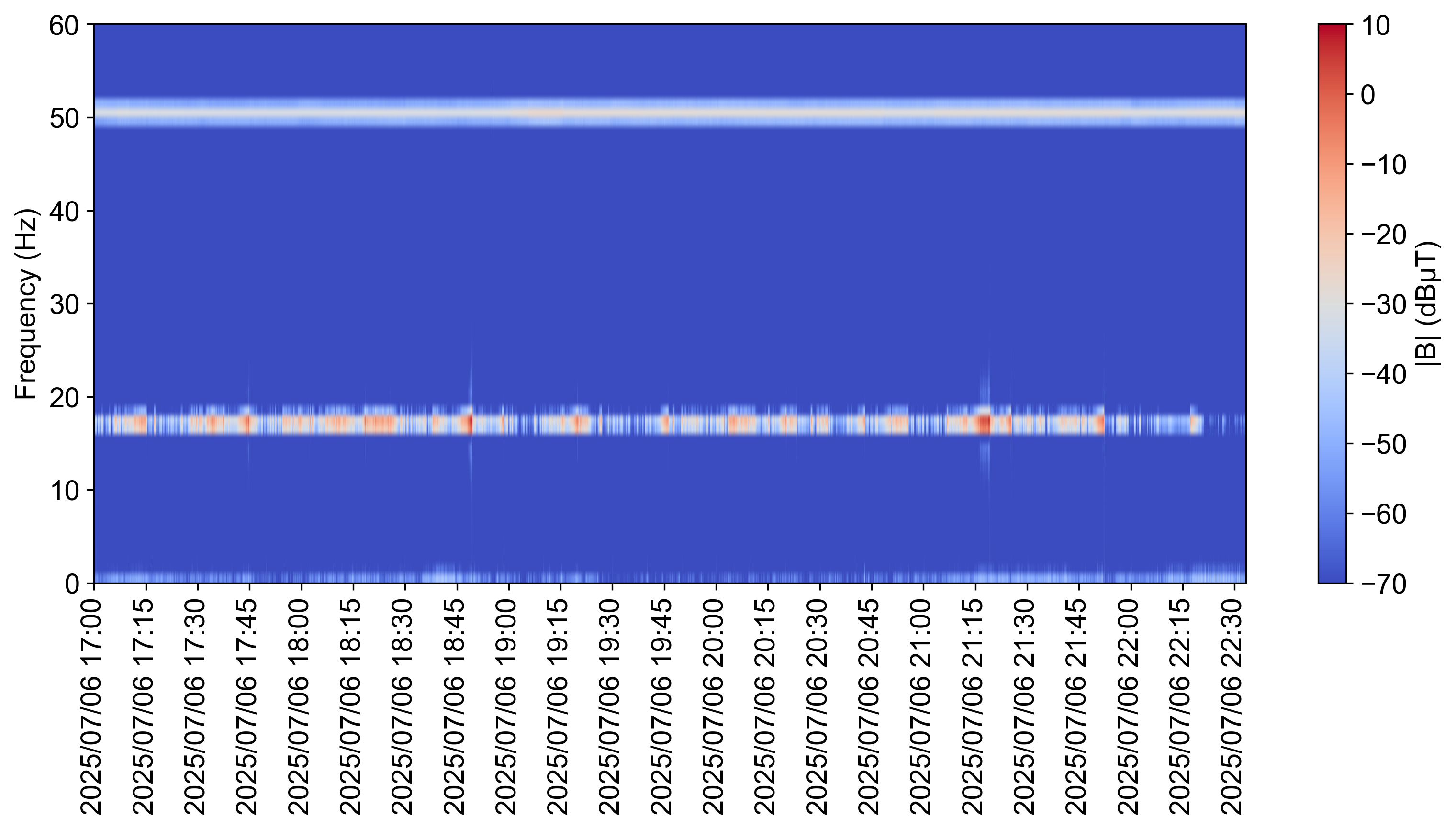}
    \caption{Heavy traffic example - spectrogram.}
    \label{fig:Bottom_spectrogram_traffic}
\end{figure}
\begin{figure}[h!]
    \centering
    \includegraphics[width=1\linewidth]{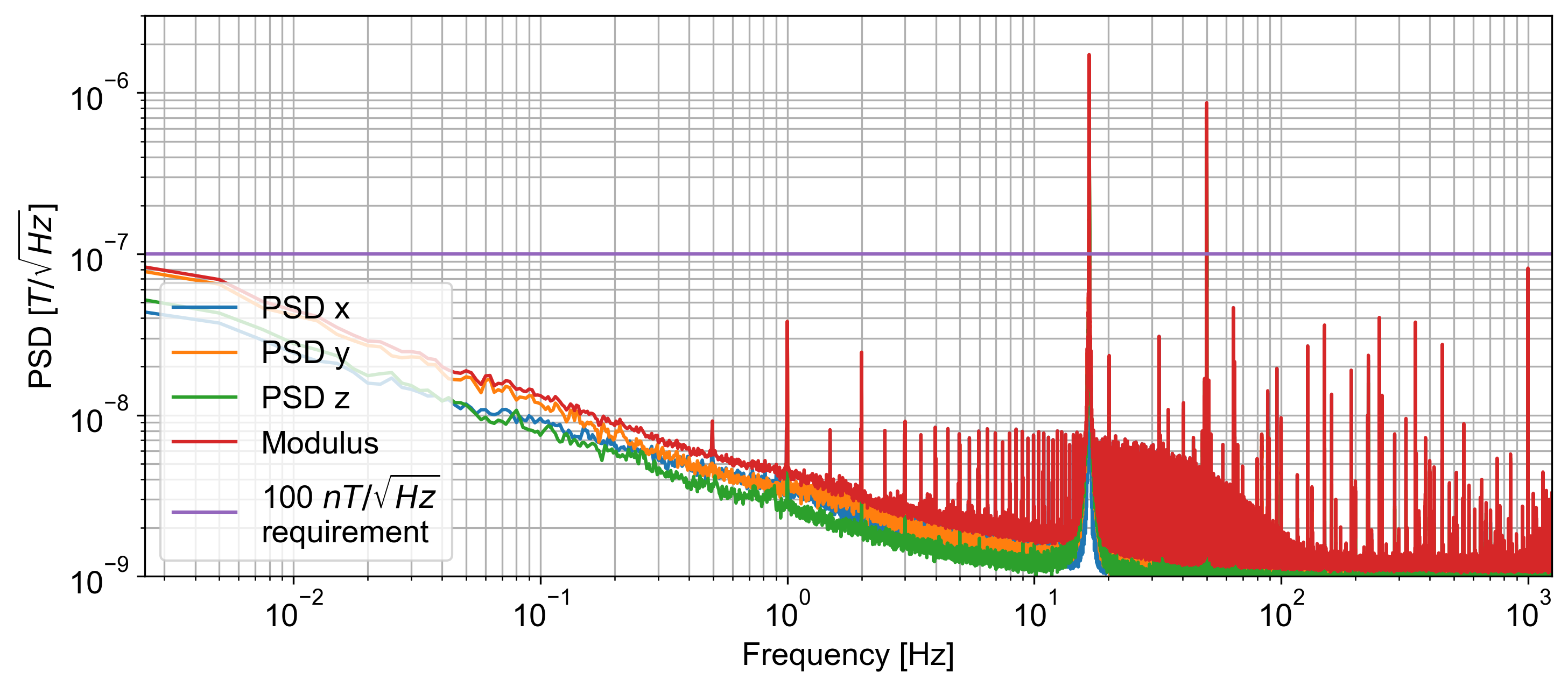}
    \caption{Heavy traffic example - PSD.}
    \label{fig:Bottom_PSD_traffic}
\end{figure}

\begin{figure}[h!]
    \centering
    \includegraphics[width=1\linewidth]{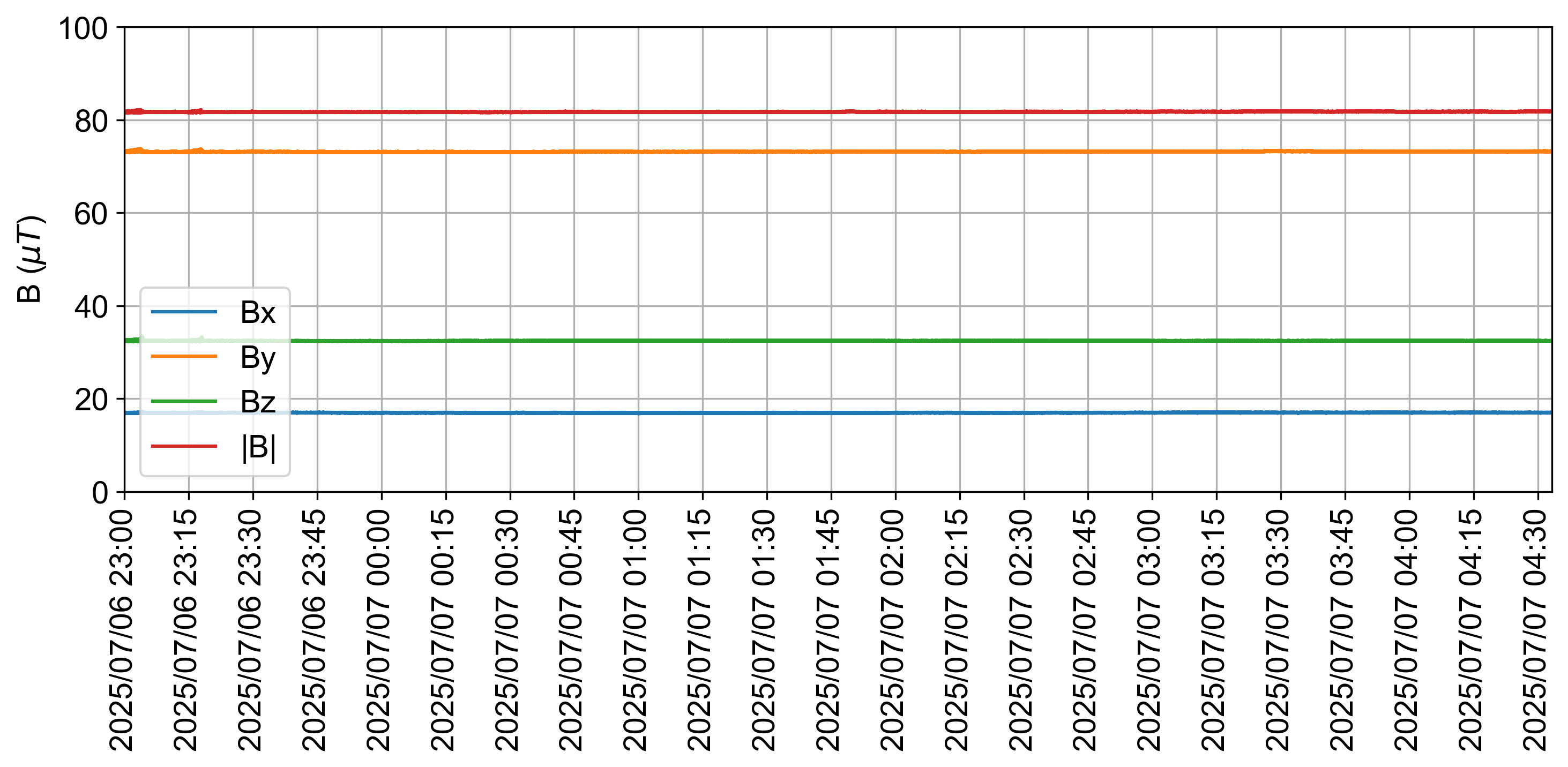}
    \caption{No traffic example - time domain measurement.}
    \label{fig:Bottom_time_notraffic}
\end{figure}
\begin{figure}[h!]
    \centering
    \includegraphics[width=1\linewidth]{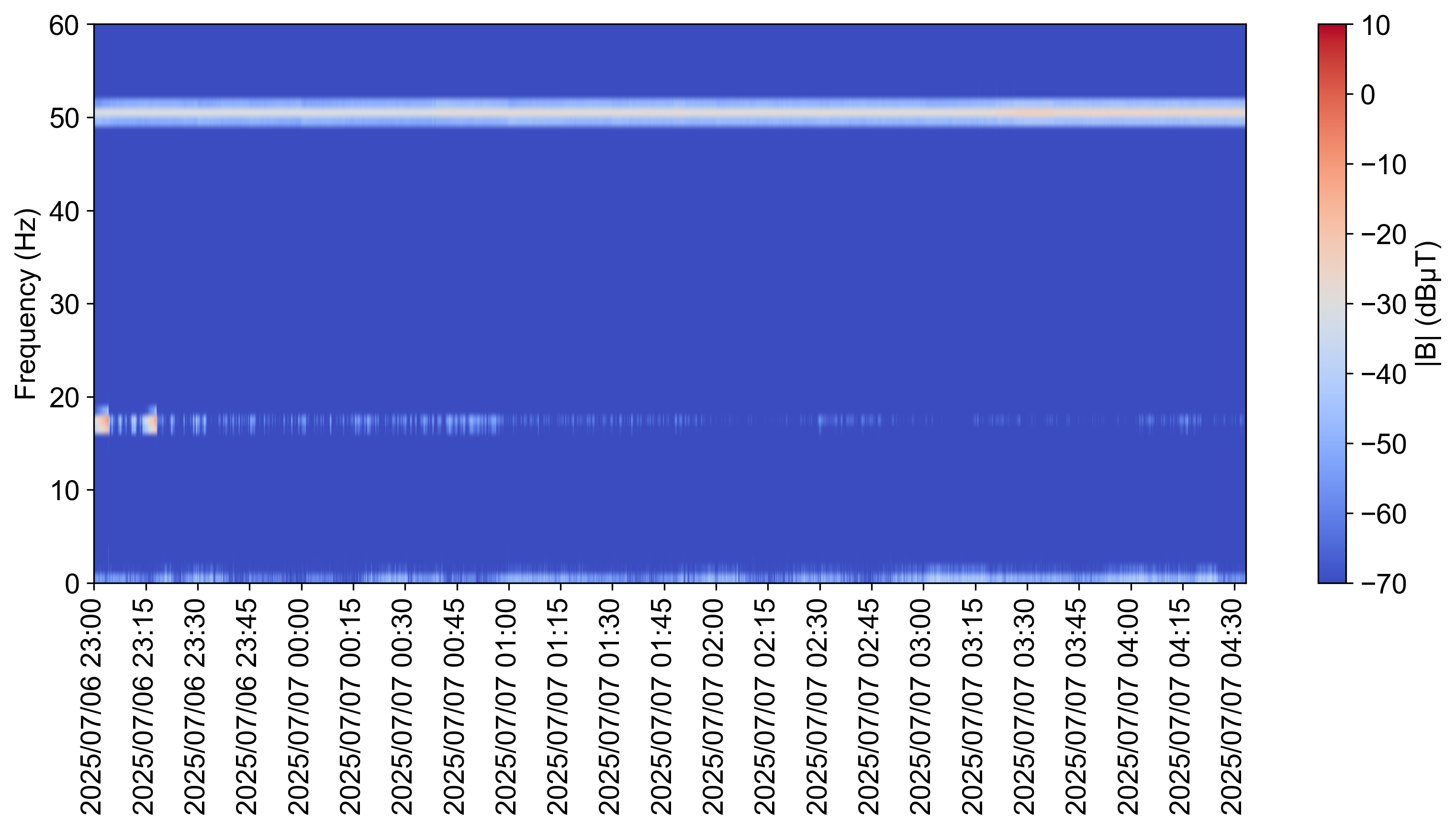}
    \caption{No traffic example - spectrogram.}
    \label{fig:Bottom_spectrogram_notraffic}
\end{figure}
\begin{figure}[h!]
    \centering
    \includegraphics[width=1\linewidth]{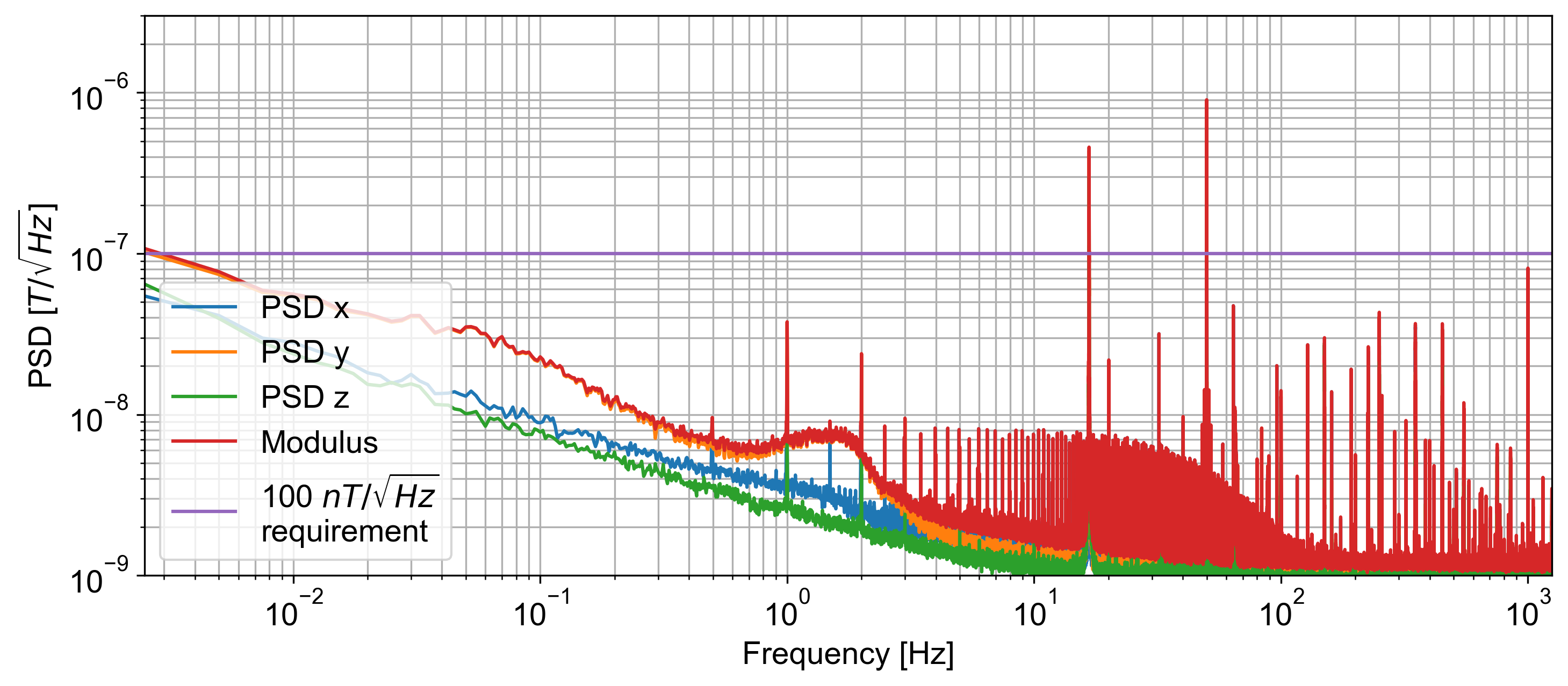}
    \caption{No traffic example - PSD.}
    \label{fig:Bottom_PSD_notraffic}
\end{figure}

% Similarly to the TOP station, we observe "quiet" periods, in other words regular operation, with no additional perturbation (case 1). Summary plots of the magnetic noise are shown in Figures \ref{fig:Bottom_quiet_minmax} and \ref{fig:Bottom_quiet_sigma}. They are based on 165 datasets out of the total of 201, i.e.,  917 hours out of 1117 observed hours for the BOTTOM station. 

\begin{figure}[h!]
    \centering
    \includegraphics[width=1\linewidth]{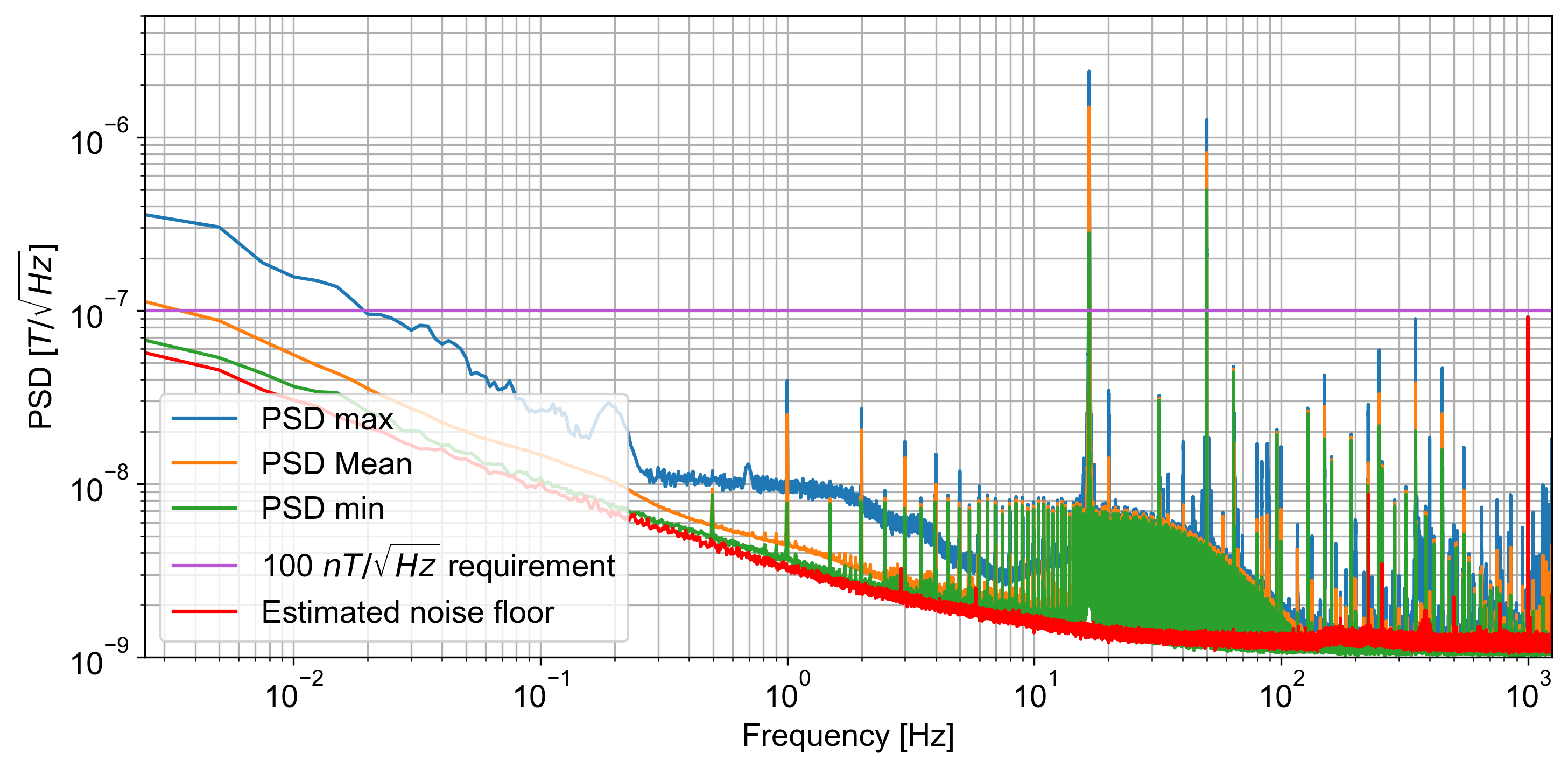}
    \caption{Minimum, mean and maximum PSDs of all 165 "quiet" datasets measured at the BOTTOM station.}
    \label{fig:Bottom_quiet_minmax}
\end{figure}
\begin{figure}[h!]
    \centering
    \includegraphics[width=1\linewidth]{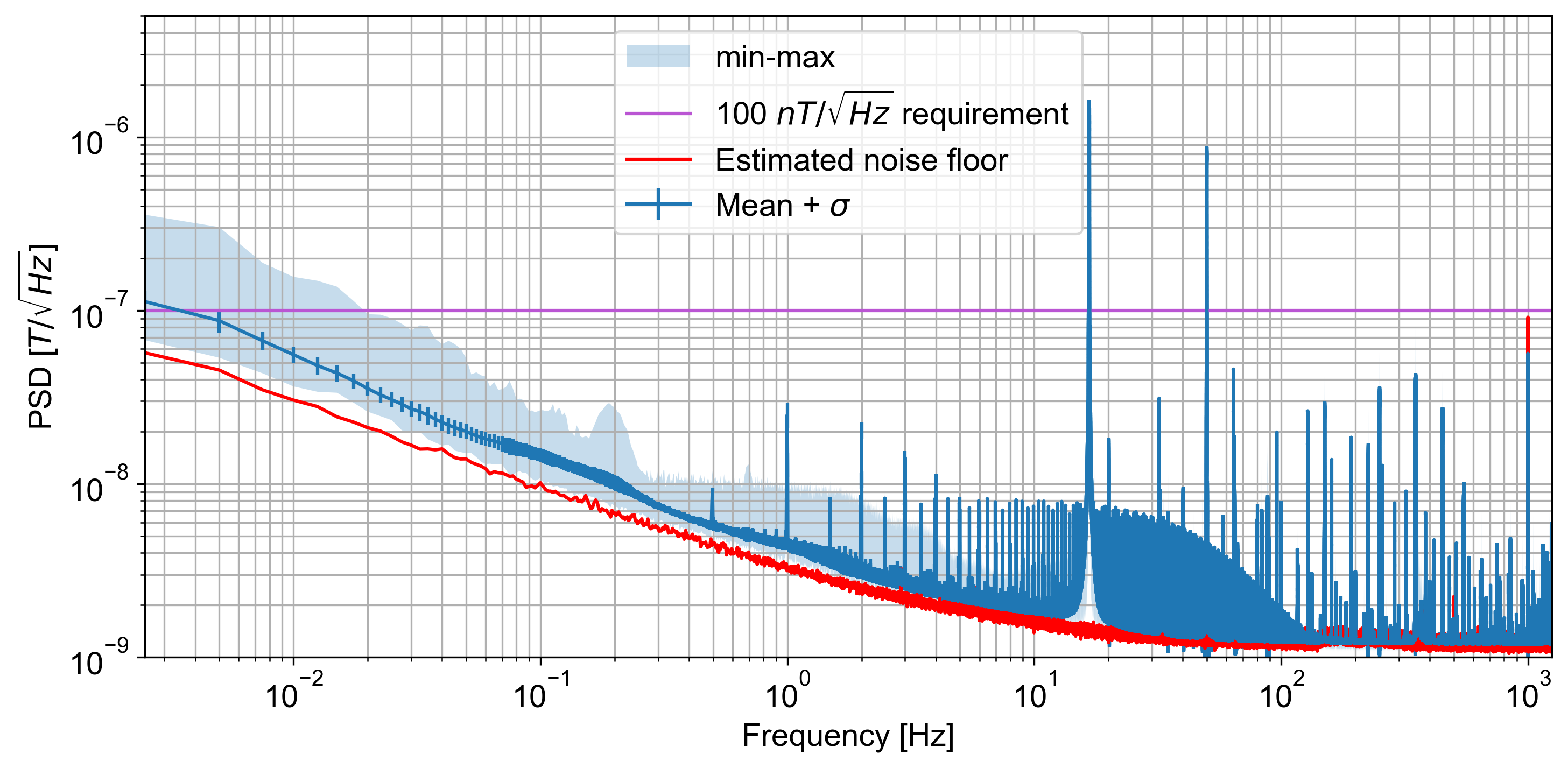}
    \caption{PSD distributions of all 165 "quiet" datasets measured at the BOTTOM station.}
    \label{fig:Bottom_quiet_sigma}
\end{figure}

% The elevator operation has a very different signature at the BOTTOM station from that the TOP station. When the elevator cage is parked there, a permanent magnetic field perturbation is present until the elevator is removed as can be seen in Fig. \ref{fig:Bottom_elevator_perturbation}.

\begin{figure}[h!]
    \centering
    \includegraphics[width=1\linewidth]{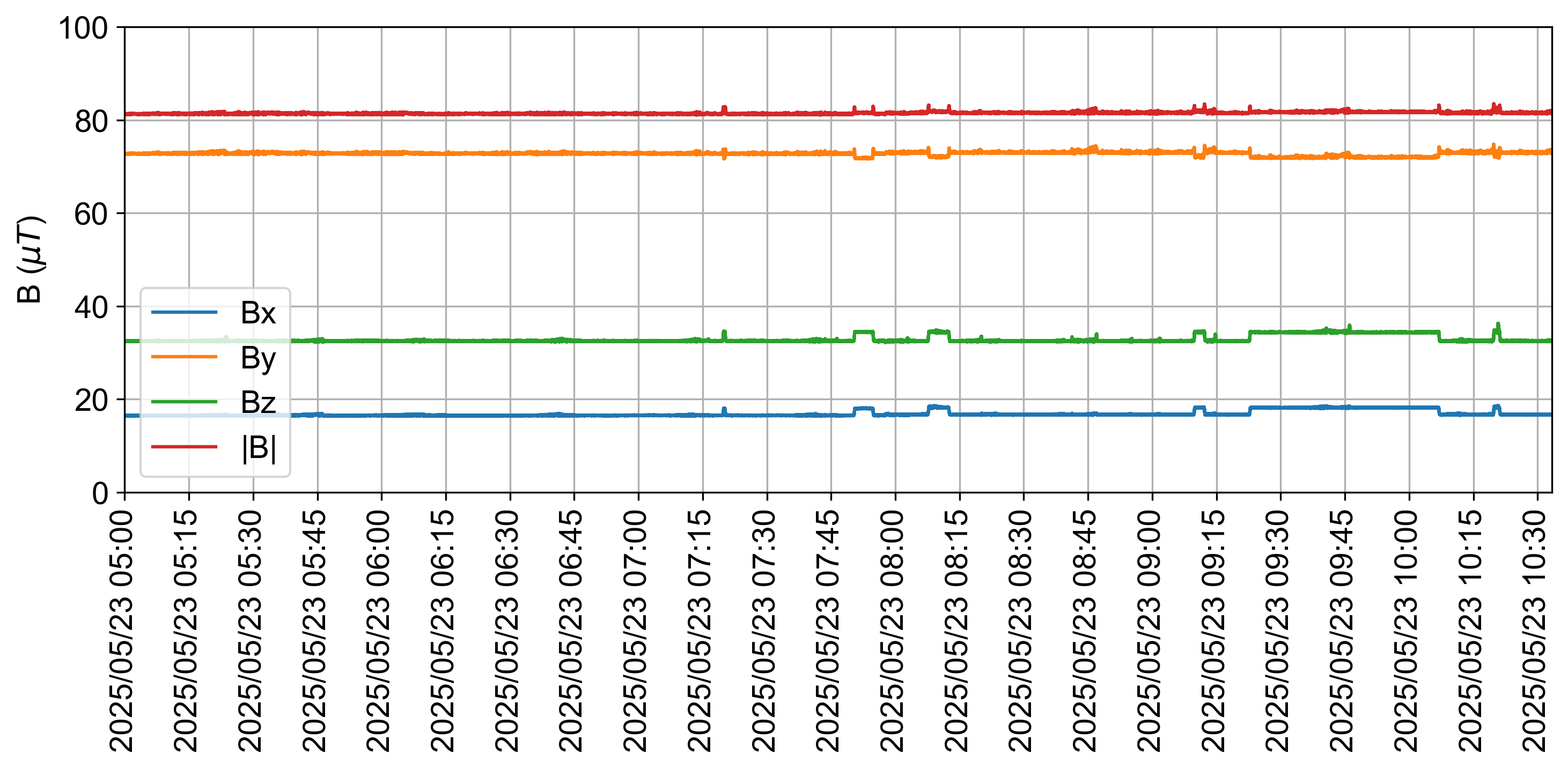}
    \caption{Perturbations of the magnetic field at the BOTTOM station when the elevator is stationary there.}
    \label{fig:Bottom_elevator_perturbation}
\end{figure}

\begin{figure}[h!]
    \centering
    \includegraphics[width=1\linewidth]{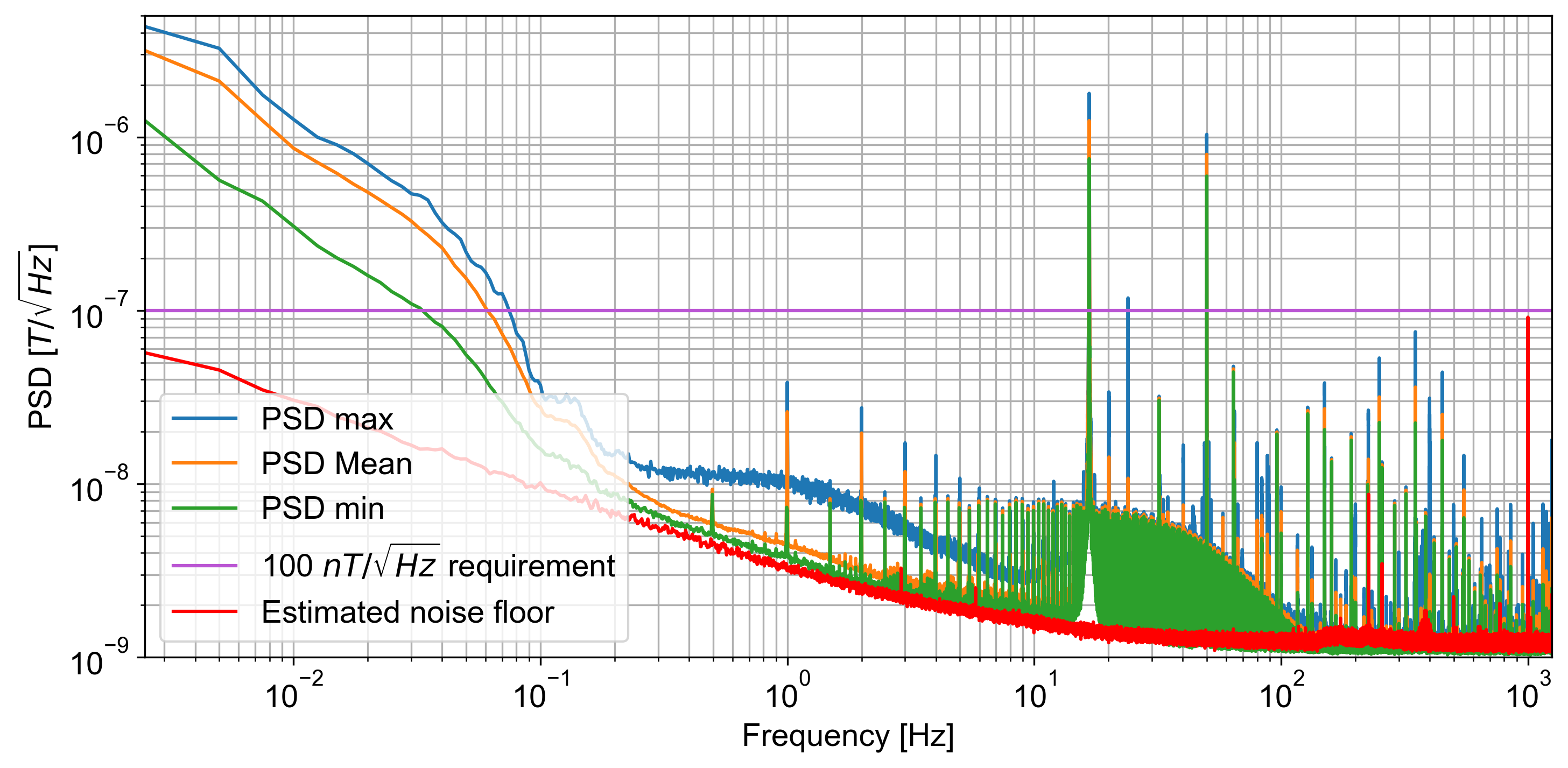}
    \caption{Minimum, mean and maximum PSDs of all  34 datasets recorded at the BOTTOM station while the MFSelevator was being operated.}
    \label{fig:Bottom_elevator_minmax}
\end{figure}
\begin{figure}[h!]
    \centering
    \includegraphics[width=1\linewidth]{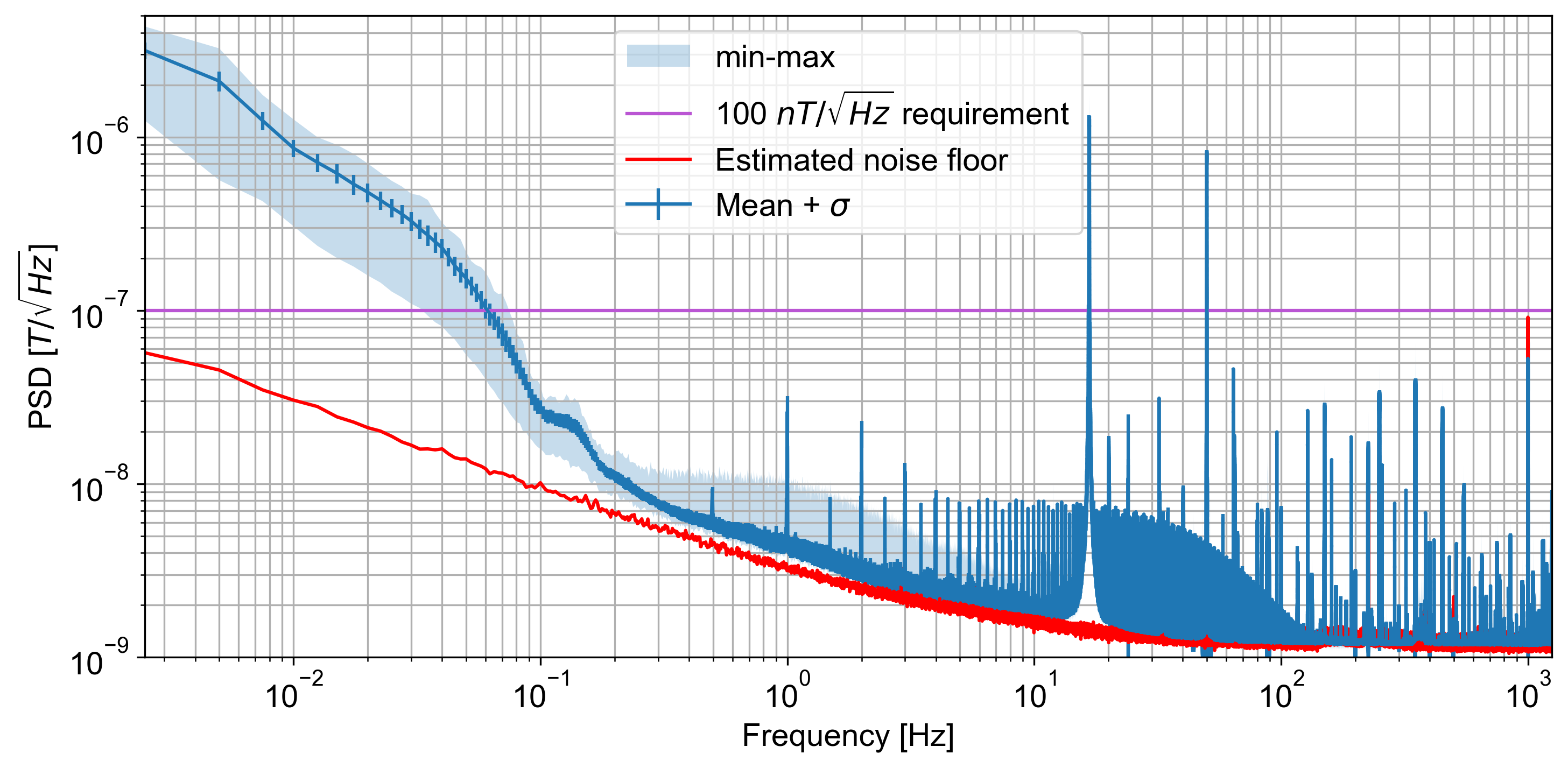}
    \caption{PSD distributions of all 34 datasets recorded at the BOTTOM station while the MFS elevator was being operated.}
    \label{fig:Bottom_elevator_sigma}
\end{figure}

\end{document}